\documentclass[twocolumn]{aastex62}
\usepackage{amsmath}     
\usepackage{wasysym}           
\usepackage{graphicx}
\usepackage{amssymb}
\usepackage{epstopdf}
\usepackage{mathrsfs}
\usepackage{anyfontsize}
\usepackage{lineno}
\usepackage{natbib}

\shorttitle{Stars Crushed by Black Holes I}
\shortauthors{Norman, Nixon \& Coughlin}

\begin{document}

\title{Stars Crushed by Black Holes.~I.~On the Energy Distribution of Stellar Debris in Tidal Disruption Events}

\correspondingauthor{C.~J.~Nixon}
\email{cjn@leicester.ac.uk}

\author{S.M.J.~Norman}
\affiliation{Department of Physics and Astronomy, University of Leicester, Leicester, LE1 7RH, UK}

\author[0000-0002-2137-4146]{C.J.~Nixon}
\affiliation{Department of Physics and Astronomy, University of Leicester, Leicester, LE1 7RH, UK}

\author[0000-0003-3765-6401]{Eric R.~Coughlin}
\affiliation{Department of Physics, Syracuse University, Syracuse, NY 13244, USA}

\begin{abstract}
The distribution of orbital energies imparted into stellar debris following the close encounter of a star with a supermassive black hole is the principal factor in determining the rate of return of debris to the black hole, and thus in determining the properties of the resulting lightcurves from such events. We present simulations of tidal disruption events for a range of $\beta\equiv r_{\rm t}/r_{\rm p}$ where $r_{\rm p}$ is the pericentre distance and $r_{\rm t}$ the tidal radius. We perform these simulations at different spatial resolutions to determine the numerical convergence of our models. We compare simulations in which the heating due to shocks is included or excluded from the dynamics. For $\beta \lesssim 8$ the simulation results are well-converged at sufficiently moderate-to-high spatial resolution, while for $\beta \gtrsim 8$ the breadth of the energy distribution can be grossly exaggerated by insufficient spatial resolution. We find that shock heating plays a non-negligible role only for $\beta \gtrsim 4$, and that typically the effect of shock heating is mild. We show that self-gravity can modify the energy distribution over time after the debris has receded to large distances for all $\beta$. Primarily, our results show that across a range of impact parameters, while the shape of the energy distribution varies with $\beta$, the width of the energy spread imparted to the bulk of the debris is closely matched to the canonical spread, $\Delta E = GM_\bullet R_\star/r_{\rm t}^2$, for the range of $\beta$ we have simulated.
\end{abstract}


\section{{Introduction}}
The accretion flares produced by tidal disruption events (TDEs) have been observed with ever-increasing frequency since their discovery by \emph{ROSAT} in the mid-1990s \citep{Komossa:2015aa,Gezari:2021aa}. Survey science is uncovering dozens of new events per year, and this rate is only expected to increase in the present decade due to wide-field, all-sky monitoring facilities such as the Rubin Observatory \citep{Ivezic:2019aa}. 

The rate at which TDEs occur was constrained through theoretical analyses of the collisional Boltzmann equation (i.e., through a kinetic treatment of a collection of stars orbiting about a supermassive black hole) by \citet{Peebles:1972aa,Bahcall:1976aa,Frank:1976aa,Lightman:1977aa} (see also \citealt{Cohn:1978aa}). Since then these seminal analyses have been expanded upon and refined (e.g., \citealt{Magorrian:1999aa,Stone:2016aa}), but the number of events per galaxy per year has been repeatedly and reliably found to be about $10^{-4}$ to $10^{-5}$; this number is consistent with observations in the optical, radio and X-ray \citep{vanVelzen:2020aa,Alexander:2020aa,Saxton:2020aa}. Radial velocity anisotropies and stellar overdensities, conceivably generated through bursts of star formation, could augment this rate in individual galaxies \citep{Stone:2018aa}, as could the presence of supermassive black hole binaries \citep{Chen:2009aa,Wegg:2011aa,Coughlin:2019ab} or eccentric nuclear discs \citep{Madigan:2018aa}.

In addition to depending on the bulk properties of the host galaxy and the mass of the black hole, the pre-marginalized TDE rate (i.e., the TDE distribution function) depends strongly on the pericentre separation between the disrupted star and the supermassive black hole, meaning that it is much more likely for TDEs with $\beta \lesssim 1$ to occur over those with $\beta \gg 1$, where $\beta = r_{\rm t}/r_{\rm p}$ is the ratio of the tidal radius ($r_{\rm t}$) to the pericentre distance of the centre of mass ($r_{\rm p}$). In the limit that the changes in the angular momentum of a star per scattering event are large relative to $\sqrt{2GM_{\bullet}r_{\rm p}}$, the probability distribution that describes the likelihood of generating a TDE with a given $\beta$ varies as $\propto \beta^{-2}$, so that the integrated probability of having a TDE with a $\beta > \beta_{\rm min}$ is $\beta_{\rm min}^{-1}$. In the limit that stars' orbits diffuse into the tidal sphere \citep{Wang:2004aa} the likelihood of having $\beta > 1$ is even smaller because the stars gradually approach the tidal radius in their pericentre.

Nevertheless, while such high-$\beta$ tidal encounters are relatively rare compared to their less-extreme, low-$\beta$ counterparts (which will often result in partial disruptions that remove only a fraction of the stellar envelope during the tidal encounter; \citealt{Guillochon:2013aa, Mainetti:2017aa, Coughlin:2019aa, Miles:2020aa, Nixon:2021ab}), the likelihood of observing such events will increase dramatically as Rubin and other all-sky monitoring facilities come online. It is therefore necessary to understand the nature of the physical processes that ensue during the tidal compression and re-expansion of the stellar material as it is strongly perturbed by the tidal field of the hole in these deeply plunging events. 

To this end \citet{Coughlin:2021aa} demonstrated through the development and implementation of a novel analytical technique that deep TDEs result in an adiabatic increase in the central density of the star until $\beta \simeq 10$, above which the inward propagation of shockwaves sets the maximum-achievable central temperature and density of the star. In this complementary paper we present and analyze the details of numerical, hydrodynamical simulations of tidal disruption events that (1) span a range of $\beta$ values, (2) include and exclude heating of the stellar debris by shocks, and (3) are performed at a broad range of spatial resolutions.

The paper is organised as follows. In Section \ref{sec:numerics} we describe the details of the numerical simulations, e.g., the initial setup, the range of parameters simulated, and the resolution employed. We also discuss our results (Section \ref{sec:results}), including the width of the energy distribution of the tidally disrupted debris (Section \ref{sec:energy}), the effects of spatial resolution (Section \ref{sec:resolution}), the importance of shock heating (Section \ref{sec:shocks}), the temporal dependence of the energy distribution (and therefore the validity of the notion that the energy of the debris is ``frozen-in'' once the disrupted material recedes to distances well outside the tidal sphere of the black hole; Section \ref{sec:time}). We provide further discussion of our results in Section~\ref{sec:discussion}, and give our conclusions in Section~\ref{sec:conclusions}.

\section{Numerical simulations}
\label{sec:numerics}
We present numerical simulations of the tidal disruption of solar-like stars by a supermassive black hole. We perform these simulations using the Smoothed Particle Hydrodynamics code {\sc phantom} \citep{Price:2018aa}, which has been used extensively for TDE simulations over the last few years \citep[e.g.][]{Coughlin:2015aa,Coughlin:2016ab,Coughlin:2016aa,Coughlin:2017aa,Bonnerot:2016aa,Dharba:2019aa,Golightly:2019ab,Golightly:2019aa,Clerici:2020aa,Miles:2020aa,Wang:2021aa}. As we wish to explore the energy distribution produced by different stellar orbits, specifically different $\beta$, and the effect of adding heating due to shocks to the gas, we model the standard TDE: we let the star be a $\gamma=5/3$ polytrope, on a parabolic orbit about the black hole, and take the supermassive black hole to have a mass of $10^6M_\odot$, modelled with Newtonian gravity. We discuss the physical limitations of these approximations in Section~\ref{sec:discussion}.

We perform simulations with varying orbital impact parameter $\beta$, particle number $N_{\rm p}$, and we repeat these simulations with two different thermodynamic approaches. In one set of simulations we model the fluid with a polytropic equation of state, where $P=K\rho^\gamma$ and the entropy function $K$ and adiabatic exponent $\gamma$ are fixed constants that do not vary with time; this is equivalent to evolving the energy equation for the gas but only including the effects of ``adiabatic'' expansion/contraction (i.e., $P{\rm d}V$ work). In the second set of simulations we evolve the energy equation and, in addition to the polytropic case, any kinetic energy that is dissipated by (e.g.) shocks is turned into heat and retained within the gas, meaning that $K$ becomes a function of both space and time. We do not impose an explicit physical viscosity in these simulations and thus any dissipation of kinetic energy--- which may be due to physical effects such as shocks, but also numerical effects such as the mixing of shearing flows where the lengthscale is smaller than the local resolution lengthscale---is mediated by the SPH numerical viscosity, which here takes the form of a constant quadratic viscosity with $\beta^{\rm AV}=2$ and a variable linear viscosity following the \cite{Cullen:2010aa} switch with $\alpha^{\rm AV}_{\rm min}=0.01$ and $\alpha^{\rm AV}_{\rm max}=1$ \citep[see][for details]{Price:2018aa}. For the simulations that include shock heating of the gas, we also include a numerical conductivity to ensure that the internal energy remains continuous in the flow; for this we employ the standard value of $\alpha_u = 1$ \citep{Price:2018aa}.\footnote{To test whether the numerical conductivity has any noticeable effect on the dynamics (for example on the energy distribution) we performed an additional set of simulations (not depicted) in which the energy equation was evolved with numerical conductivity included, but with heating from the numerical viscosity excluded. The results in this case should be the same as those in which a polytropic equation of state is enforced exactly, with any differences coming from a combination of errors associated with numerical integration of the energy equation and the effects of numerical conductivity. We found no discernible differences between these sets of simulations.}

\begin{figure*}
	\centering
	\includegraphics[width=0.4\textwidth]{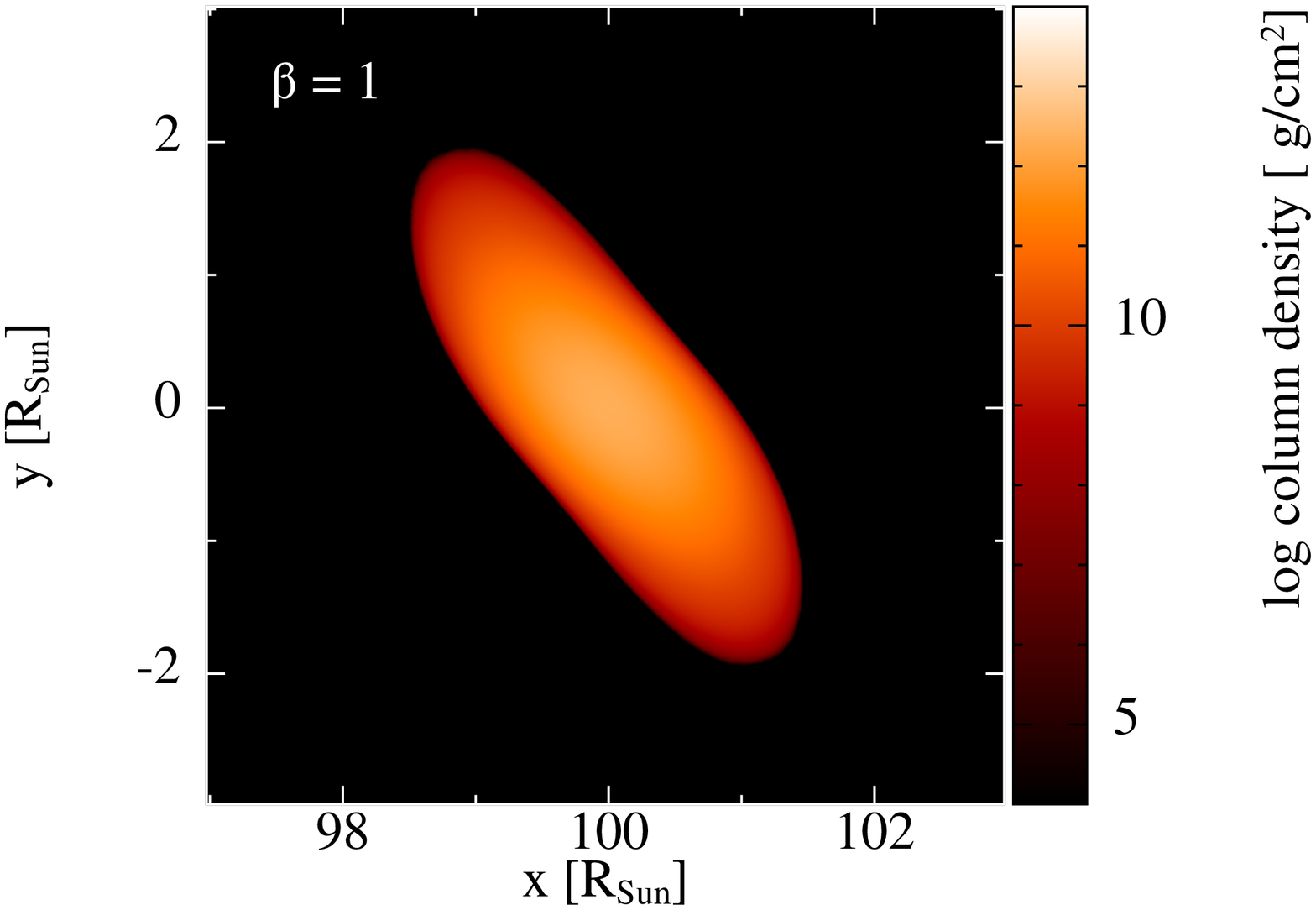}\hspace{0.3in}
	\includegraphics[width=0.4\textwidth]{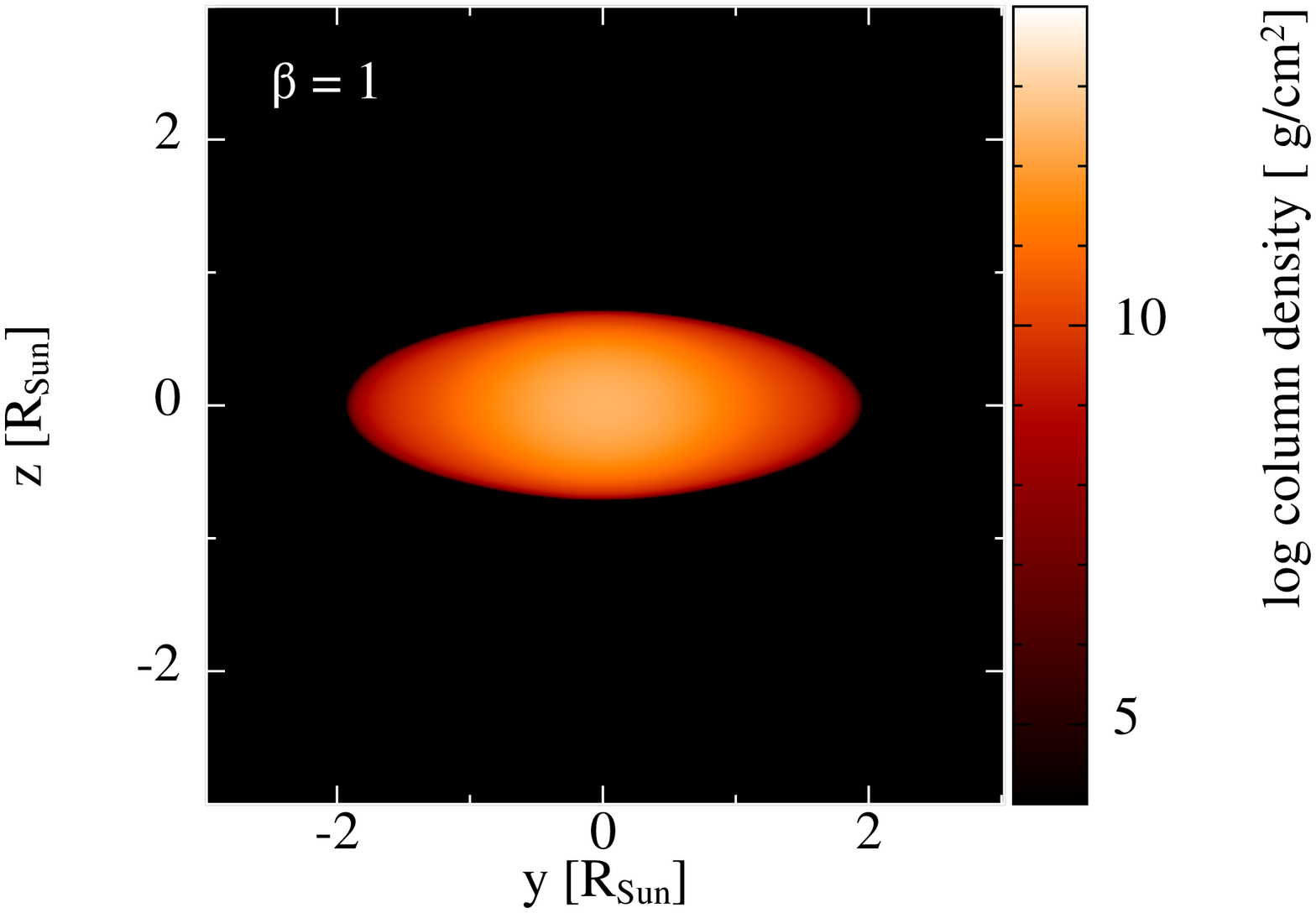}
	\includegraphics[width=0.4\textwidth]{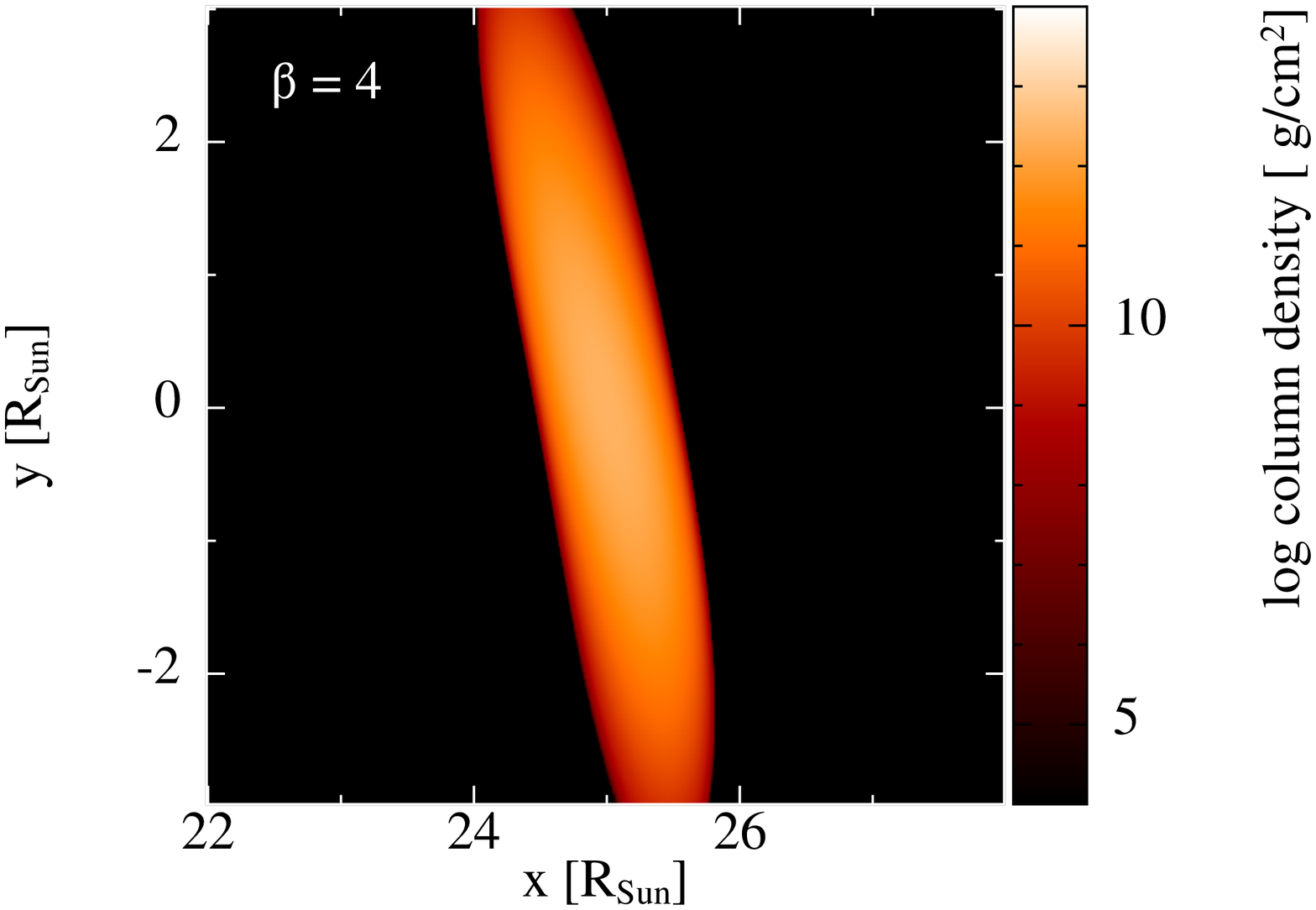}\hspace{0.3in}
	\includegraphics[width=0.4\textwidth]{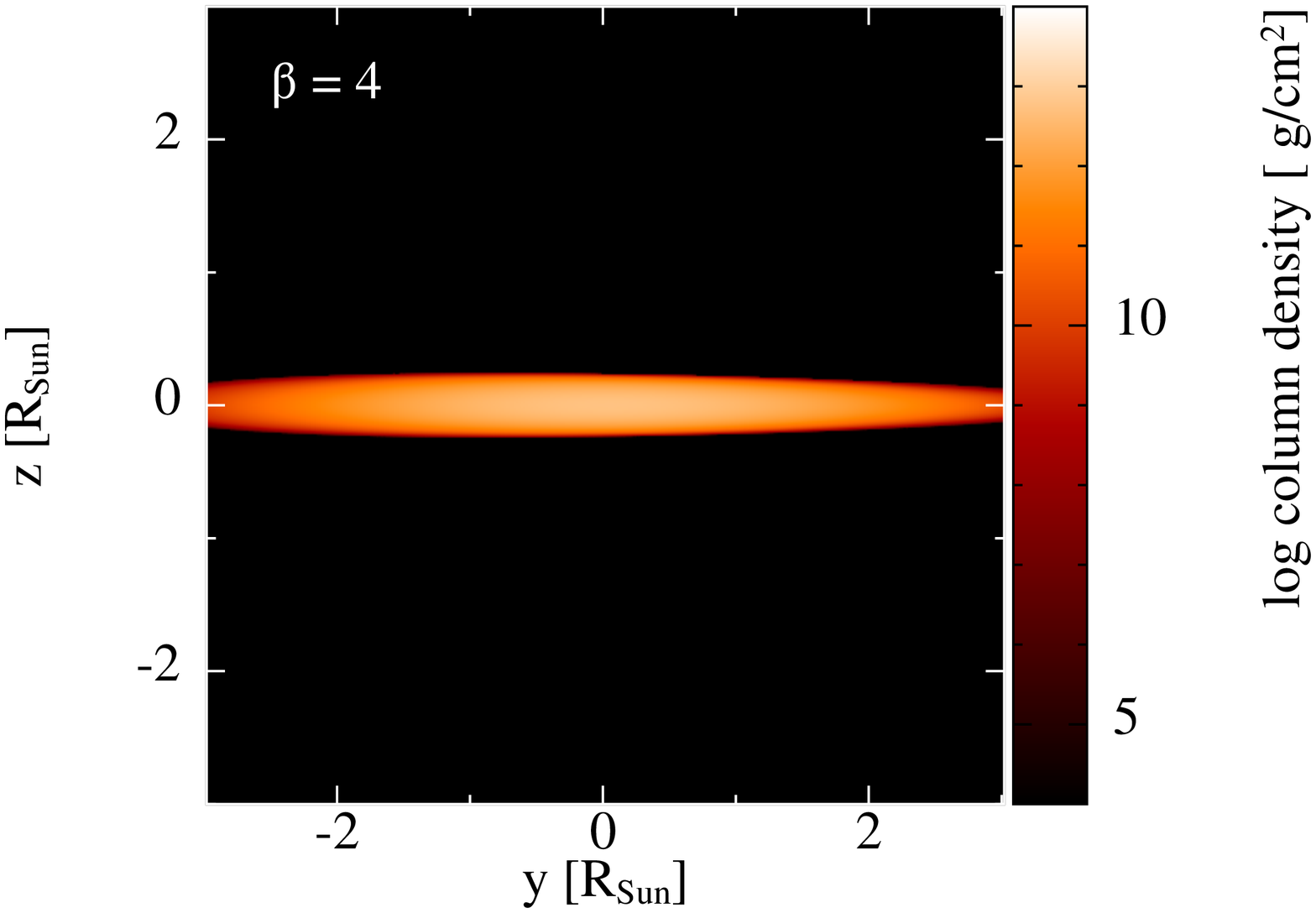}
	\includegraphics[width=0.4\textwidth]{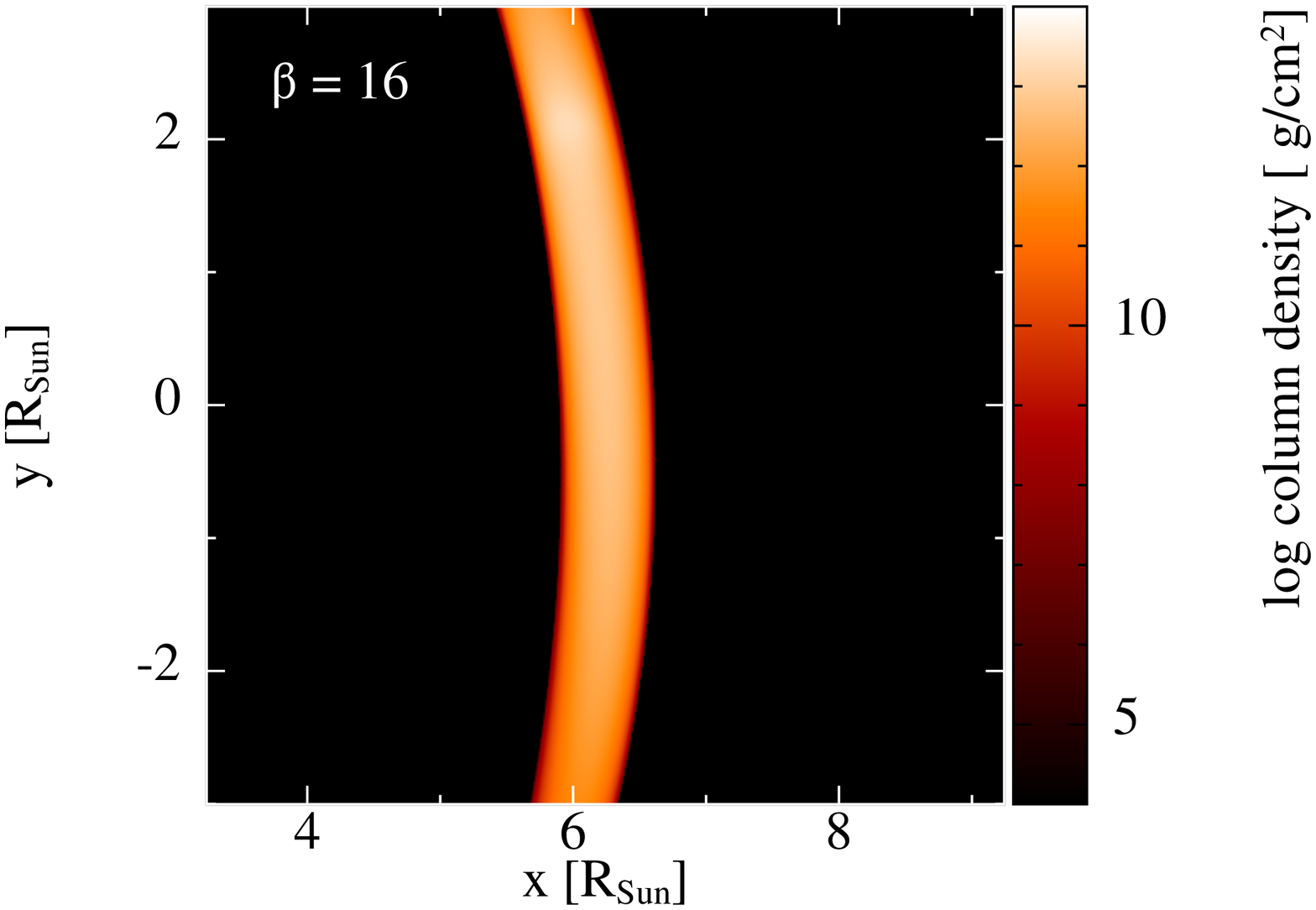}\hspace{0.3in}
	\includegraphics[width=0.4\textwidth]{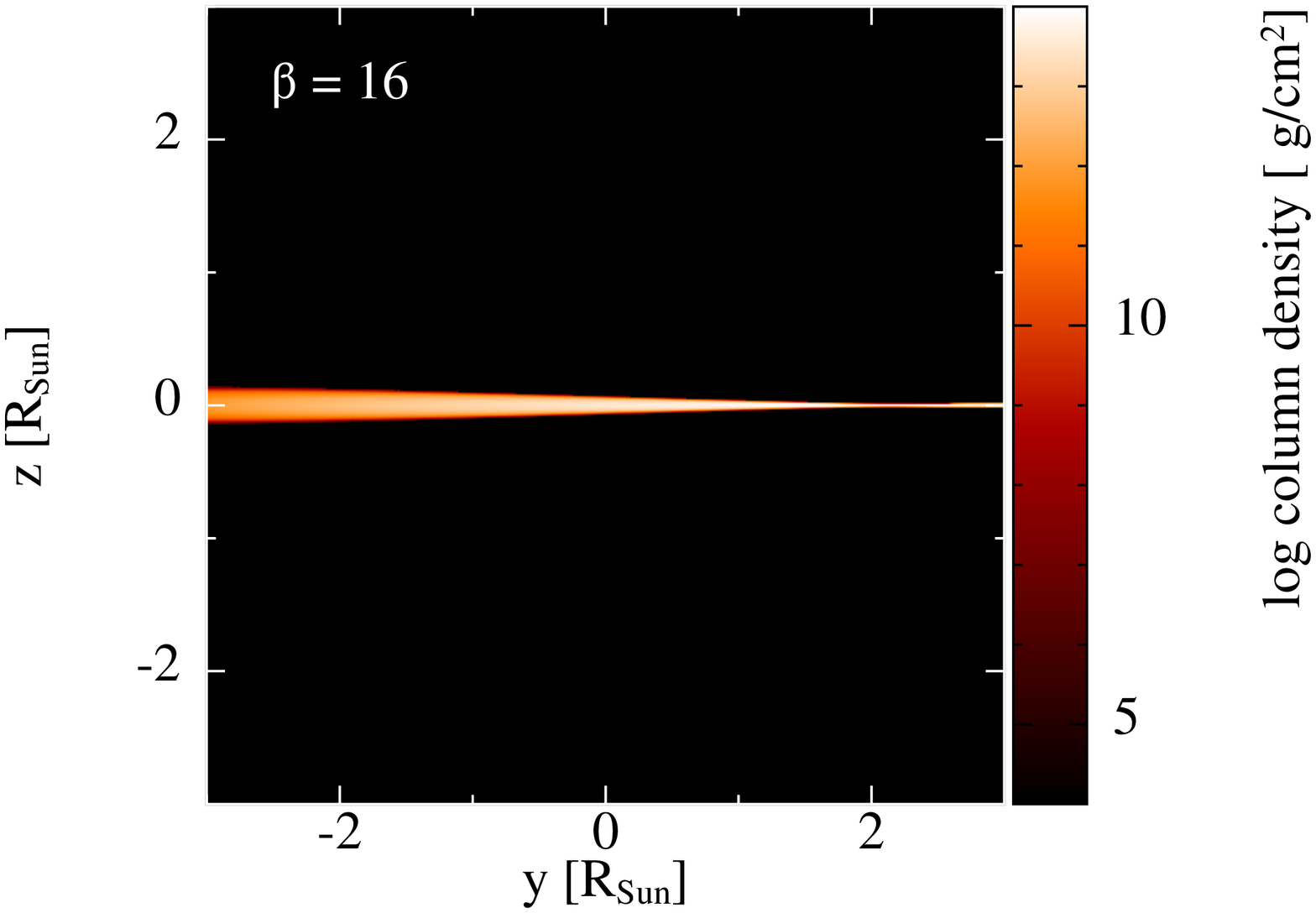}
	\caption{Column density renderings of the $\beta=1$, $4$ and $16$ simulations performed with $128$M particles when the centre-of-mass orbit for the star reaches pericentre. The left column shows the distribution in the orbital plane, and the right column shows the view across the orbital plane. The centre of mass of the star starts in the negative $x$ and $y$ quadrant, and reaches pericentre with a positive $x$ position and $y=0$. Stellar material at positive $y$ values therefore has already passed pericentre, and negative $y$ values are yet to reach the pericentre of the original stellar orbit. The top panels correspond to $\beta=1$, the middle panels to $\beta=4$, and the bottom panels to $\beta=16$. In each case we have restricted the view to the position of the centre of mass $\pm 3R_\odot$, which for larger $\beta$ does not quite encompass the entire debris stream, with larger $\beta$ corresponding to more stretching of the star by the time pericentre is reached. As $\beta$ is increased the star is increasingly crushed into the orbital plane, with the thickness for $\beta=16$ being $\sim 0.01-0.1\,R_\odot$ (the spread in vertical height is due to the fact that the point of maximum compression occurs slightly post-pericentre, and most of the star has not yet reached this point).}
	\label{fig0}
\end{figure*}

We simulate five different values of $\beta$ corresponding to $\beta = 1$, $2$, $4$, $8$ and $16$. We vary the resolution of the simulations corresponding to $N_{\rm p} = 250$k, $2$M, $16$M and $128$M particles.\footnote{As the star is initially cut from a cube of particles before being stretched to the desired density profile \citep{Price:2018aa}, the actual numbers of particles are 250,045, 2,000,491, 16,003,537 \& 128,024,080 respectively.} In Fig.~\ref{fig0} we show the star as the centre of mass reaches pericentre for $\beta = 1$, $4$ and $16$, each at $128$M particles. For small $\beta$ the star is significantly distorted at pericentre, while for large $\beta$ the star is crushed into the orbital plane by the tidal field of the black hole. In the following subsections we report our results with respect to (1) the width of the energy distribution with varying impact parameter measured at different times post-pericentre, (2) the convergence of the numerical simulations in terms of the energy distribution measured when the zero-energy orbit has reached a distance of $5r_{\rm t}$ from the black hole, (3) the evolution of the energy distributions measured at different times after the pericentre passage of the stars, (4) the impact of shock heating on the energy distribution at different $\beta$ values, and (5) the entropy generation and maximum density and temperature attained in the simulations as the star passes pericentre as a function of impact parameter and resolution.

\subsection{Results}
\label{sec:results}

\subsubsection{The shape of the energy distribution}
\label{sec:energy}
\begin{figure*}
	\centering
	\includegraphics[width=0.33\textwidth]{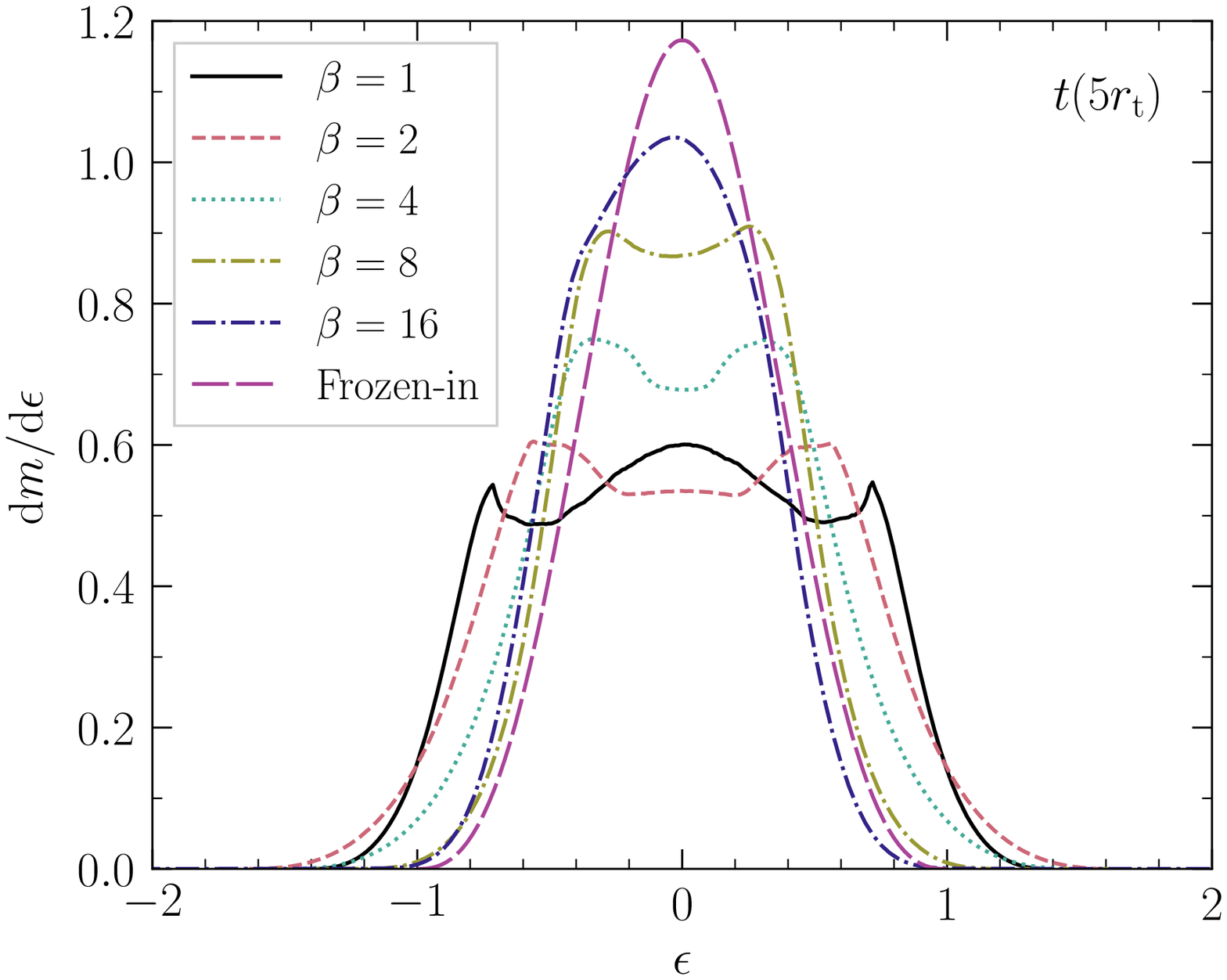}\hfill
	\includegraphics[width=0.33\textwidth]{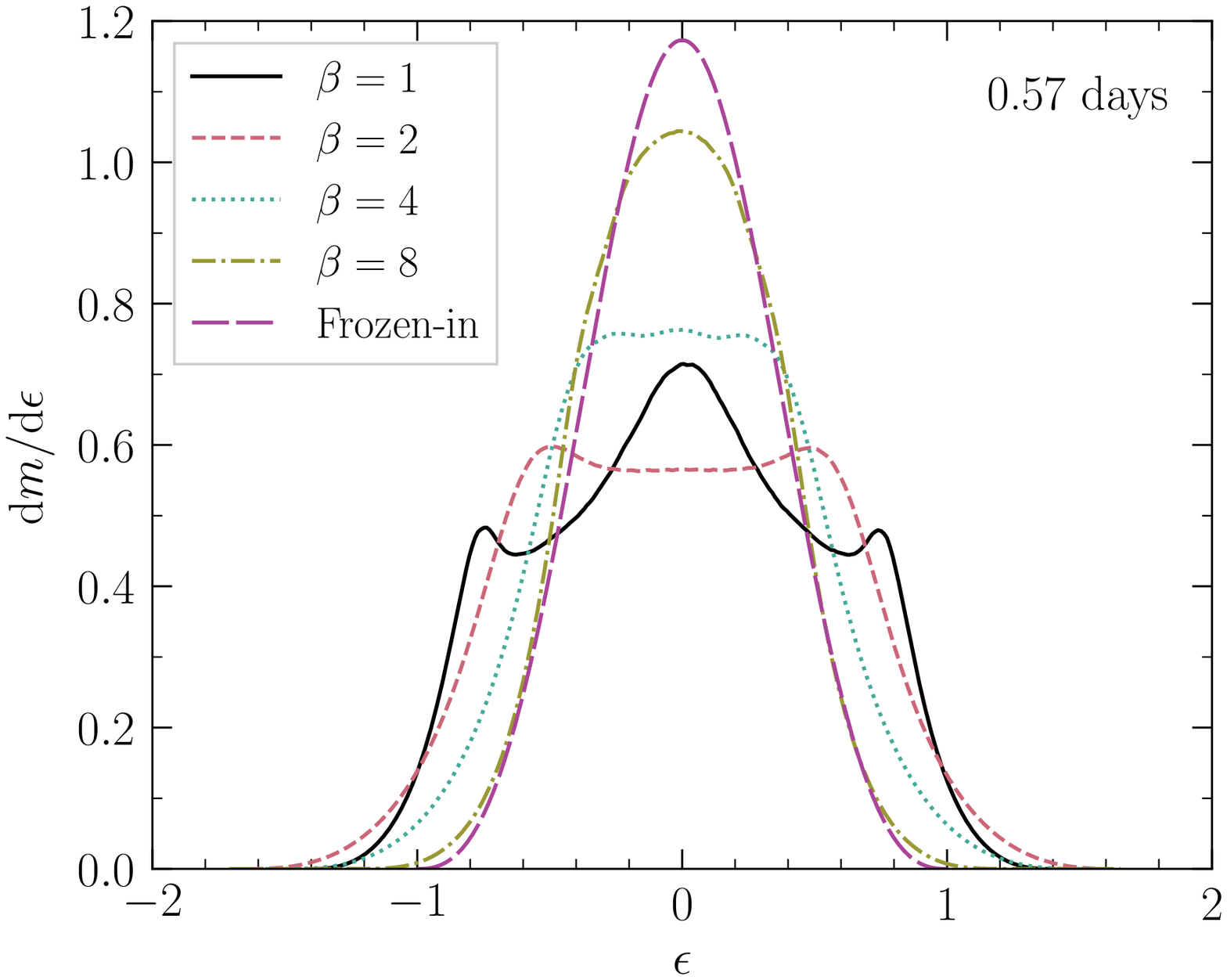}\hfill
	\includegraphics[width=0.33\textwidth]{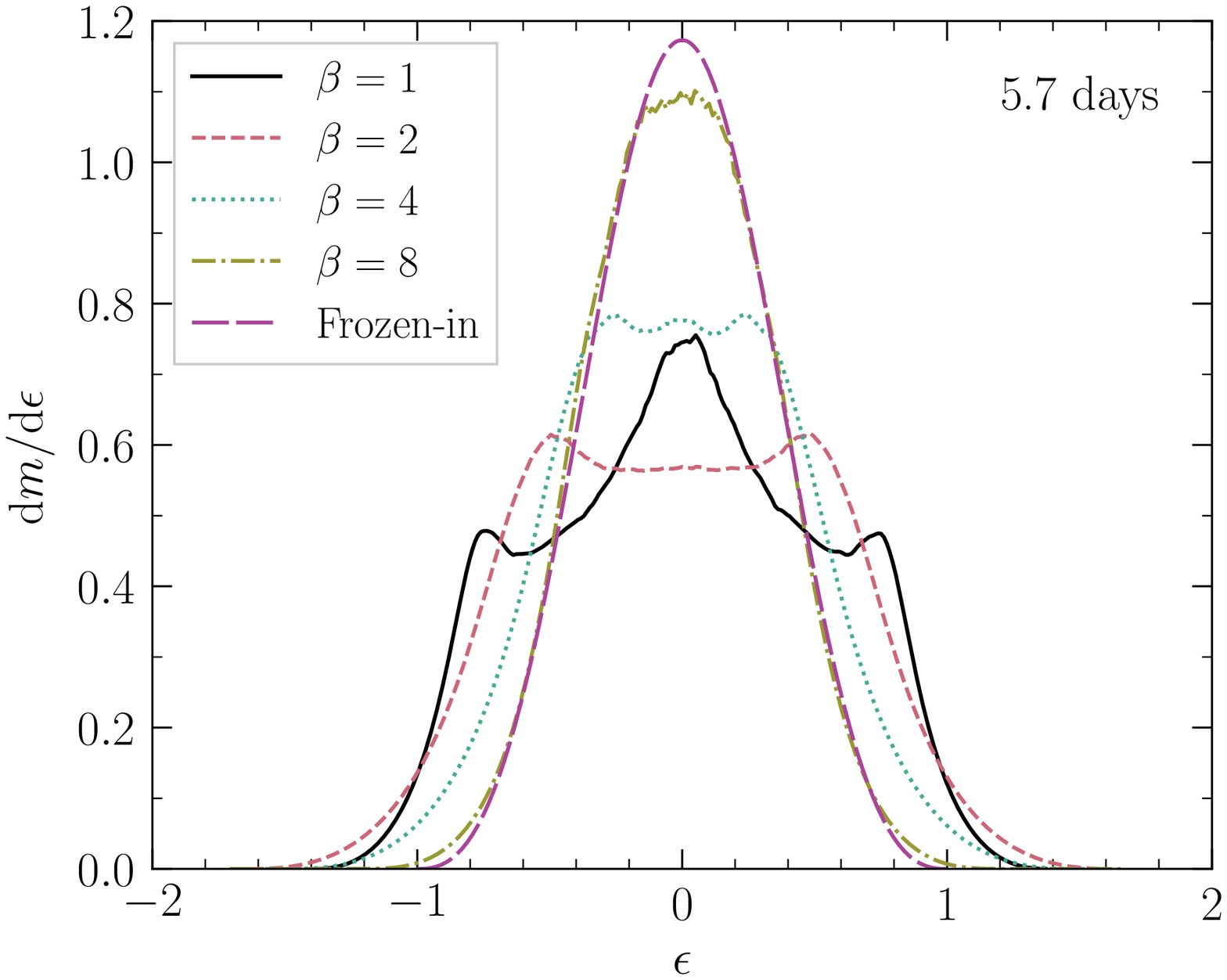}
	\caption{Comparison of the energy distributions for the simulations with $N_{\rm p} = 128$M and a polytropic equation of state for different $\beta$ values. The Keplerian energy $E$ has been normalised by the energy spread predicted by the frozen-in approximation with $\epsilon = E/\Delta E$ where $E = v^2/2-GM_\bullet/r$ is the Keplerian energy, and $\Delta E = GM_\bullet R_\star/r_{\rm t}^2$. A similar normalisation is applied to the vertical axis with ${\rm d}m/{\rm d}\epsilon = (\Delta E / M_\star){\rm d}M/{\rm d}E$. In each panel the time at which the energy distribution is computed is given in the top right corner, with $t=t(5r_{\rm t})$ referring to the time at which the zero-energy orbit has receded to a distance of $5r_{\rm t}$ from the black hole; this occurs between 2.9-2.4\,hr post-pericentre for these $\beta$. The later times of 0.57\,d and 5.7\,d are significantly larger than the dynamical timescale $\sqrt{R_\star^3/GM_\star}\approx 27$\,minutes. On each panel the value of $\beta$ corresponding to each line colour is given in the legend in the top left of the panel. In each case the majority of the debris is confined to a region $|\epsilon| \lesssim 1$, with the full range having $|\epsilon| \lesssim 2$ for these $\beta$ values.}
	\label{fig1}
\end{figure*}

The energy distributions for different values of the impact parameter $\beta$ and at several times post-pericentre are given in Fig.~\ref{fig1}. The left panel corresponds to a time at which the zero-energy orbit has receded to a distance of $5r_{\rm t}$ from the black hole, the middle panel corresponds to a time of $10^4 GM/c^3 \approx 14$\,hr and the right panel corresponds to a time of $10^5 GM/c^3 \approx 5.7$\,days. We have normalised the energy by the canonical energy spread, i.e., $\epsilon = E/\Delta E$ where $E = v^2/2-GM_\bullet/r$ is the Keplerian energy\footnote{We note that while a Keplerian orbital energy is always a well-defined quantity, the orbital dynamics is not Keplerian when one accounts for the self-gravity of the material (which also includes the possible existence of a self-bound core; \citealt{Coughlin:2019aa}) and hence the Keplerian orbital energy is not a conserved quantity. We show this explicitly below.} and $\Delta E = GM_\bullet R_\star/r_{\rm t}^2$. Note that $2\Delta E$ is the spread of orbital energies across the star if each fluid element moves precisely with the centre of mass when the centre of mass crosses the tidal sphere \citep{Lacy:1982aa}. A similar normalisation is applied to the energy distribution with  ${\rm d}m/{\rm d}\epsilon = (\Delta E / M_\star){\rm d}M/{\rm d}E$. We also include on the plot the prediction of the ``frozen-in'' model; this model is most similar to the high-$\beta$ cases and while it provides a reasonable order-of-magnitude estimate for lower-$\beta$ it is generally not correct in detail.

This figure shows that, in all cases, the energy distribution is closely centred around $E=0$ and is approximately symmetric about this point (except for $\beta=16$ which displays a noticeable asymmetry). The asymmetry for $\beta=16$ that is apparent in the left panel of this figure is driven by a combination of (1) the differing times at which different fluid elements reach their point of maximal compression---this occurs at approximately the same true anomaly which is reached by different parts of the stream at different times---and (2) the influence of self-gravity, which is particularly strong for the maximally compressed material. As the leading part of the star reaches the point of maximum compression first, it is this region that maximises the effect of self-gravity and thus `steals' some of the mass that would otherwise be distributed more evenly in the energy distribution; we would expect this asymmetry to persist at higher $\beta$. Further, Fig.~\ref{fig1} shows that the breadth of the energy distribution is $\approx \Delta E$ largely independent of $\beta$.\footnote{This result is not likely to hold for partial disruptions with $\beta \ll \beta_{\rm c}$ in which the core of the star is not disrupted during pericentre passage as the gravitational tidal influence is typically not strong enough to overcome the stellar gravity and therefore to impart this energy spread in this case. See, e.g., \cite{Nixon:2021ab}.} However, we can also see that there is a dependence on $\beta$ to the overall shape of the energy distribution. Typically larger $\beta$ corresponds to more concentrated distributions with more debris with, for example, $|E| < 0.5\Delta E$, while smaller $\beta$ show stronger self-gravitating ``shoulders'' that persist to later times \citep[see also][]{Coughlin:2015aa}. 

\subsubsection{Varying the spatial resolution}
\label{sec:resolution}
\begin{figure*}
	\includegraphics[width=0.28\textwidth]{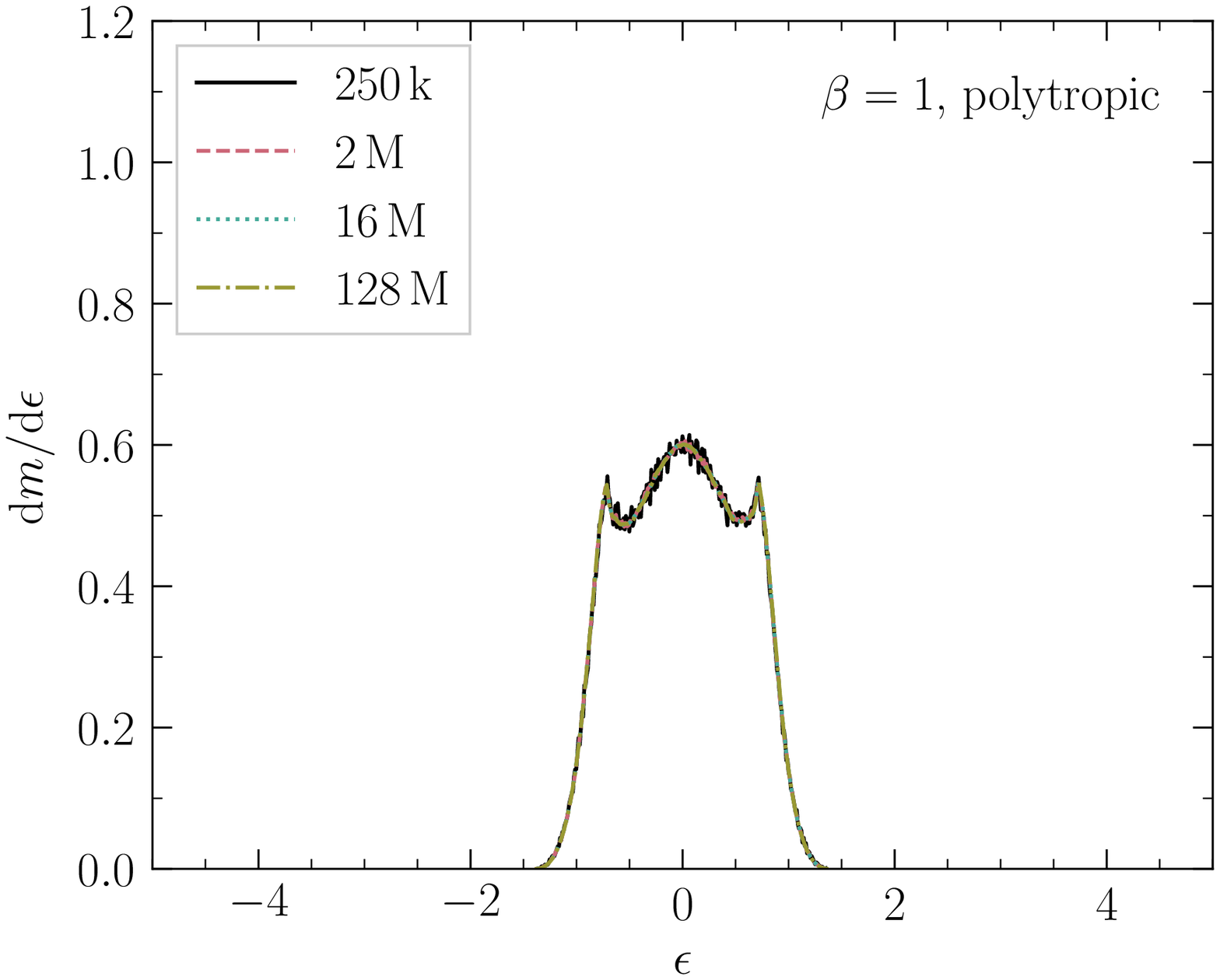}\hfill
	\includegraphics[width=0.28\textwidth]{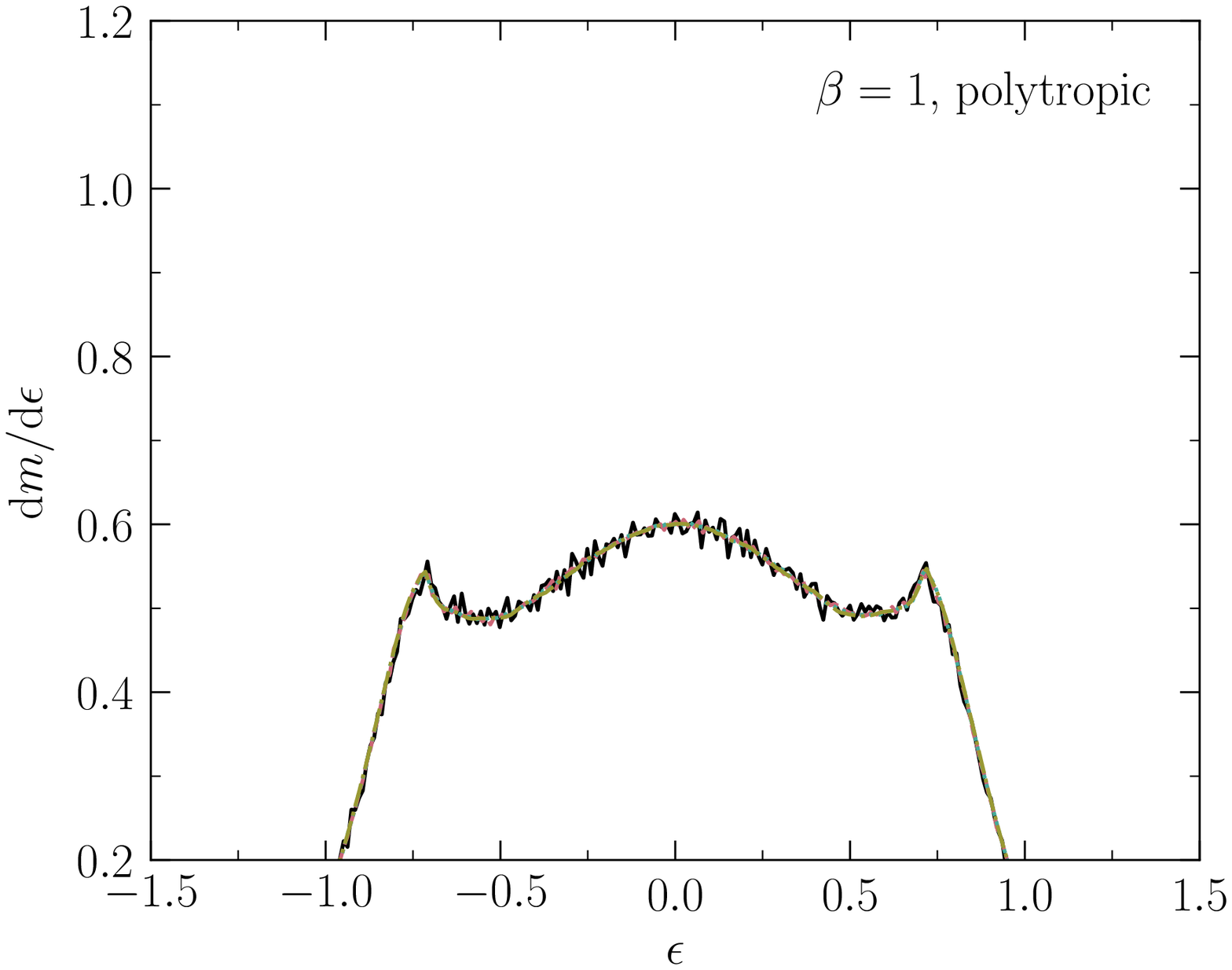}\hfill
	\includegraphics[width=0.28\textwidth]{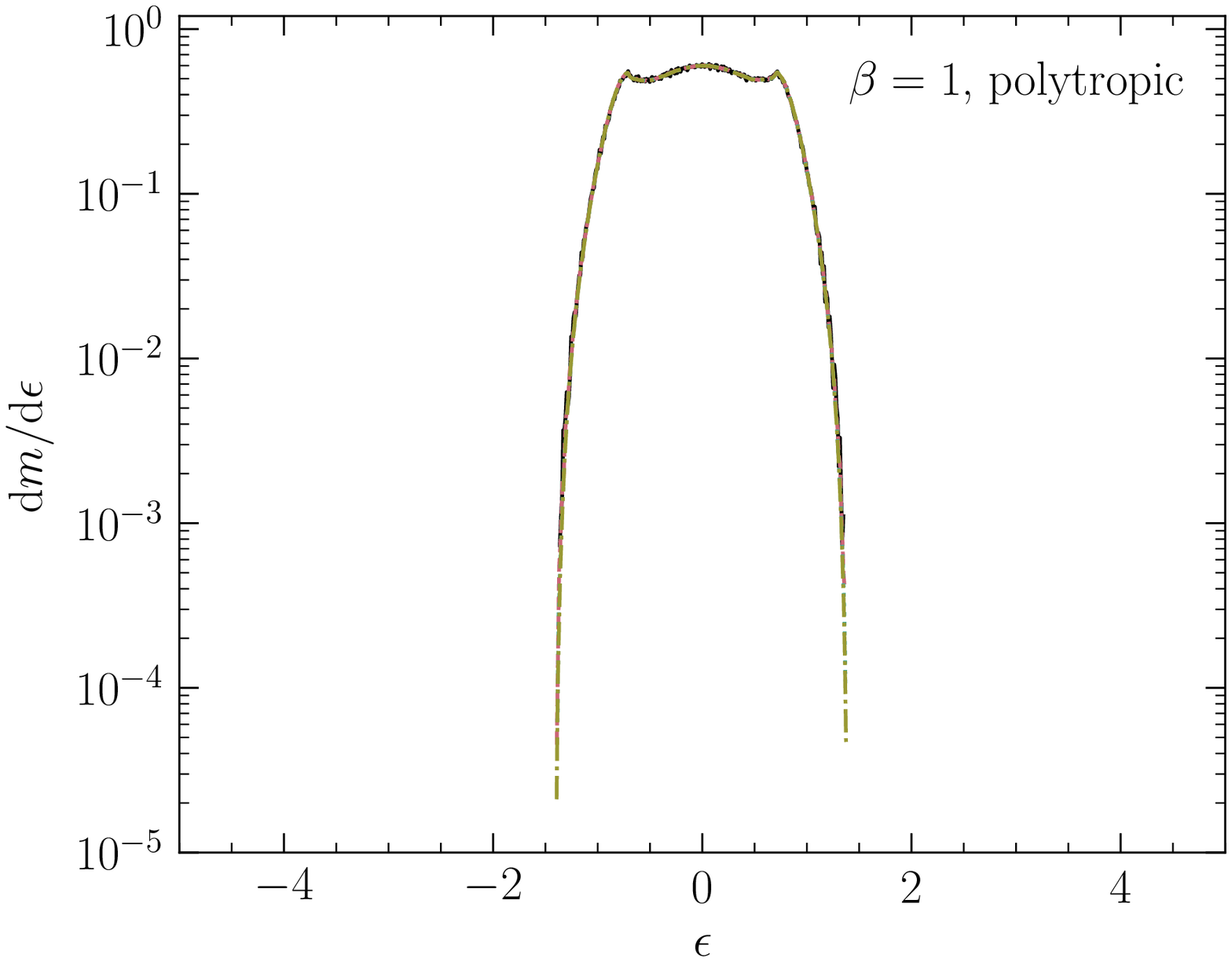}
	\includegraphics[width=0.28\textwidth]{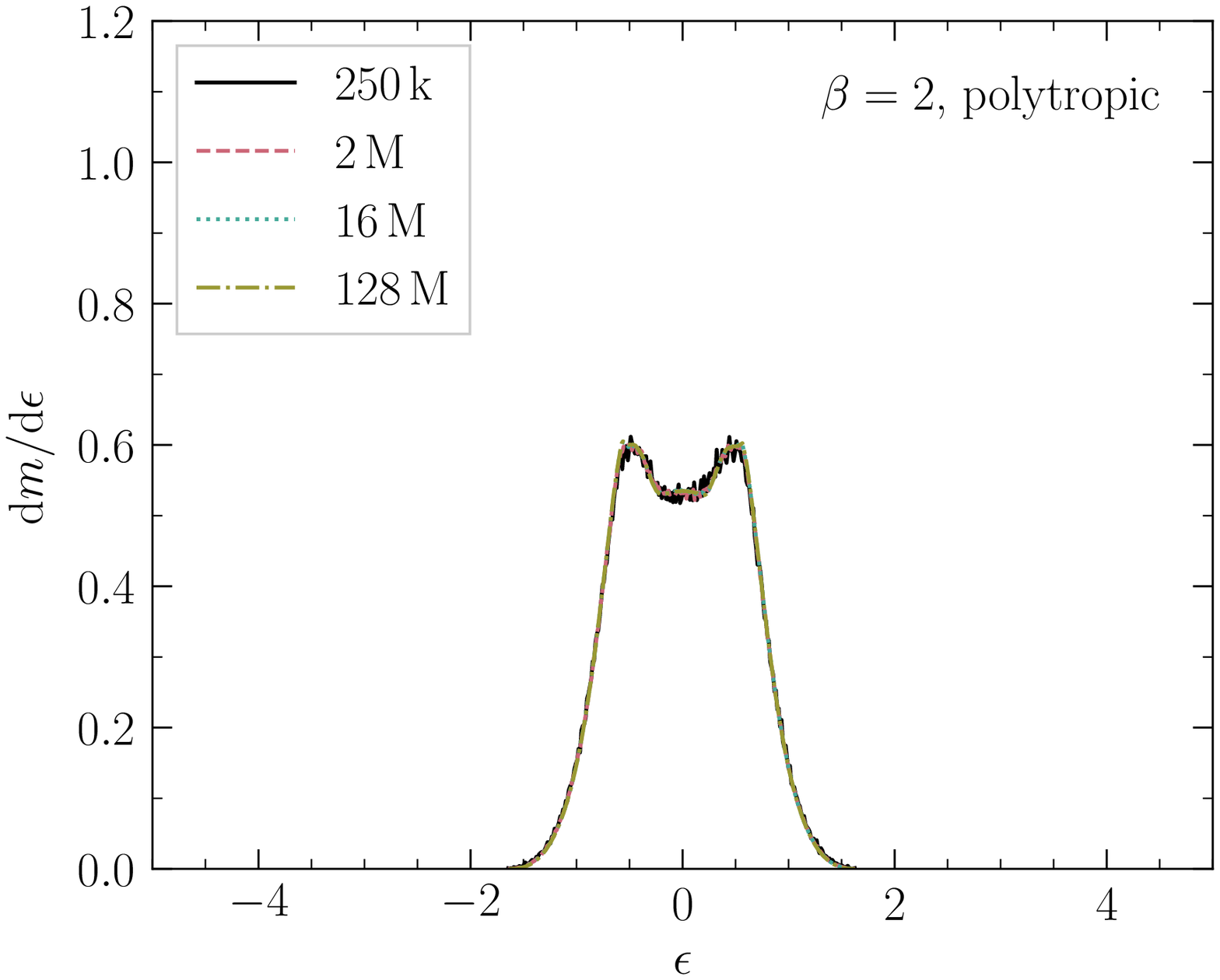}\hfill
	\includegraphics[width=0.28\textwidth]{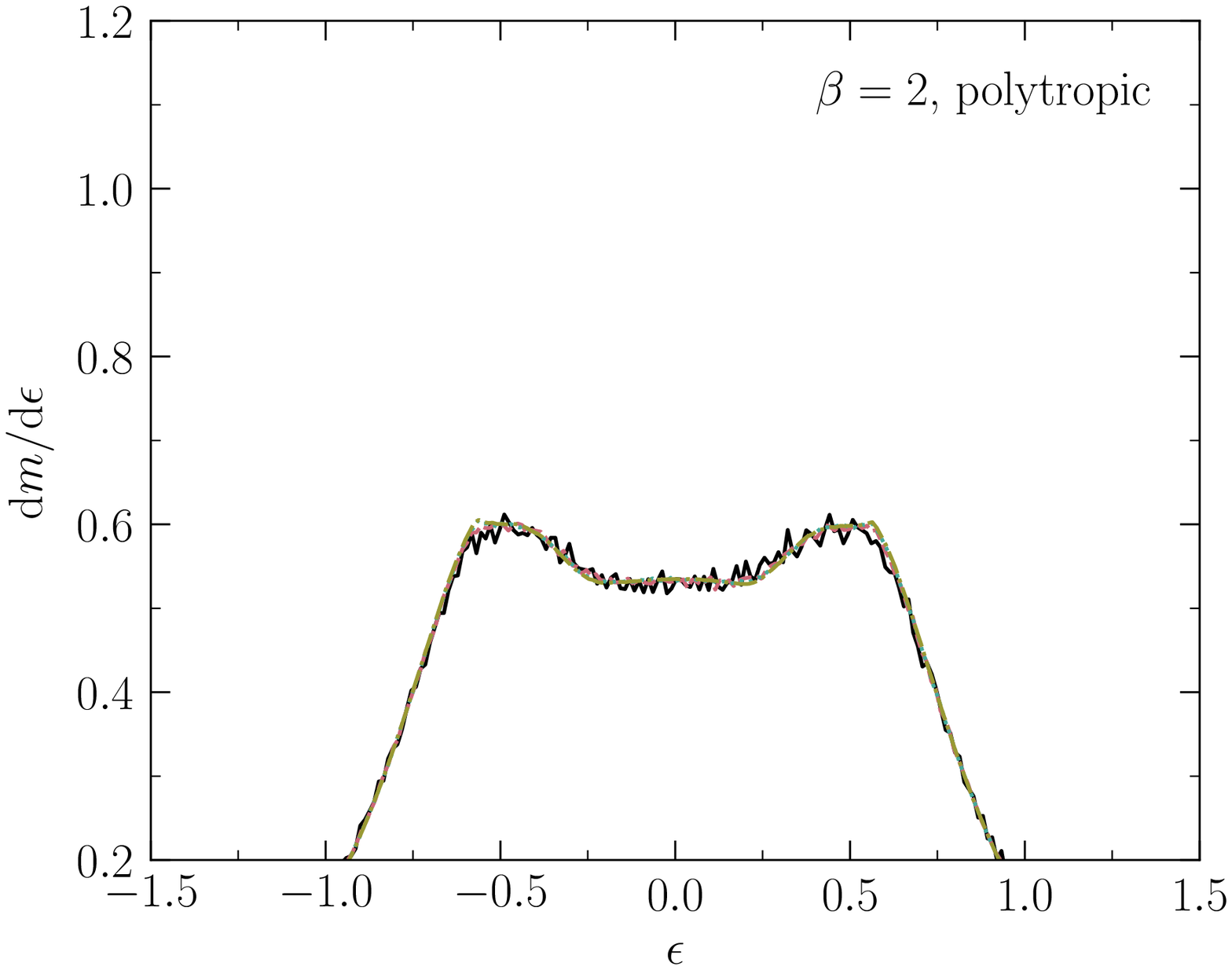}\hfill
	\includegraphics[width=0.28\textwidth]{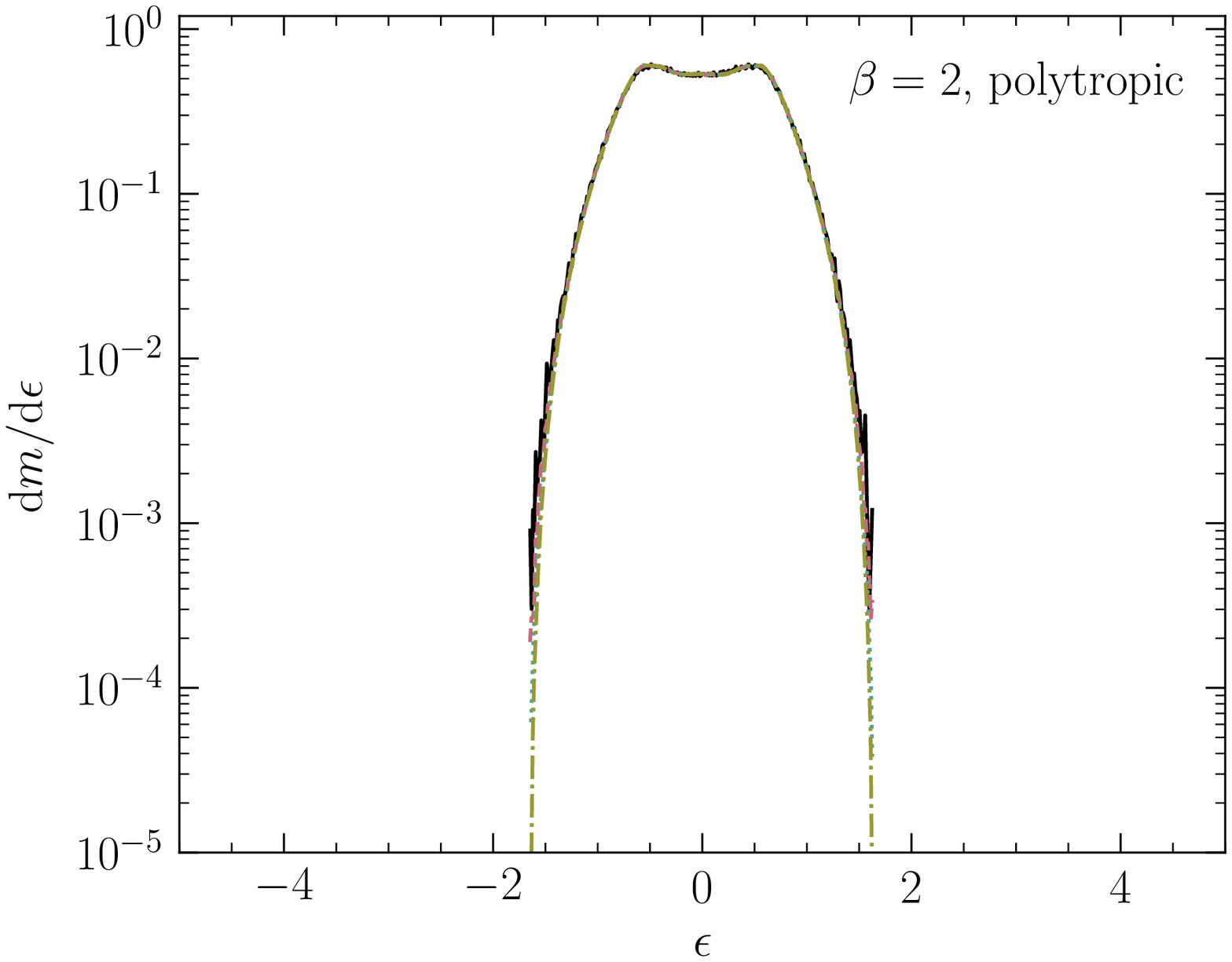}
	\includegraphics[width=0.28\textwidth]{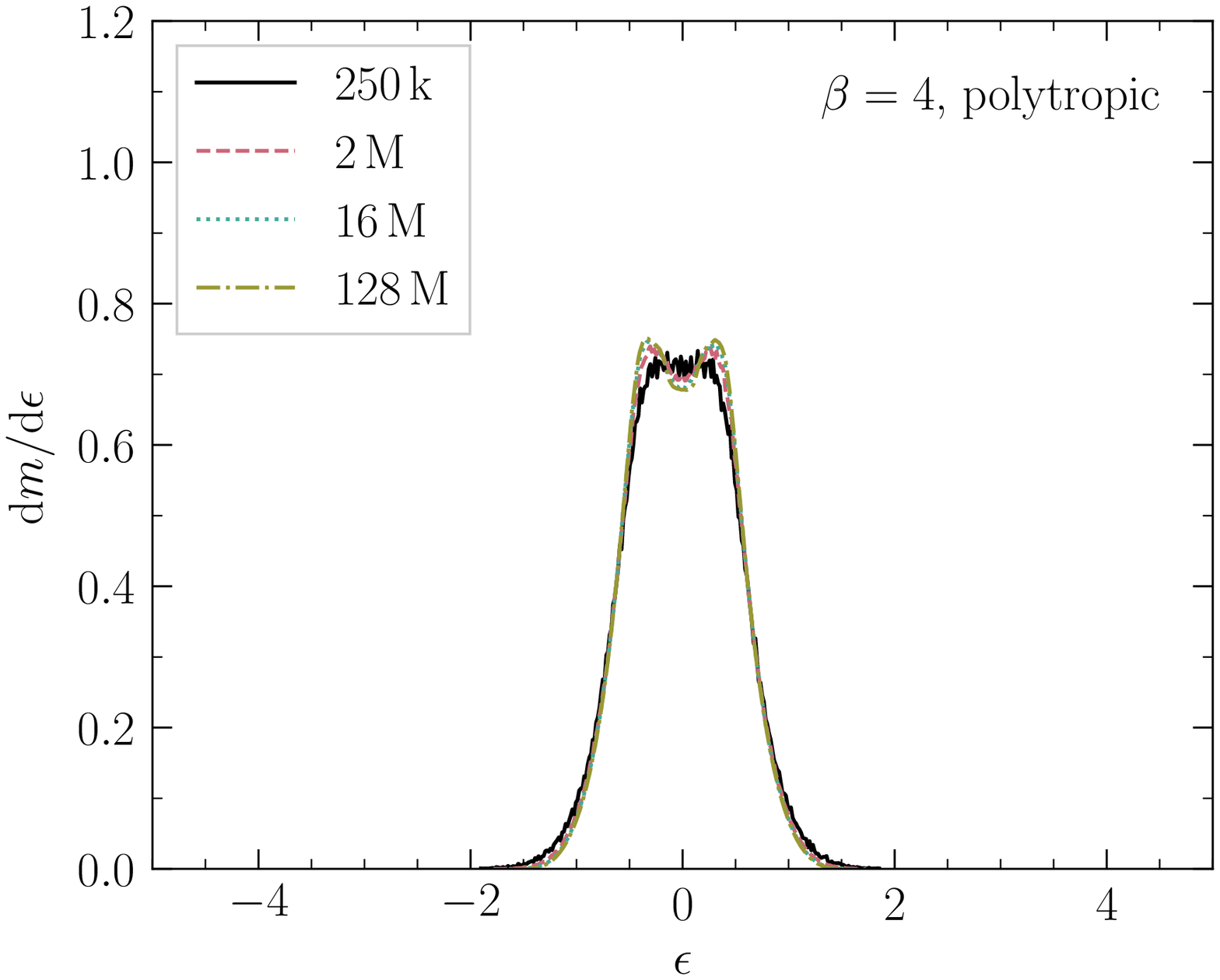}\hfill
	\includegraphics[width=0.28\textwidth]{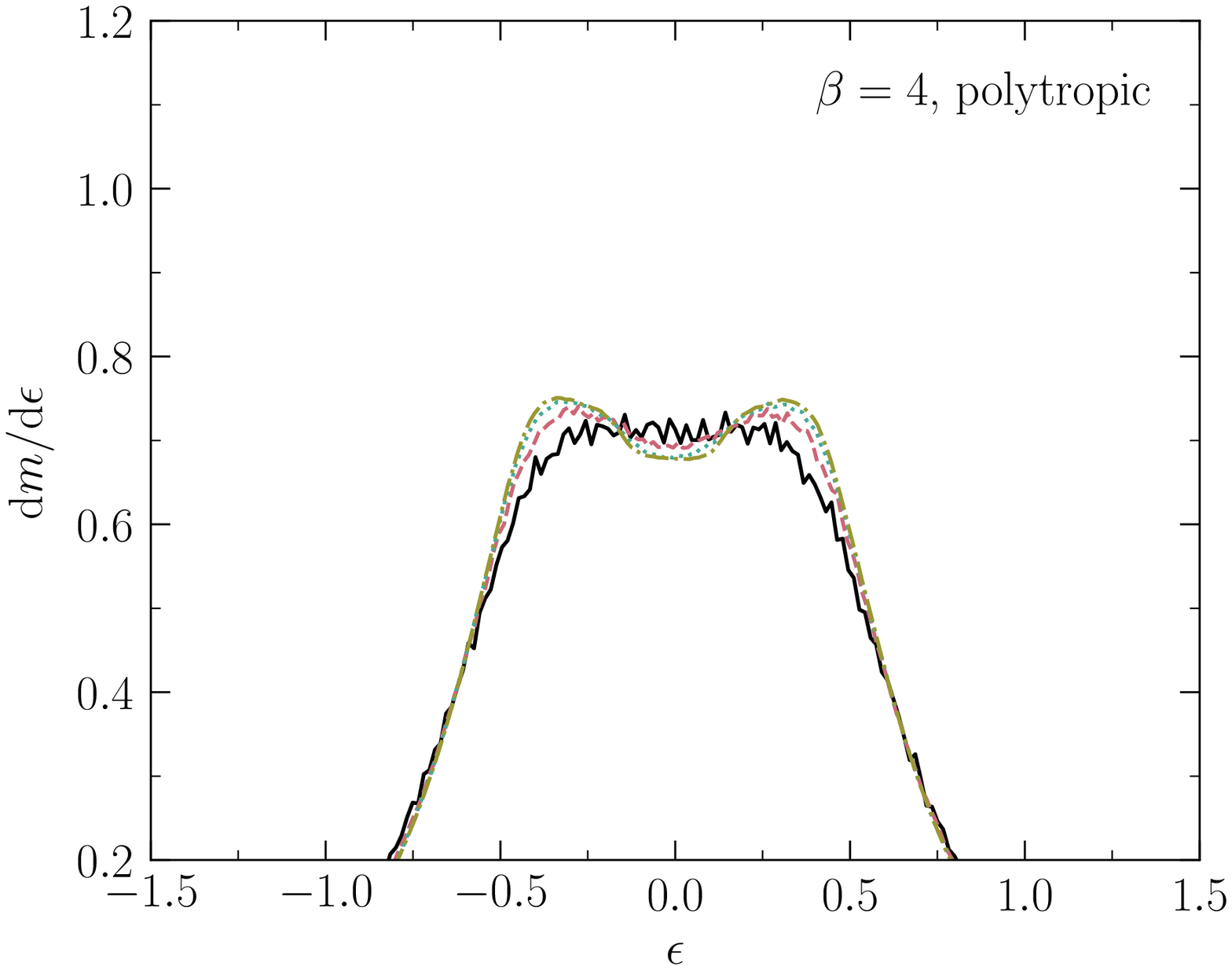}\hfill
	\includegraphics[width=0.28\textwidth]{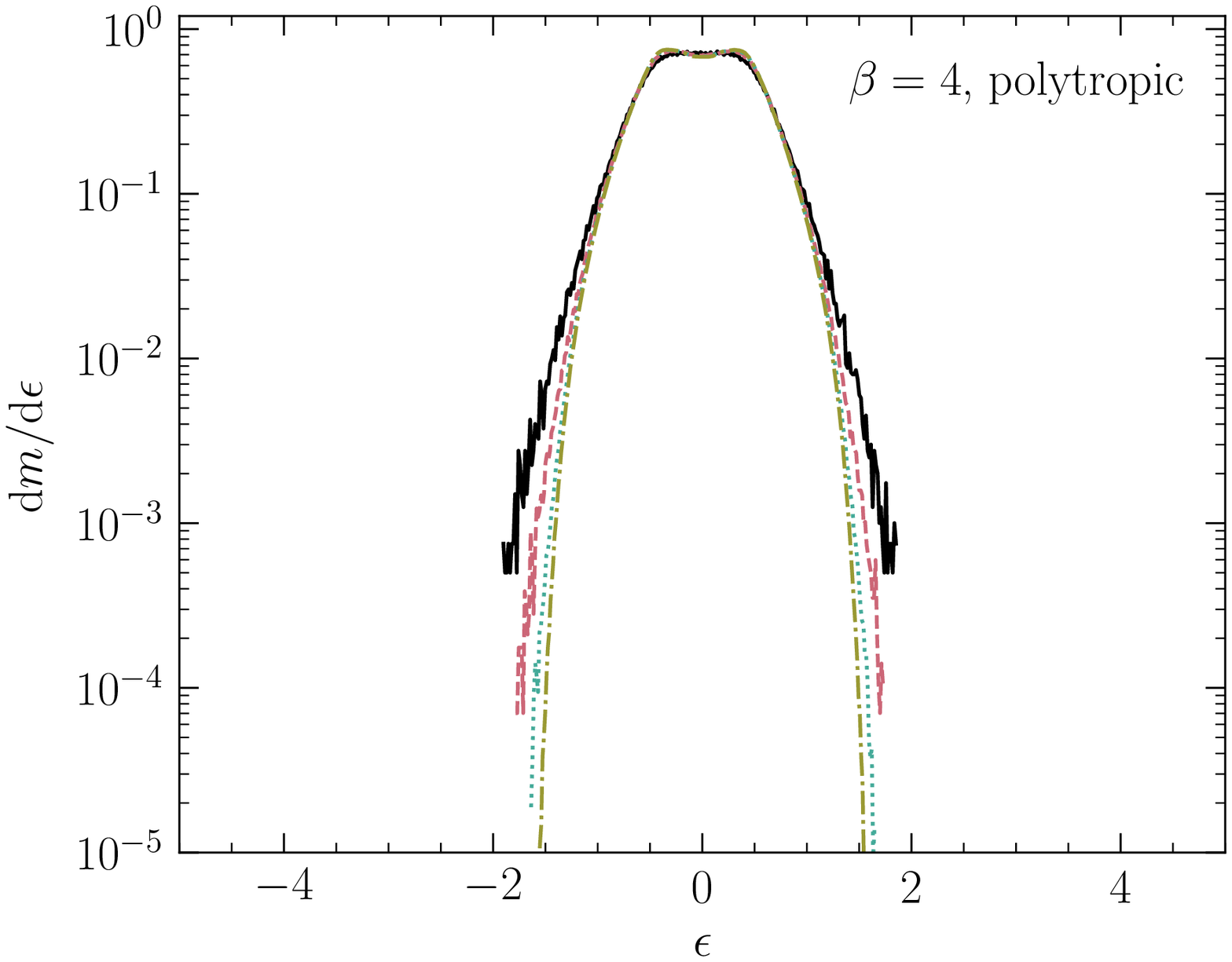}
	\includegraphics[width=0.28\textwidth]{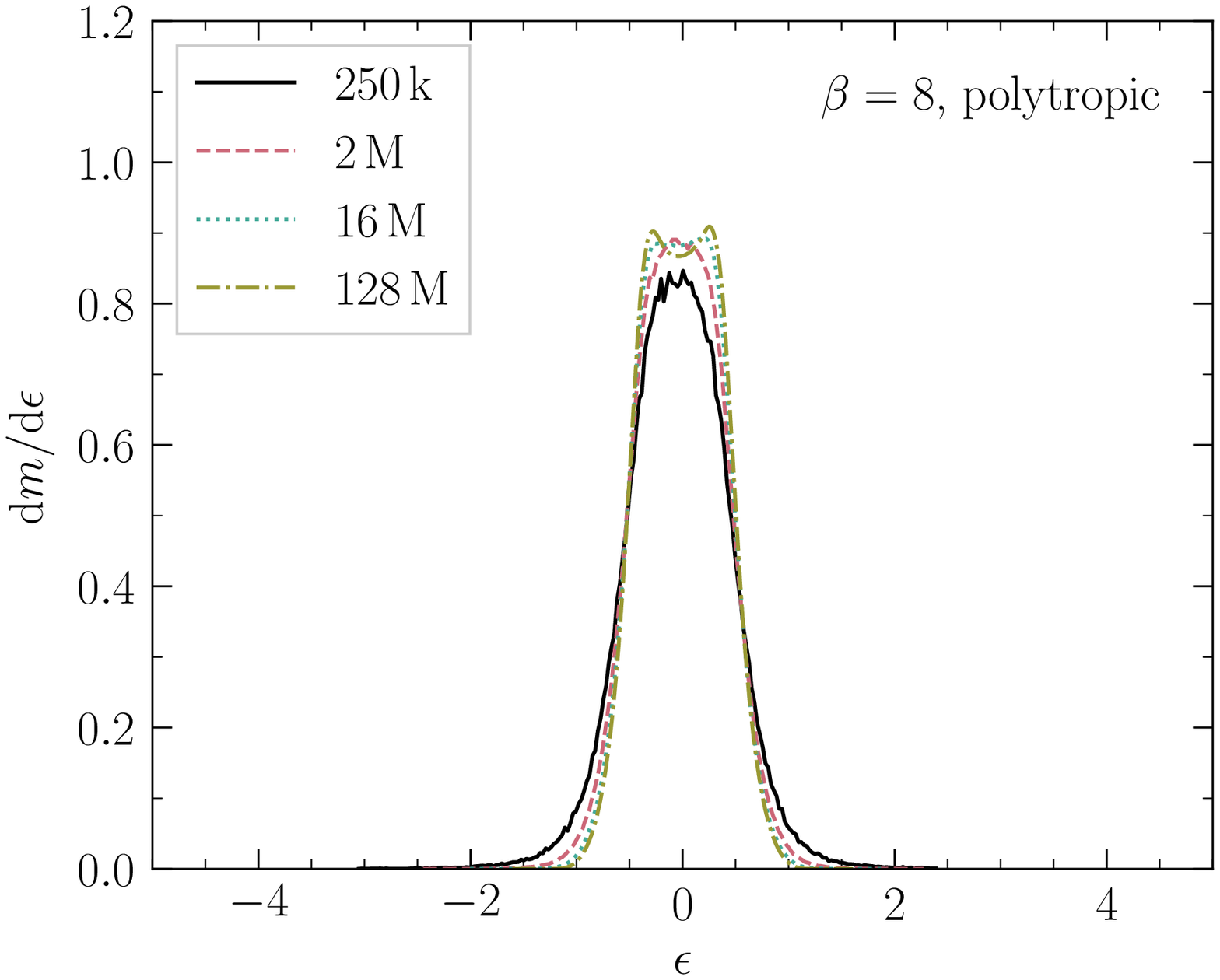}\hfill
	\includegraphics[width=0.28\textwidth]{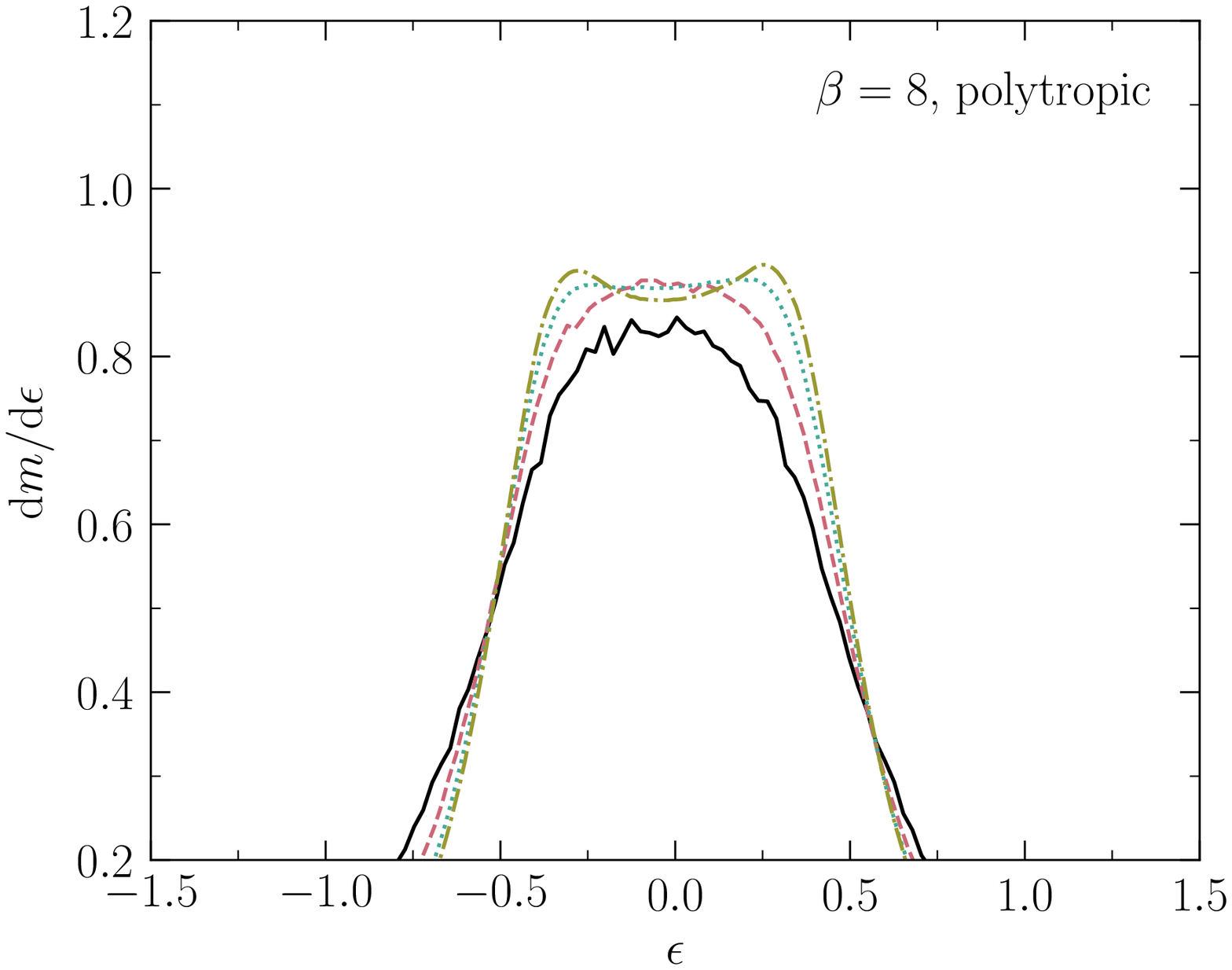}\hfill
	\includegraphics[width=0.28\textwidth]{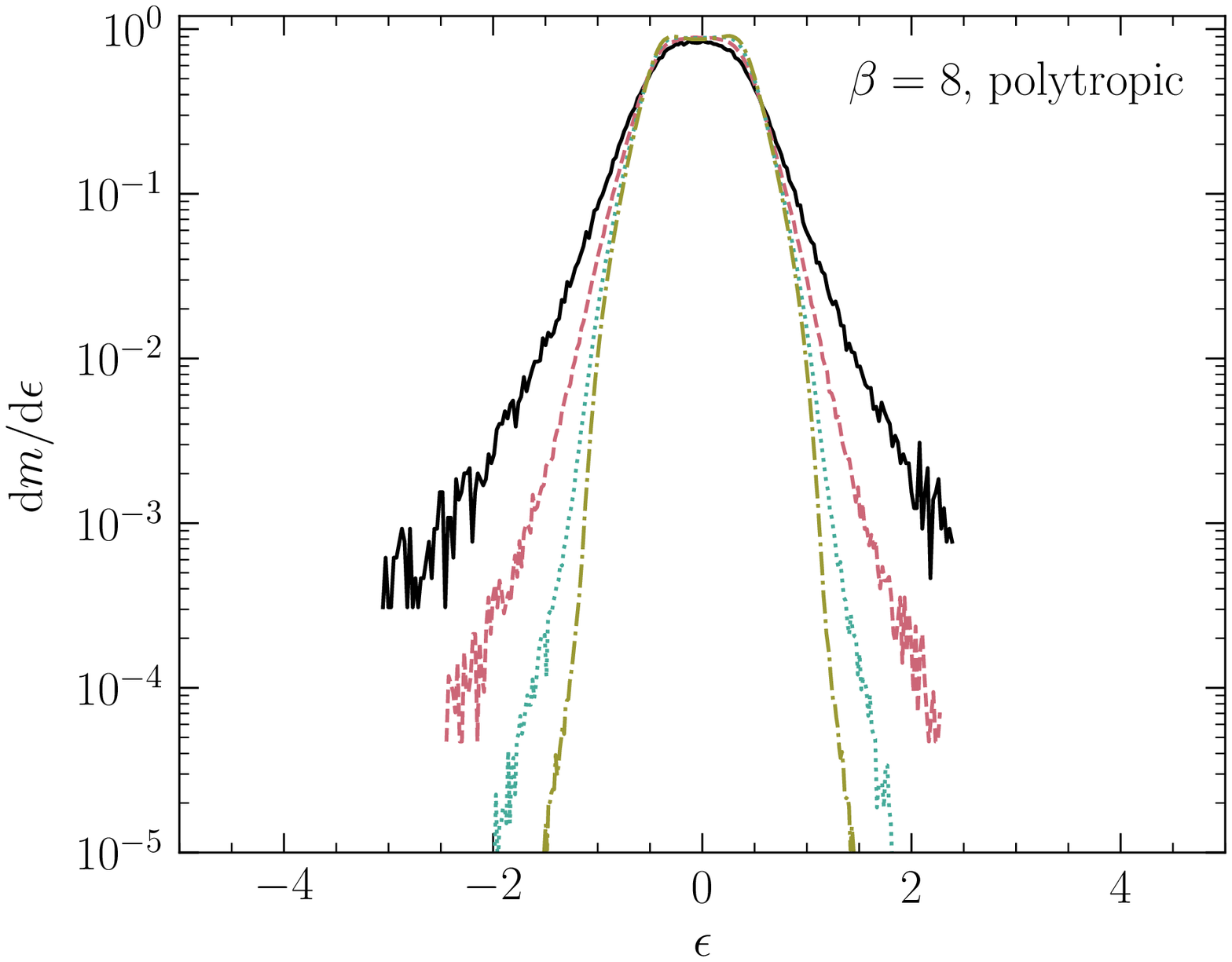}
	\includegraphics[width=0.28\textwidth]{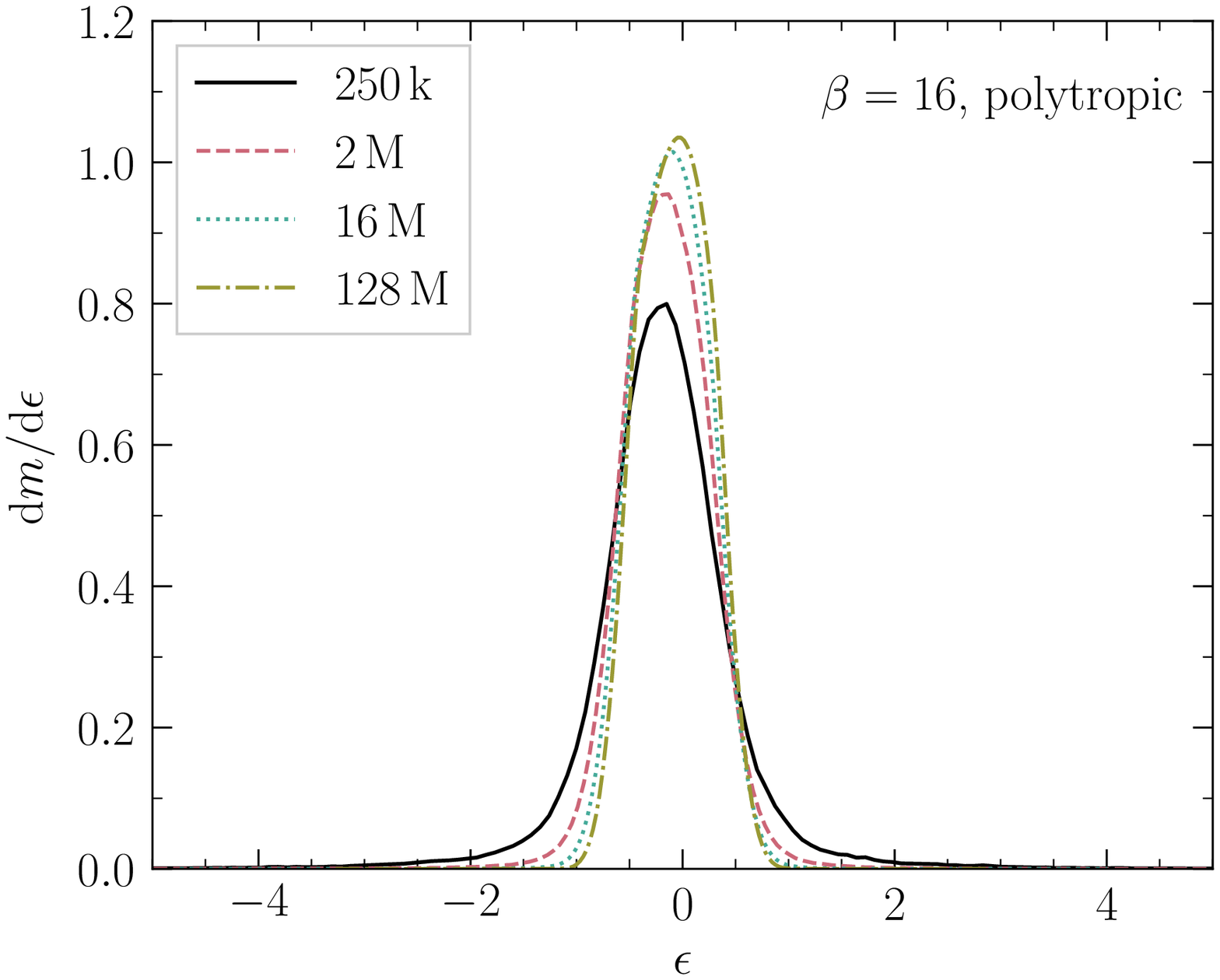}\hfill
	\includegraphics[width=0.28\textwidth]{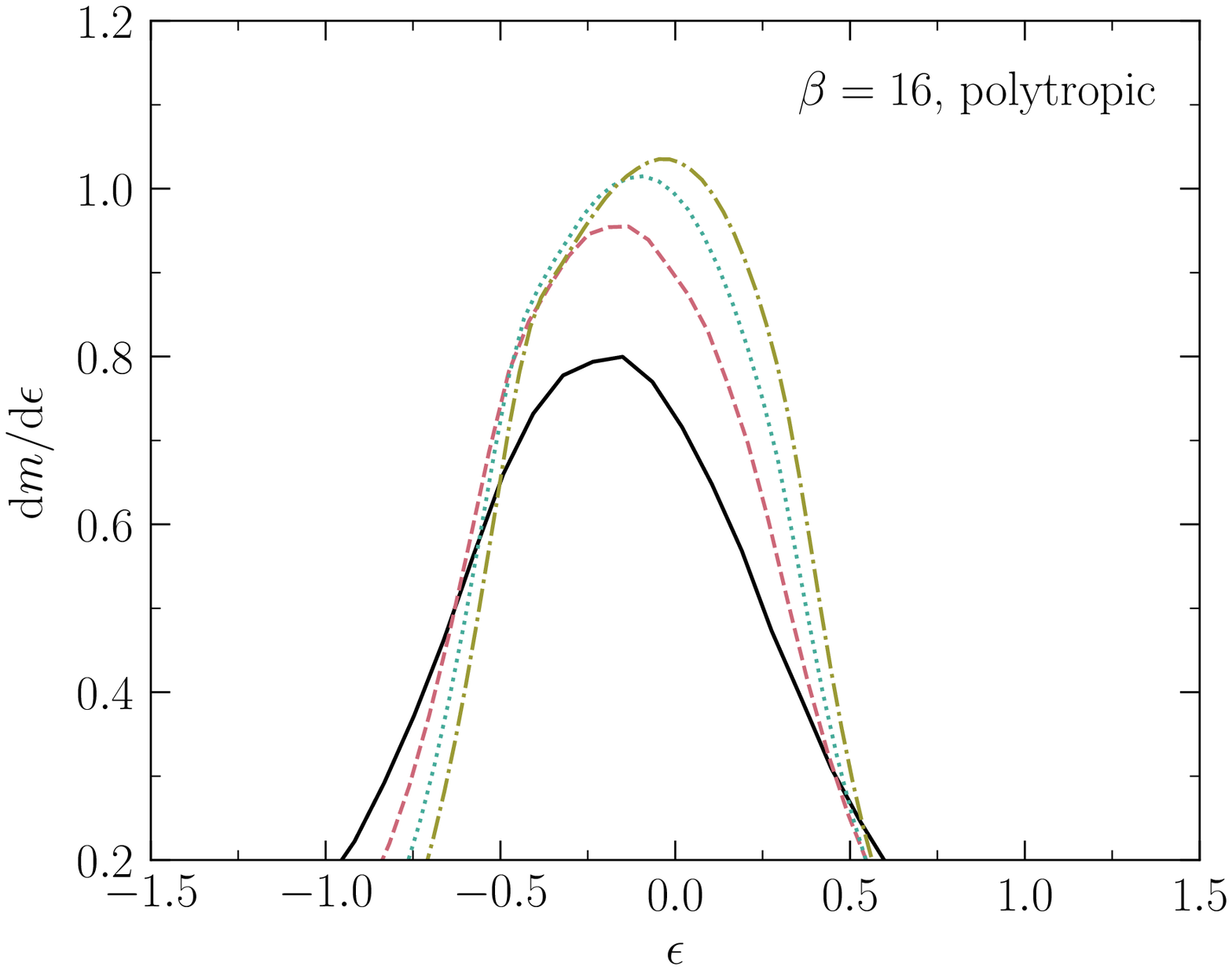}\hfill
	\includegraphics[width=0.28\textwidth]{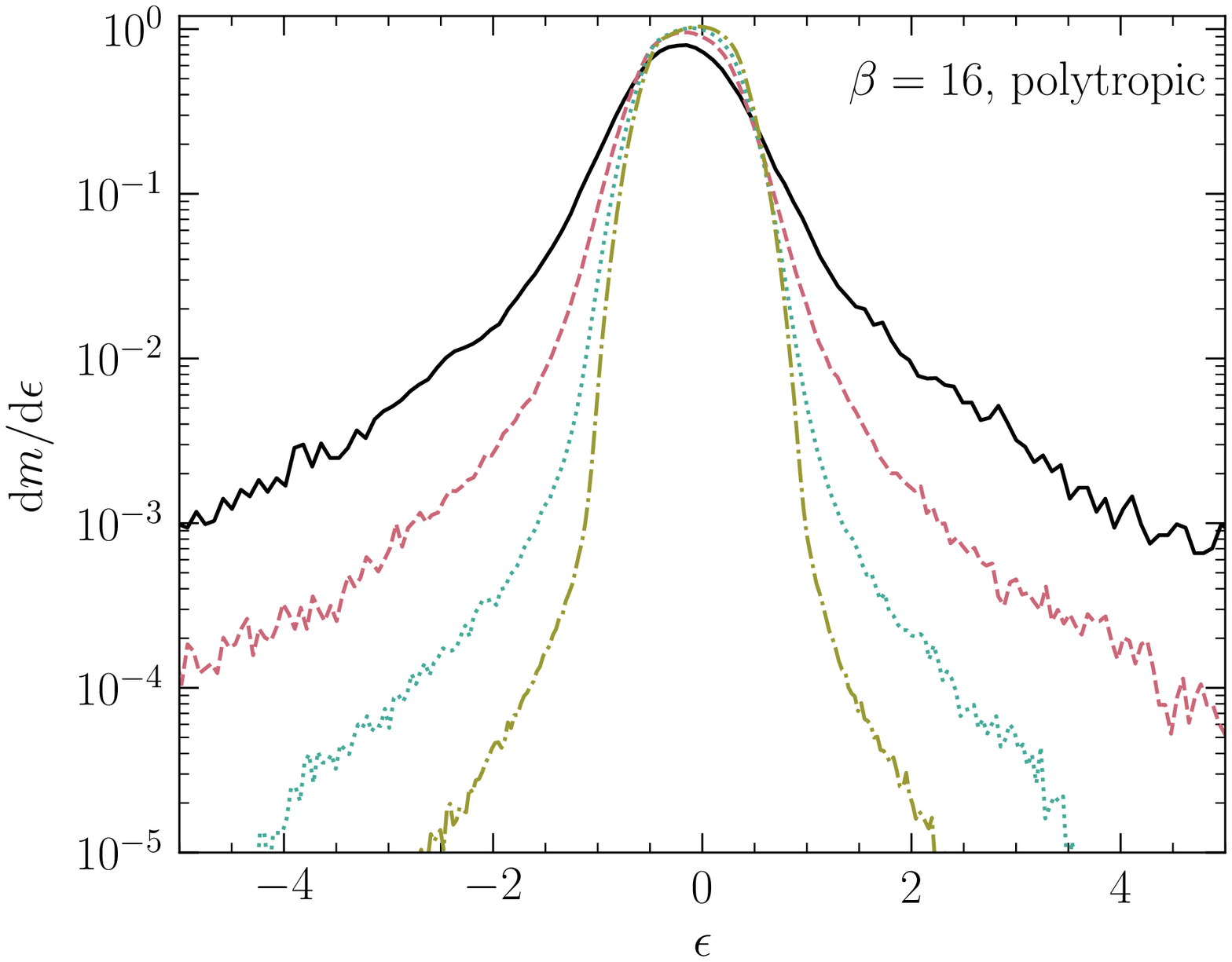}
	\caption{Energy distributions for all of the simulations with a polytropic equation of state, with the $x$-axis being the orbital energy normalised by the canonical energy spread $\epsilon = E/\Delta E$ and the $y$-axis being the mass-energy distribution normalised by the canonical energy spread and the stellar mass ${\rm d}m/{\rm d}\epsilon = (\Delta E / M_\star){\rm d}M/{\rm d}E$. The left column shows the energy distribution with linear axes, the middle column shows the same but with the axes zoomed in on the peak, and the right column shows the energy distribution with the $y$-axis on a log scale. The value of $\beta$ is given in the panel title and increases from the top row to the bottom row. In each column the axis ranges are the same to ease comparison between different $\beta$ values. On each panel the line colour corresponds to the resolution of the simulations as given in the legend in the left column.}
	\label{fig2}
\end{figure*}

SPH simulations of TDEs with $\beta \approx 1$ have yielded qualitatively and quantitatively similar results in terms of the energy distribution of the debris measured not long after pericentre passage with modest numbers of particles, $N_{\rm p} \approx 10^4-10^5$ \citep{Evans:1989aa,Lodato:2009aa}. However, the numerical convergence of such simulations at larger $\beta$ is not so well-established. Here we aim to assess the level of numerical convergence of the energy distribution measured at the same time in the post-disruption debris for the range of impact parameters and particle numbers we have simulated. For this, we evolve the stellar debris until the zero-energy orbit has reached a distance of $5r_{\rm t}$ post-pericentre. We plot in Fig.~\ref{fig2} the resulting energy distributions on a linear-linear scale (left column), a zoom-in on the peak (middle column), and a log-linear scale (right column), with each row corresponding to a different value of $\beta$ increasing from the top to the bottom. On each panel four lines are present indicating the simulation results at different resolutions corresponding to particle numbers ranging from 250\,k to 128\,M. Across this range of resolution, corresponding to a factor of 512 in particle number and thus a factor of 8 in spatial resolution, we see no difference for the simulations at $\beta=1$ and $\beta=2$. At $\beta=4$ there is a noticeable change from 250k to 2M, but by 16M particles this case is also well-converged. For $\beta=8$ the simulation results are very similar between 16M and 128M, albeit with minor differences in the shape of the distribution at the peak and the breadth of the energy distribution for the very low-mass tail ($\lesssim 0.1-1$\% of the mass); however, at lower resolution the shape and width of the energy distribution are not well-converged, indicating that simulating $\beta$ at least this high requires substantial numerical resolution (corresponding to particle numbers in excess of $\sim 10$\,M). The $\beta=16$ case shows significant discrepancies, particularly at low-resolution: For low resolution this case is skewed towards negative energies with the peak occurring at around $\epsilon = -0.2$; this is no longer present for the highest-resolution case (although the distribution remains clearly asymmetric). Also, the energy distributions at low-resolution are very broad. It is clear in the bottom-right panel of Fig.~\ref{fig2} that this breadth is not physical; by 128M particles almost all of the mass ($\gtrsim 99.9$\%) is confined to $|E| \lesssim \Delta E$. It is not currently feasible to perform simulations with $N_{\rm p} \gg 128$M, but with the trends seen in the right panels appropriate to $\beta = 4$ and $\beta = 8$, it is clear that going to even higher resolution would reduce the breadth of the energy spread in the low ${\rm d}m/{\rm d}\epsilon$ wings ($\lesssim 10^{-4}$) even further. Given this we are reasonably confident that the properties of the vast majority of the debris in the $\beta=16$ case are accurate for the 128\,M particle simulation (i.e., the region above ${\rm d}m/{\rm d}\epsilon \sim 10^{-3}$ in the bottom-right panel of Fig.~\ref{fig2}). This statement is further supported by the excellent agreement between the results of this simulation and the analytical predictions of \citet[][see their Fig.~17]{Coughlin:2021aa}.

To provide a quantitative measure of the convergence of the simulations we calculate the L2 error norm, or root mean square error, for the lower resolution simulations, with respect to the highest resolution simulation (128M), given by
\begin{equation}
\label{L2}
L_2(N_{\rm p}) = \left[\frac{1}{N_i f_{\rm max}^2}\sum_{i=1}^{N_i}e_i^2\right]^{1/2}\,,
\end{equation}
where $f_{\rm max}$ is the maximum value of ${\rm d}m/{\rm d}\epsilon$, $N_i=200$ is the number of bins used for the ${\rm d}m/{\rm d}\epsilon$ array, and $e_i$ is the difference between the value of ${\rm d}m/{\rm d}\epsilon$ for the simulation with $N_{\rm p}$ particles and the value for the simulation with 128M particles for each energy bin. We restrict the sum to the energy range corresponding to the full range of the 128\,M simulation. The L2 error norm value for each of the simulations is given in Table~\ref{tab}. For smooth flow the numerical scheme employed in the {\sc phantom} SPH code is formally second-order accurate in space, while for flows containing discontinuities such as shocks the accuracy is reduced to first order (this also occurs where the particles reach the ceiling of the switch for the linear artificial viscosity term, $\alpha_{\rm max}^{\rm AV}$). Therefore we expect the error to scale roughly as $\propto h^2$ for cases where the flow is smooth and as $\propto h$ where shocks play a significant role. In our 3D simulations this suggests the error should scale between $\propto N^{-1/3}$ and $\propto N^{-2/3}$, and thus for each factor of 8 increase in particle number we expect the error to be reduced by a factor of 2-4. The $\alpha^{\rm AV}$ values for the particles are typically large during the pericentre passage of the star, with the fraction that are at the ceiling increasing with increasing $\beta$. This appears to be why the lower $\beta$ simulations show stronger convergence at these resolutions. However, in each case the error is reduced by a factor $\gtrsim 2$ for a factor of $8$ increase in particle number. Thus we can conclude that the simulations are converging appropriately. It is also worth remarking that the errors for the higher $\beta$ values at the higher resolutions are approaching the errors for the lower $\beta$ values at lower resolutions (i.e. the values towards the bottom-right of Table~\ref{tab} are similar to the values towards the top-left). This further substantiates our suggestion at the end of the previous section that the high-$\beta$ simulations are accurate for the majority of the stellar debris at 128\,M particles.

\begin{table}\centering
\begin{tabular}{|c|c|c|c|}
\hline
$\beta$ & $N_{\rm p} = 250$\,k& $N_{\rm p} = 2$\,M& $N_{\rm p}=16$\,M\\
\hline
1 & $1.2\times 10^{-2}$ & $4.3\times 10^{-3}$ & $2.2\times 10^{-3}$\\
\hline
2 & $1.3\times 10^{-2}$ & $5.2\times 10^{-3}$ & $2.1\times 10^{-3}$\\
\hline
4 & $4.0\times 10^{-2}$ & $1.7\times 10^{-2}$ & $5.4\times 10^{-3}$\\
\hline
8 & $8.8\times 10^{-2}$ & $4.5\times 10^{-2}$ & $1.7\times 10^{-2}$\\
\hline
16 & $1.1\times 10^{-1}$ & $6.6\times 10^{-2}$ & $3.2\times 10^{-2}$\\
\hline
\end{tabular}
\caption{L2 error norms for the energy distributions from the simulations with a polytropic equation of state, calculated from equation~\ref{L2} with the 128M simulation taken as the reference value. As the resolution is increased, the error decreases. For larger $\beta$ the error is larger at the same number of particles.}
\label{tab}
\end{table}

\subsubsection{{The effect of shock heating}}
\label{sec:shocks}
\begin{figure*}
	\includegraphics[width=0.28\textwidth]{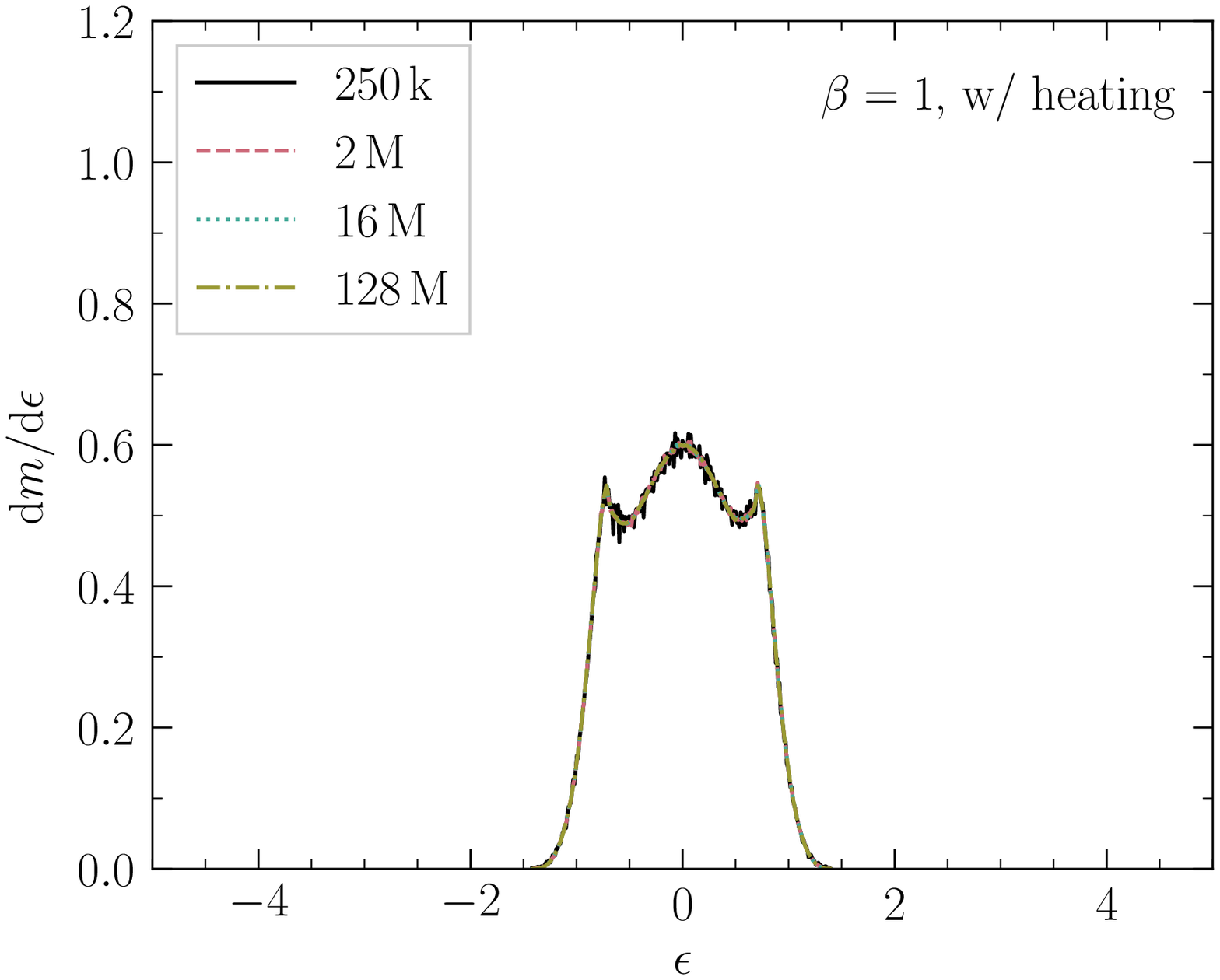}\hfill
	\includegraphics[width=0.28\textwidth]{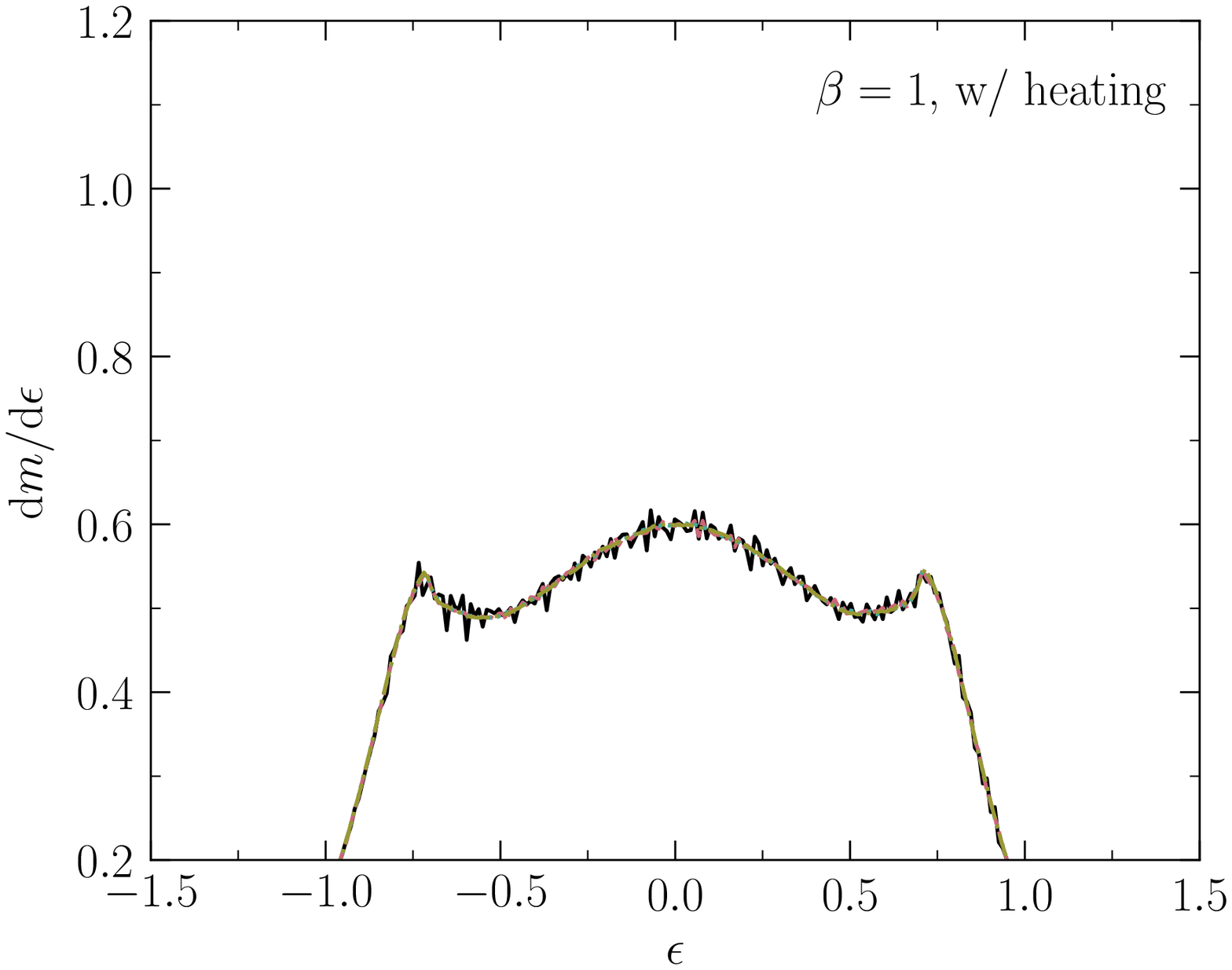}\hfill
	\includegraphics[width=0.28\textwidth]{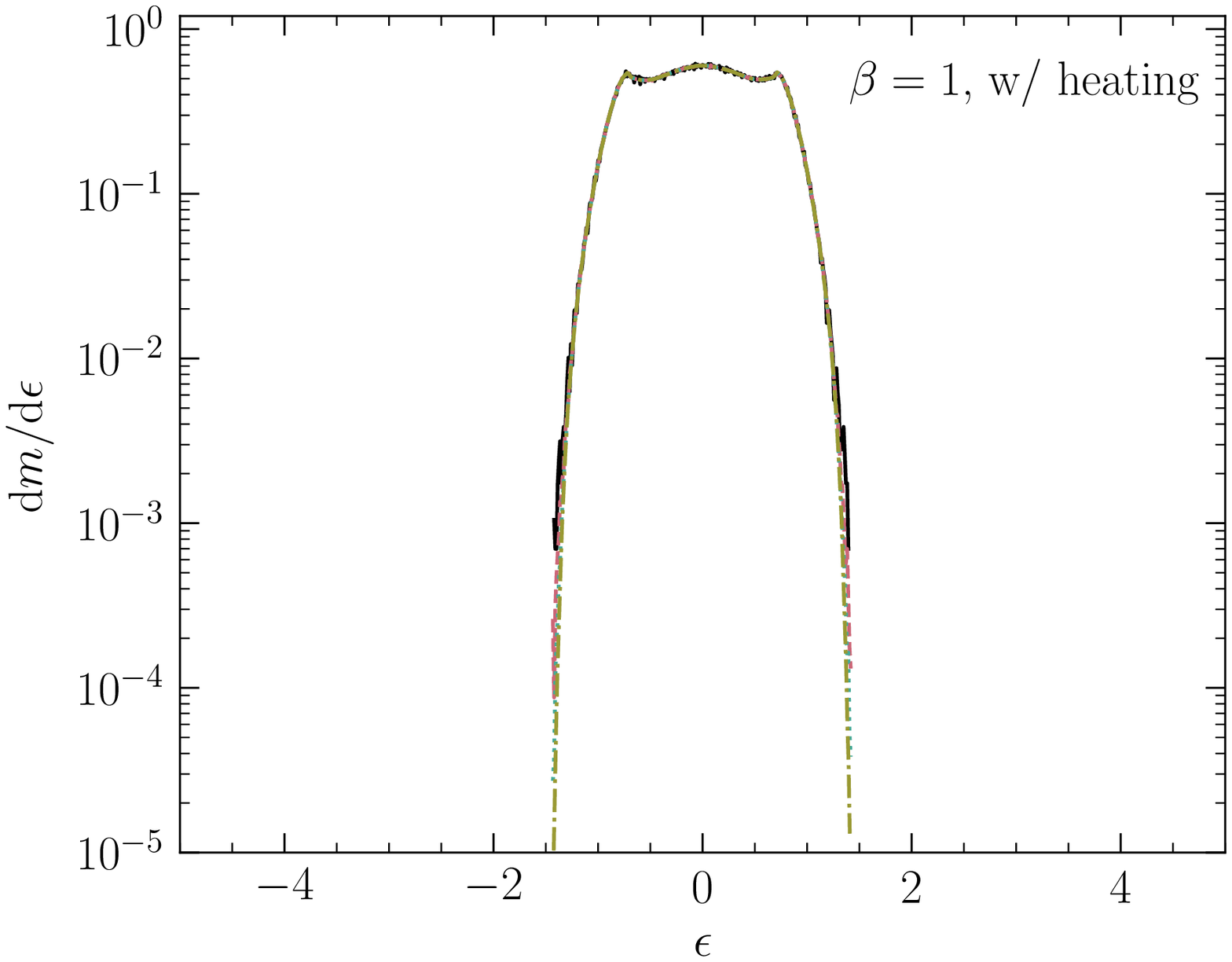}
	\includegraphics[width=0.28\textwidth]{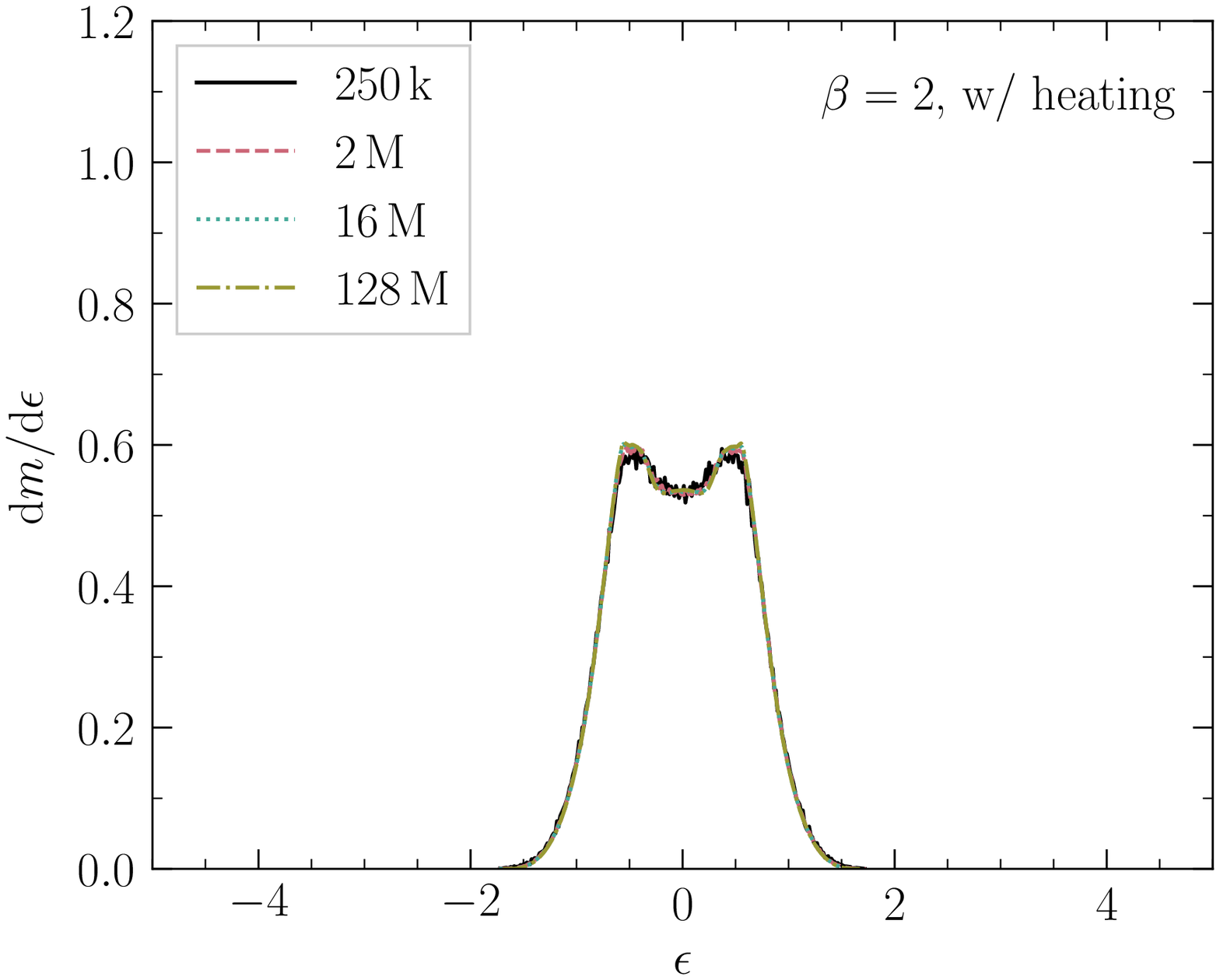}\hfill
	\includegraphics[width=0.28\textwidth]{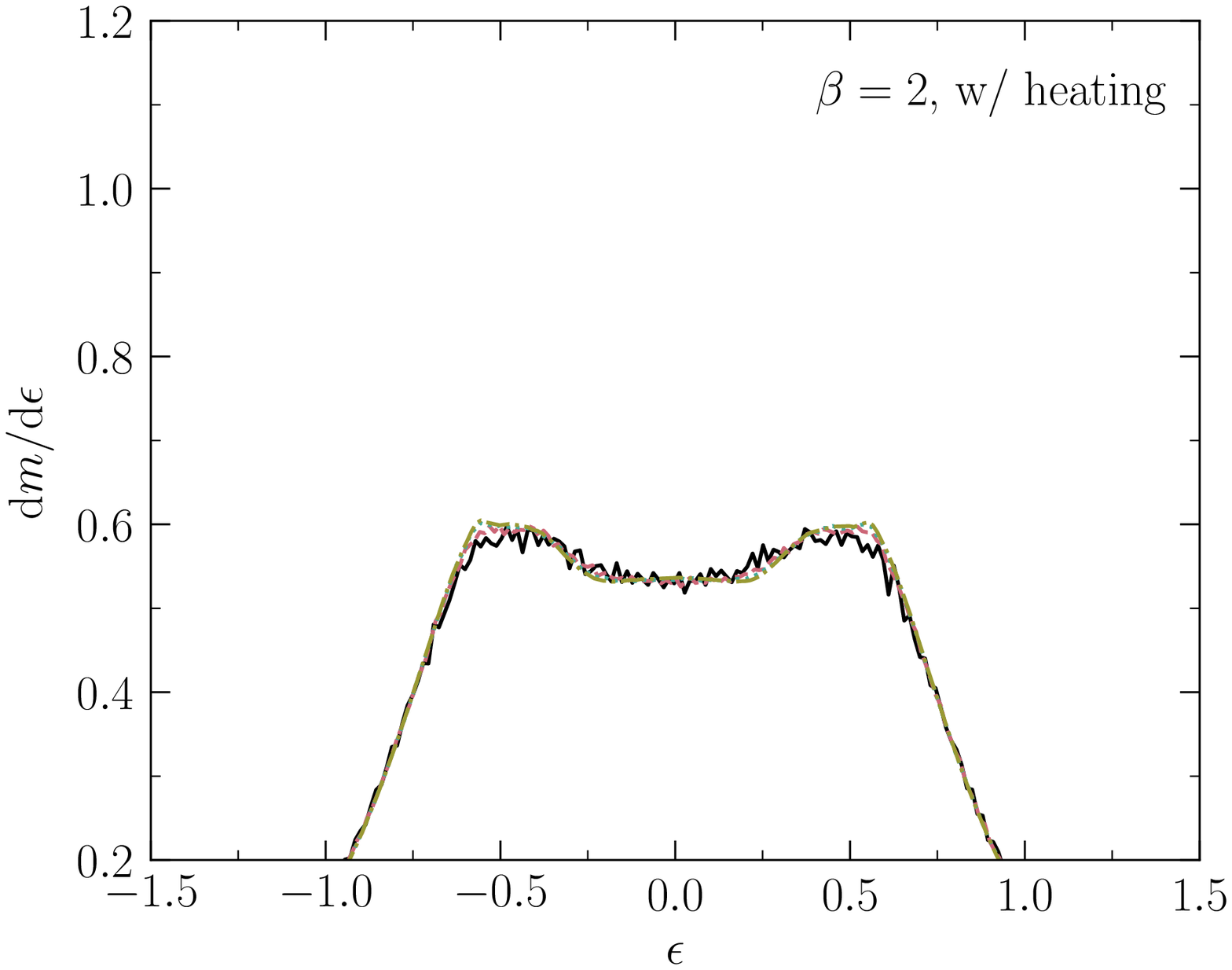}\hfill
	\includegraphics[width=0.28\textwidth]{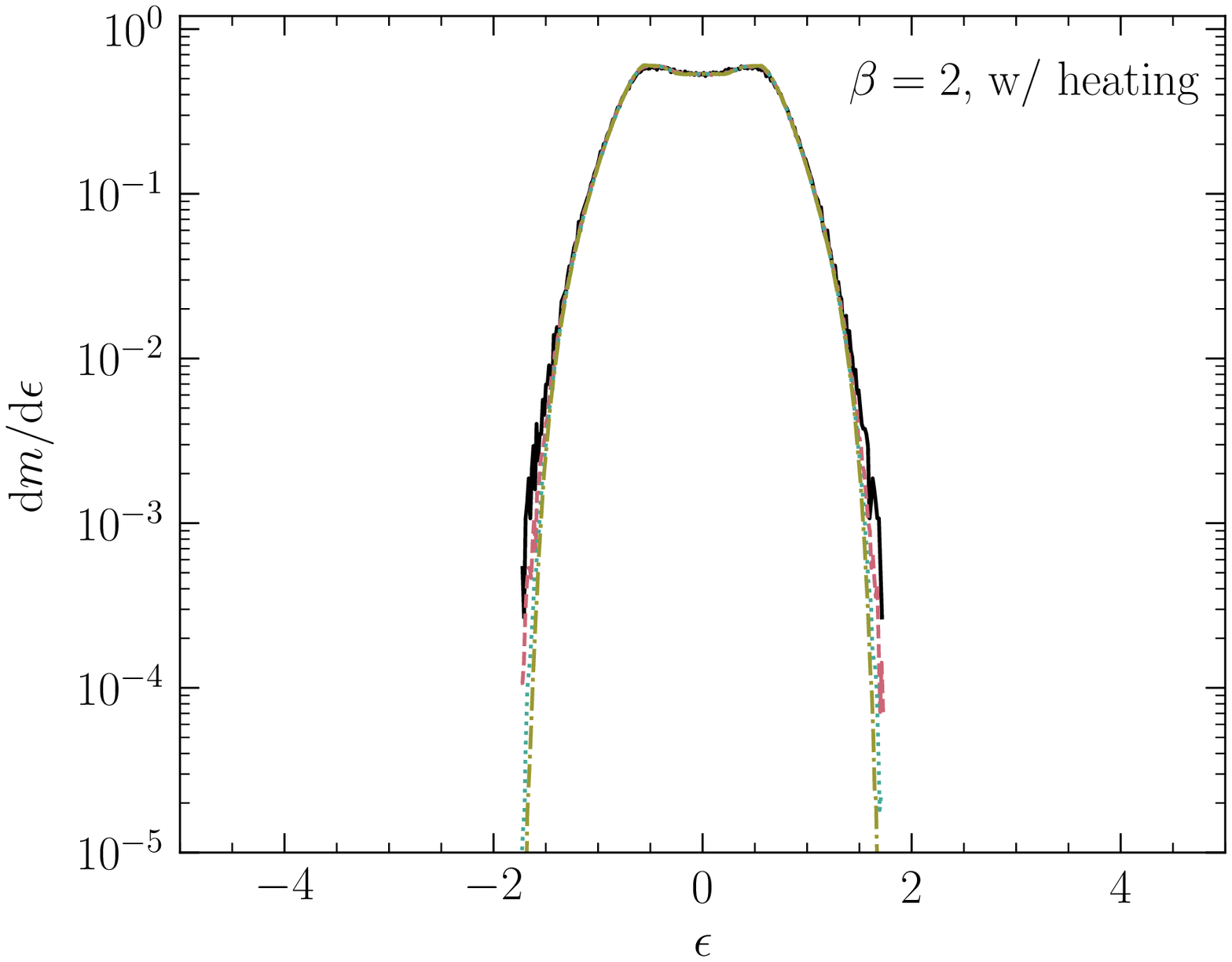}
	\includegraphics[width=0.28\textwidth]{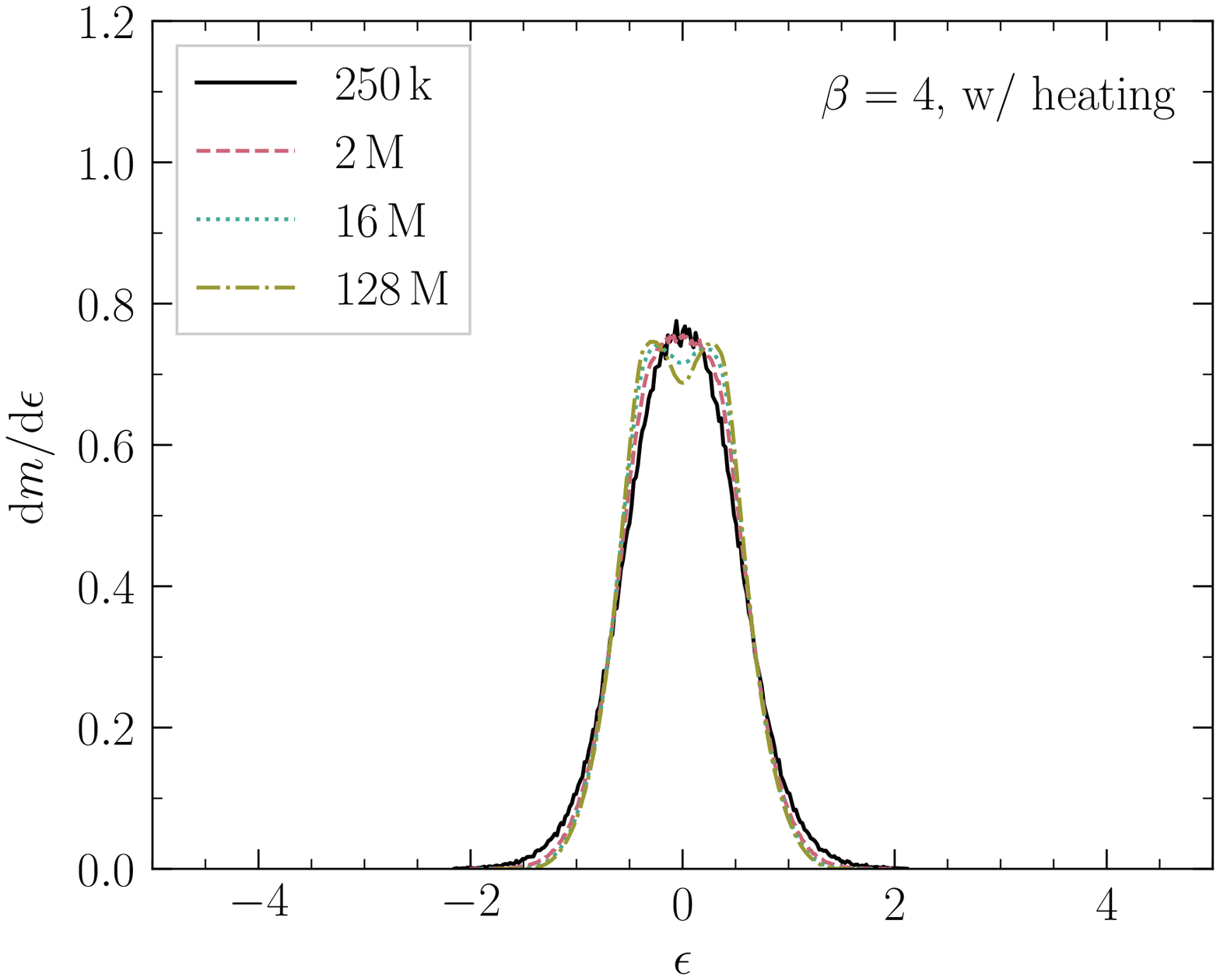}\hfill
	\includegraphics[width=0.28\textwidth]{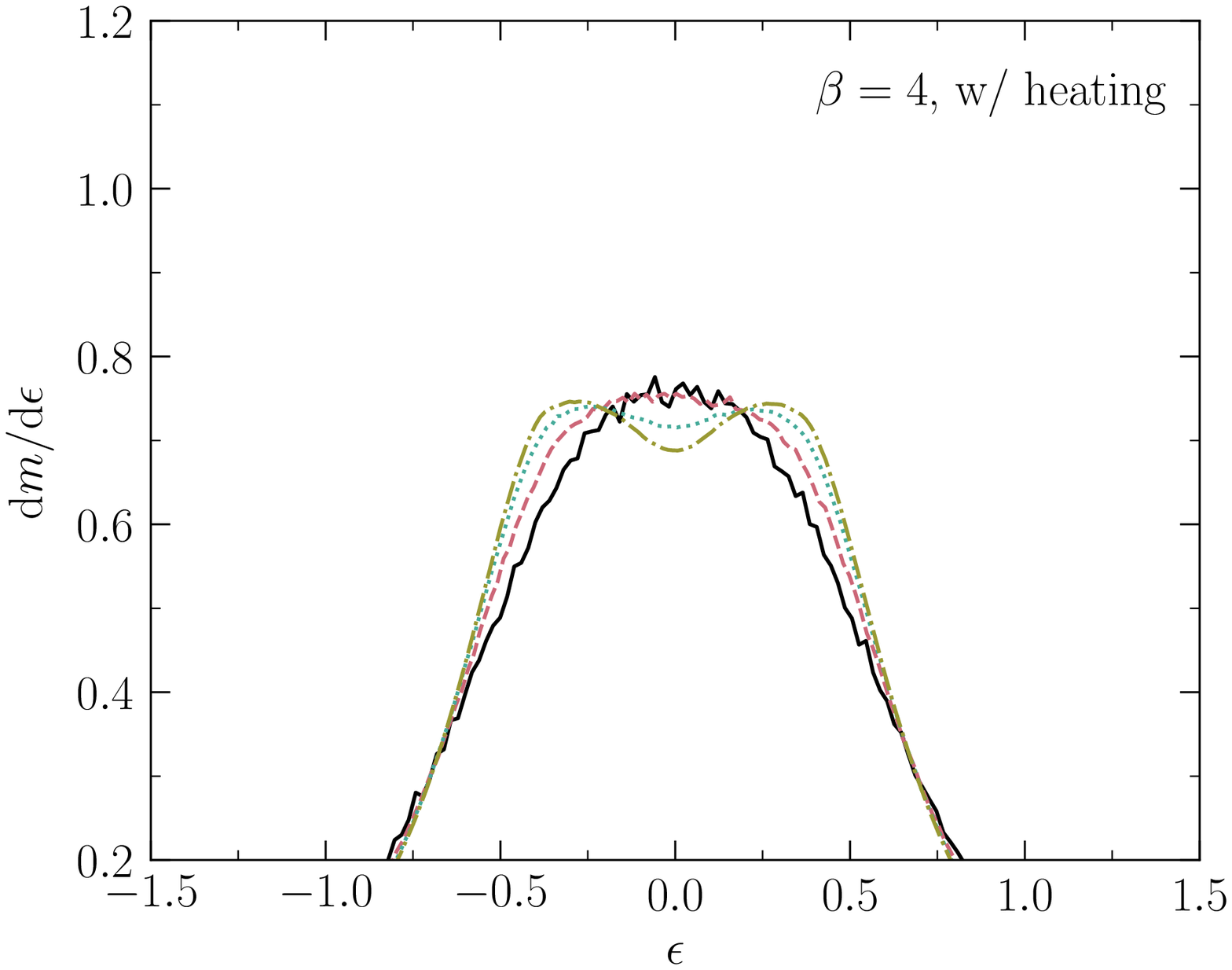}\hfill
	\includegraphics[width=0.28\textwidth]{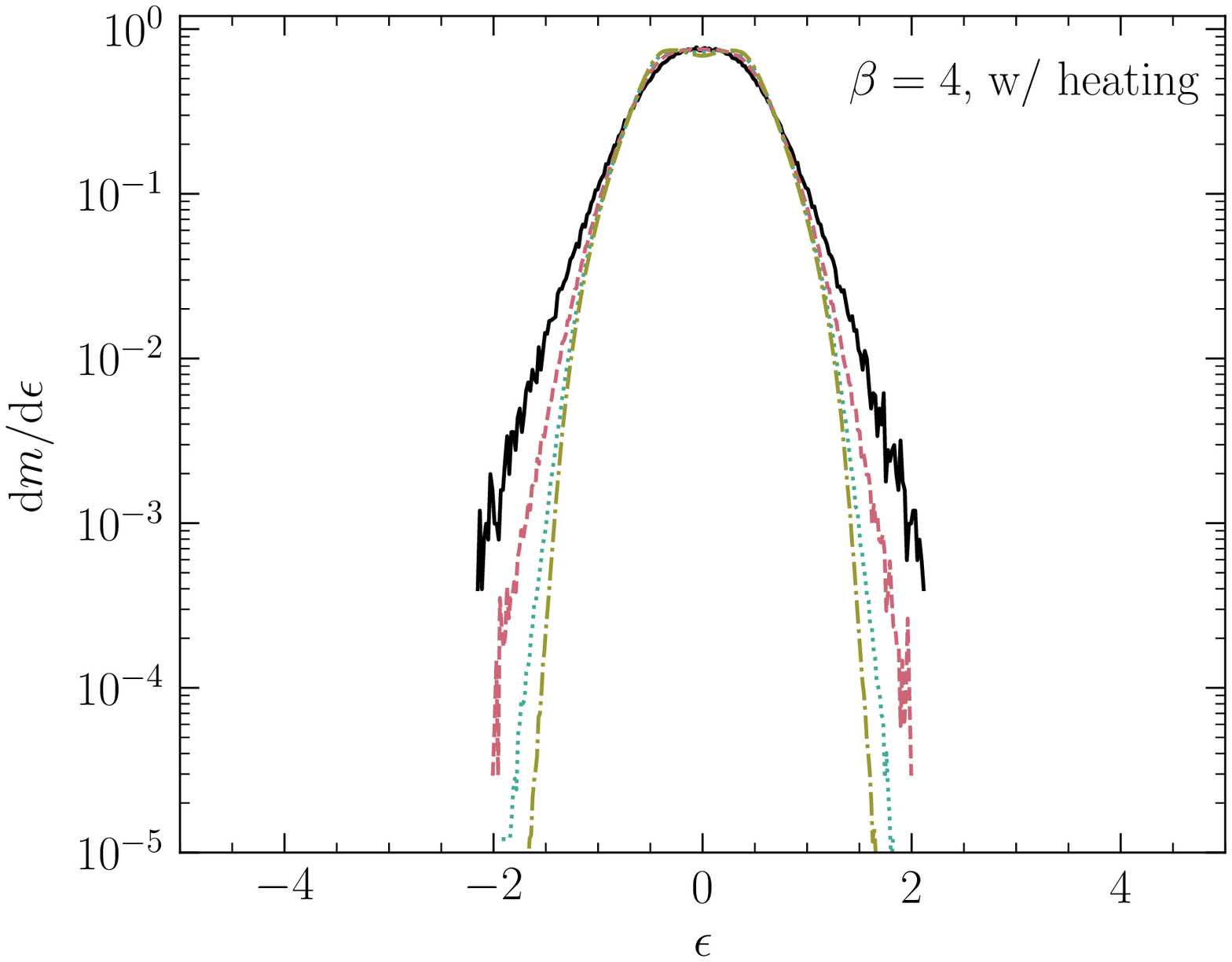}
	\includegraphics[width=0.28\textwidth]{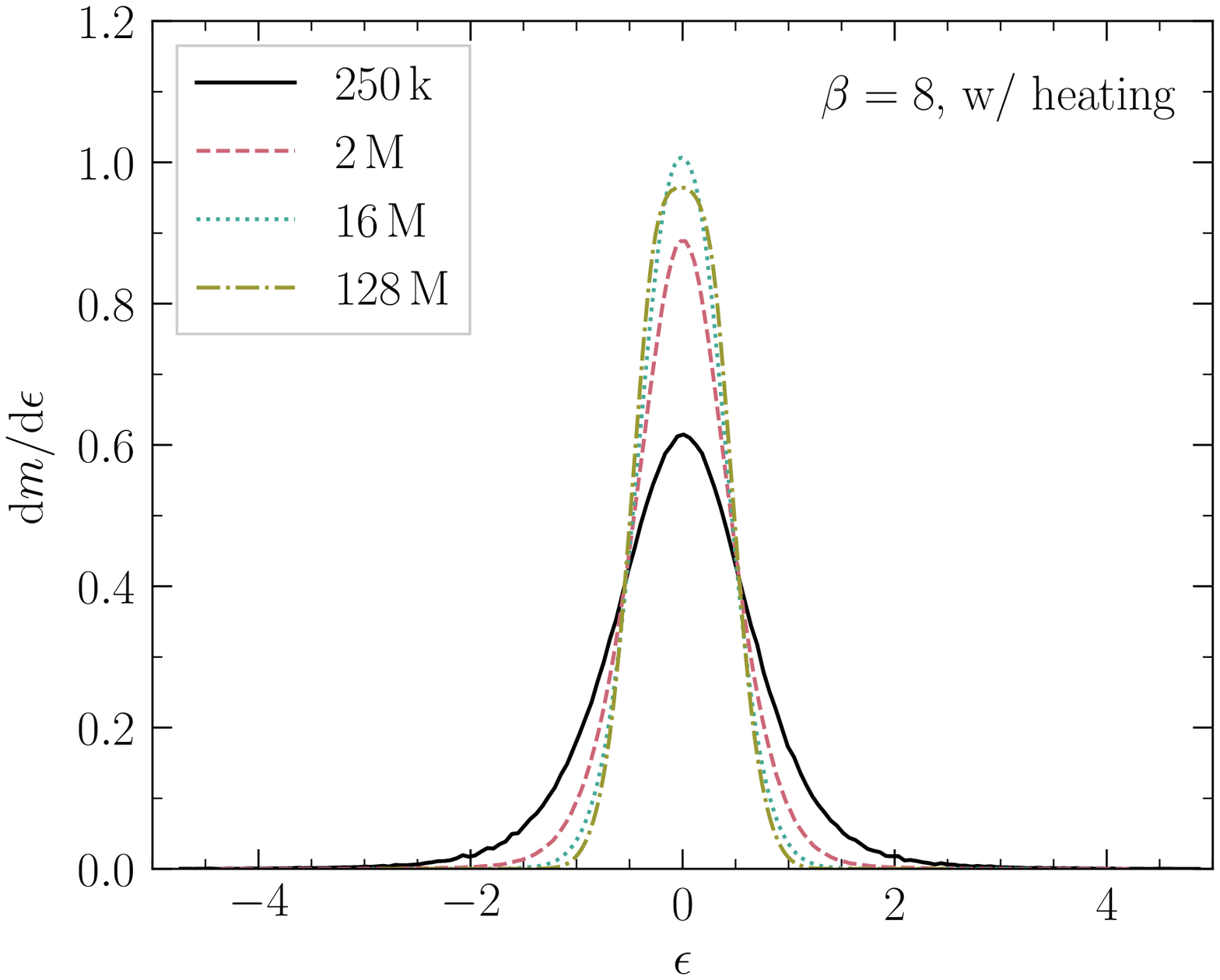}\hfill
	\includegraphics[width=0.28\textwidth]{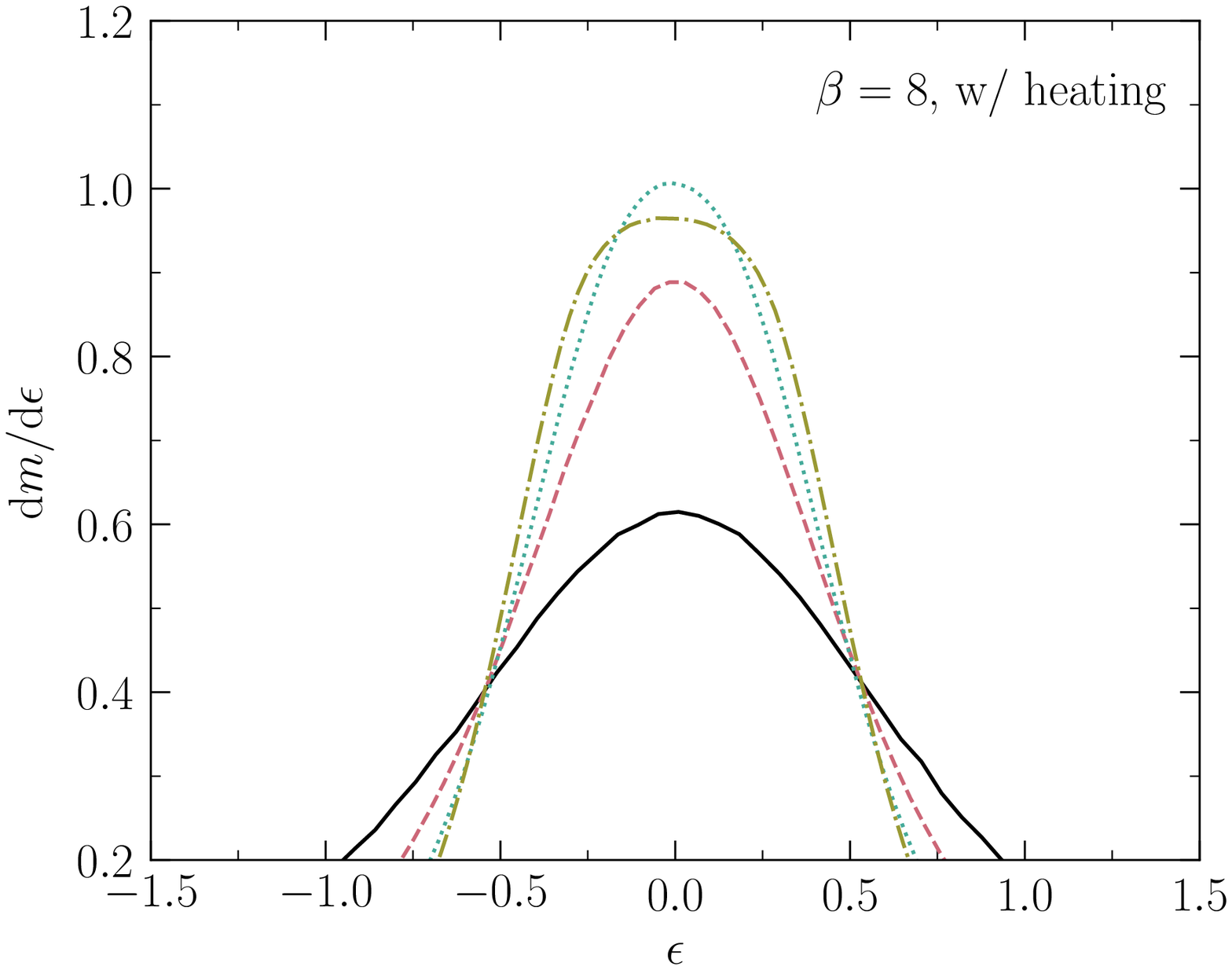}\hfill
	\includegraphics[width=0.28\textwidth]{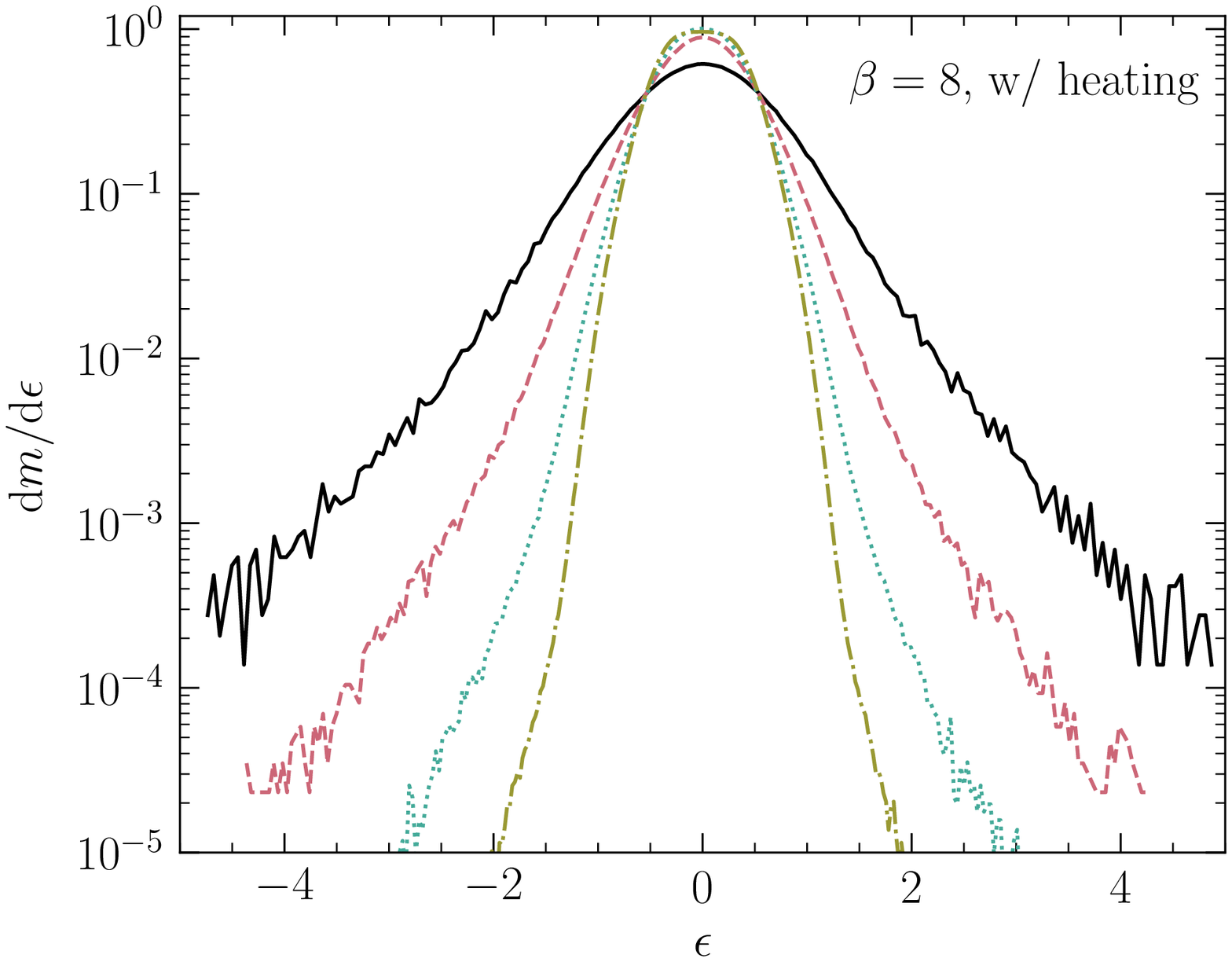}
	\includegraphics[width=0.28\textwidth]{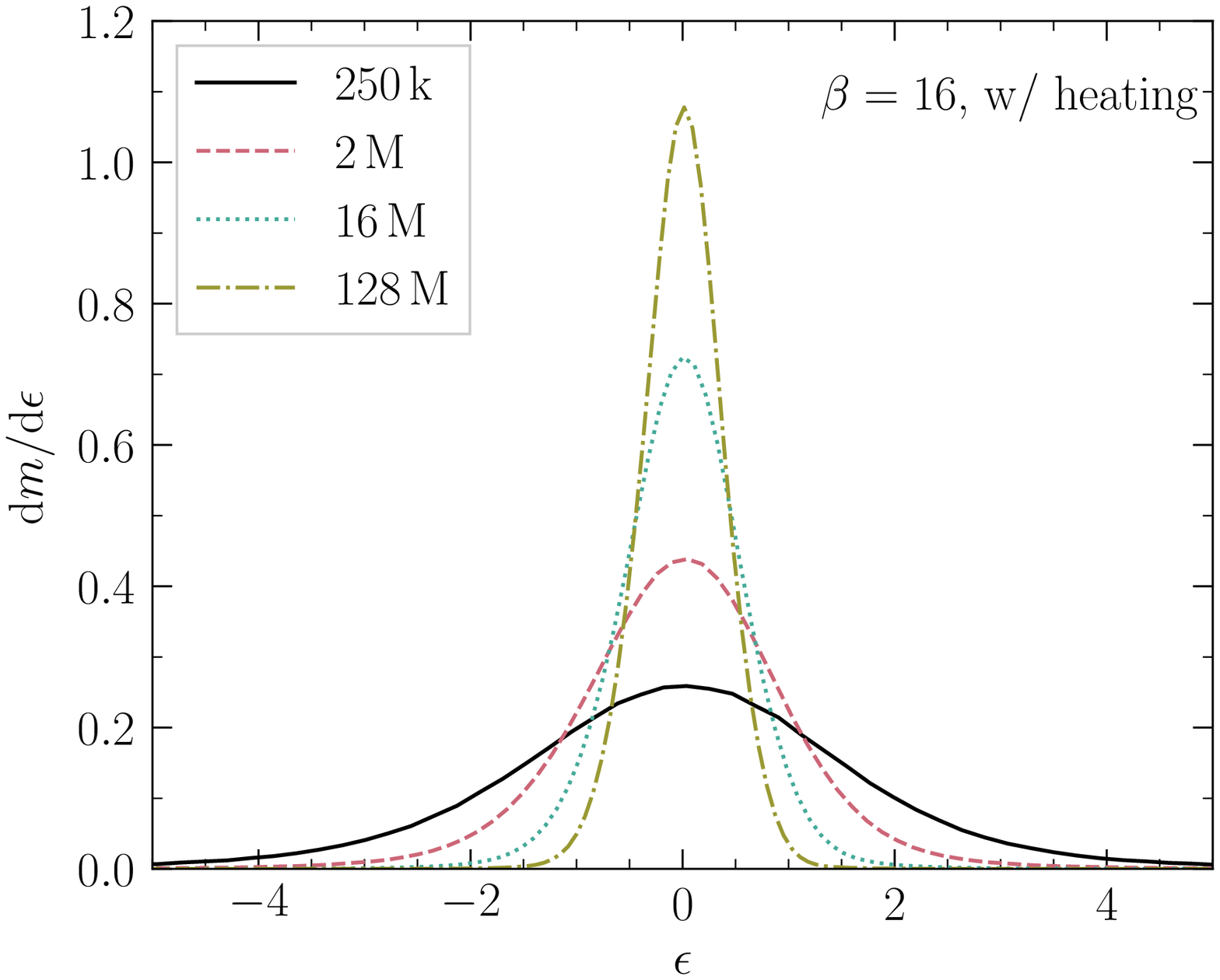}\hfill
	\includegraphics[width=0.28\textwidth]{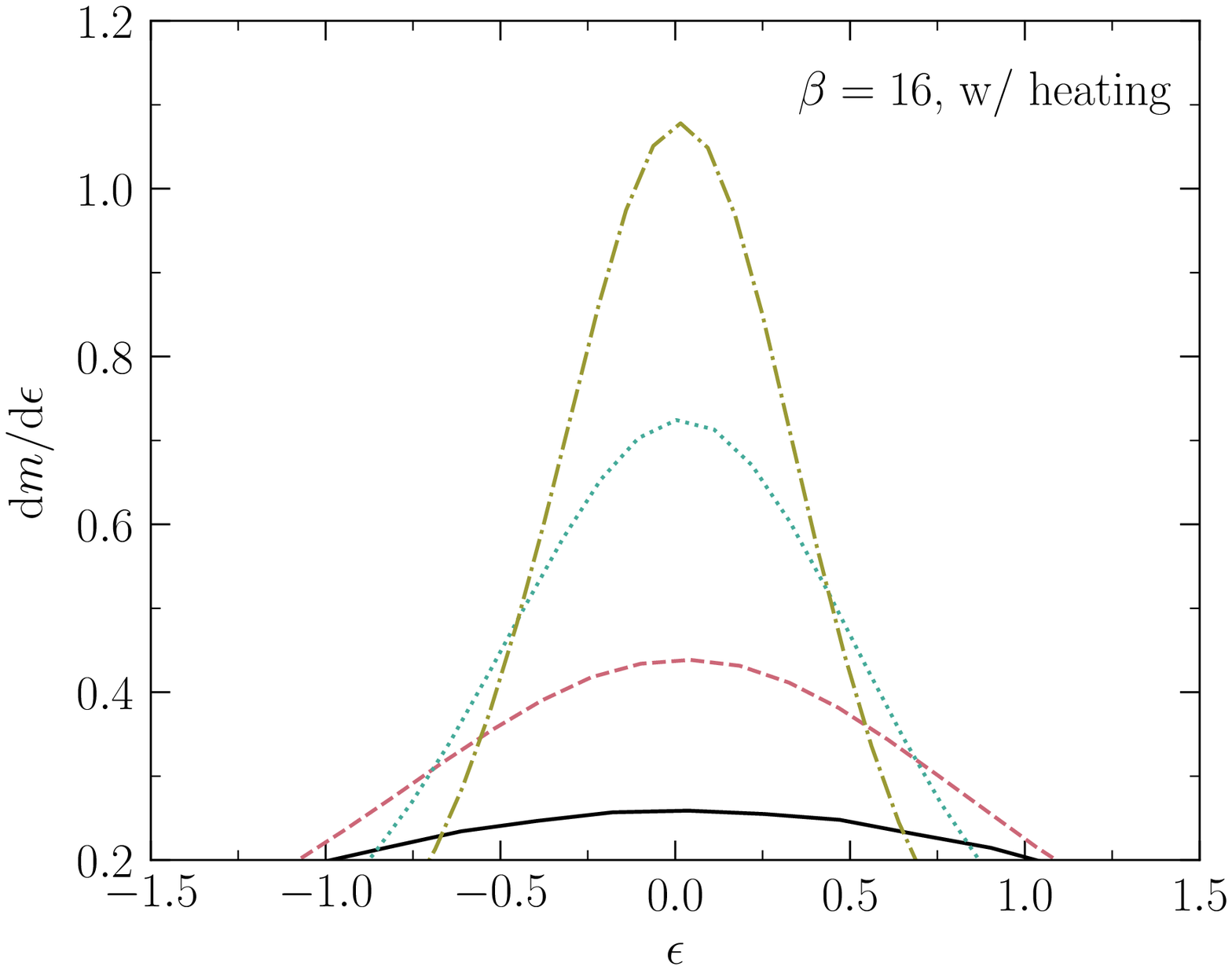}\hfill
	\includegraphics[width=0.28\textwidth]{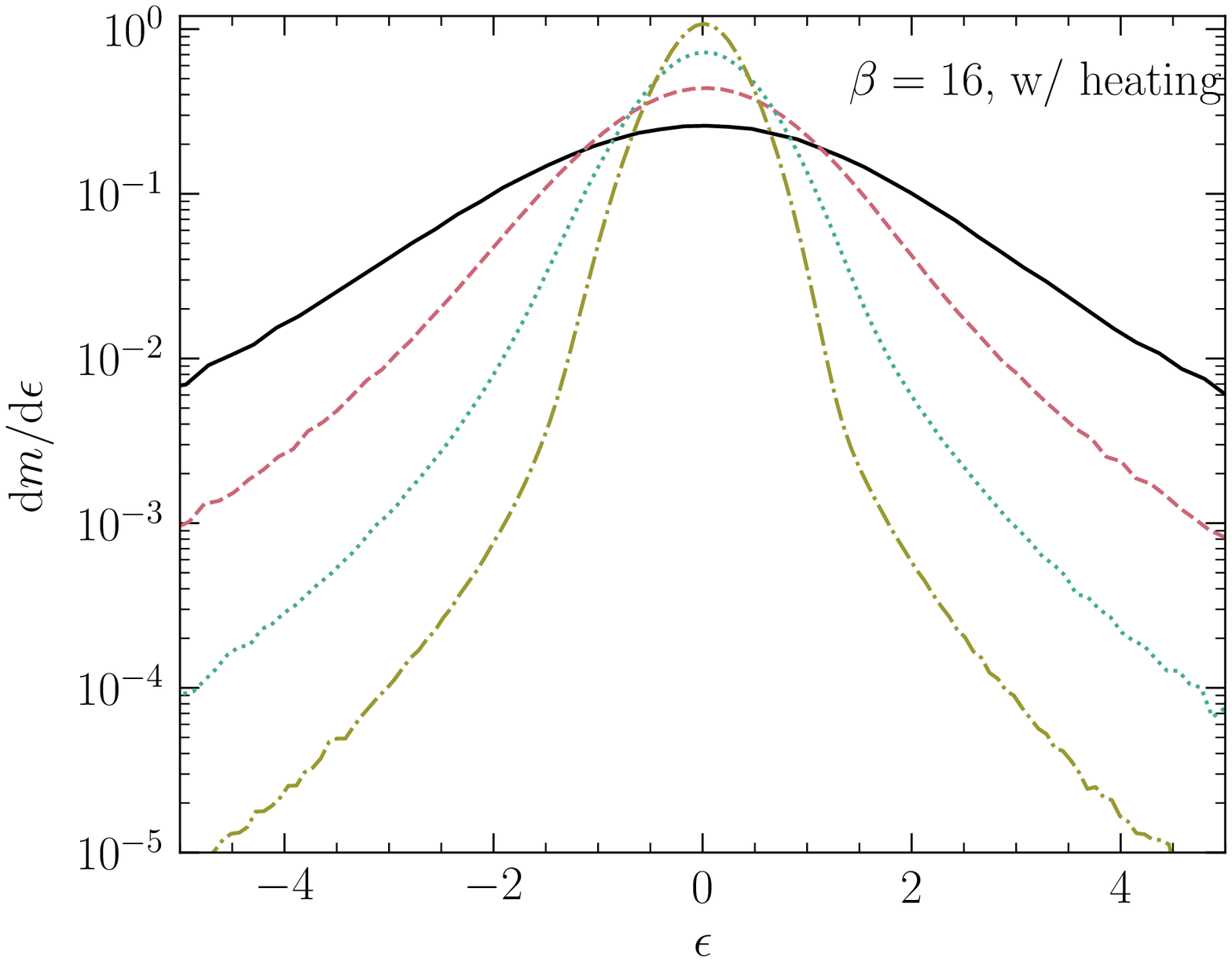}
	\caption{This figure follows the same format as Fig.~\ref{fig2}, but the data plotted here are for the simulations in which shock heating of the gas has been included.}
	\label{fig3}
\end{figure*}

So far we have explored the debris energy distribution in simulations that enforce a polytropic equation of state where $P=K\rho^\gamma$ and $\gamma=5/3$ and $K$ is a global constant determined by hydrostatic equilibrium of the initial star. In Fig.~\ref{fig3} we show the set of plots that are analogous to those presented in Fig.~\ref{fig2}, but in this case the heating of the gas due to dissipation of kinetic energy mediated by numerical viscosity is included in the dynamics. In the limit of infinite resolution, this heating would accurately reflect the presence of shocks within the flow. However, at finite resolution there is always present some additional numerical heating, even when the velocity profile is relatively smooth (i.e., nowhere discontinuous). From Fig.~\ref{fig3} it is apparent that at low $\beta$ ($\lesssim 4$), heating via shocks has a negligible effect on the energy spread. This is consistent with the predictions of \citet{Coughlin:2021aa} who show that shocks occur for $\beta\gtrsim 3$ and when they do occur they are typically weak. At larger $\beta$ ($\gtrsim8$), the simulations show that shock heating can broaden the distribution of energy, but as the resolution is increased the degree of broadening decreases to the point at which the energy distributions are similar to the cases in which the heating is excluded (see also Figure \ref{fig6} below). This demonstrates that simulations with inadequate spatial resolution can lead to inaccurate inferences for the breadth and shape of the energy distribution. 

\begin{figure}
   \centering
   \includegraphics[width=0.475\textwidth]{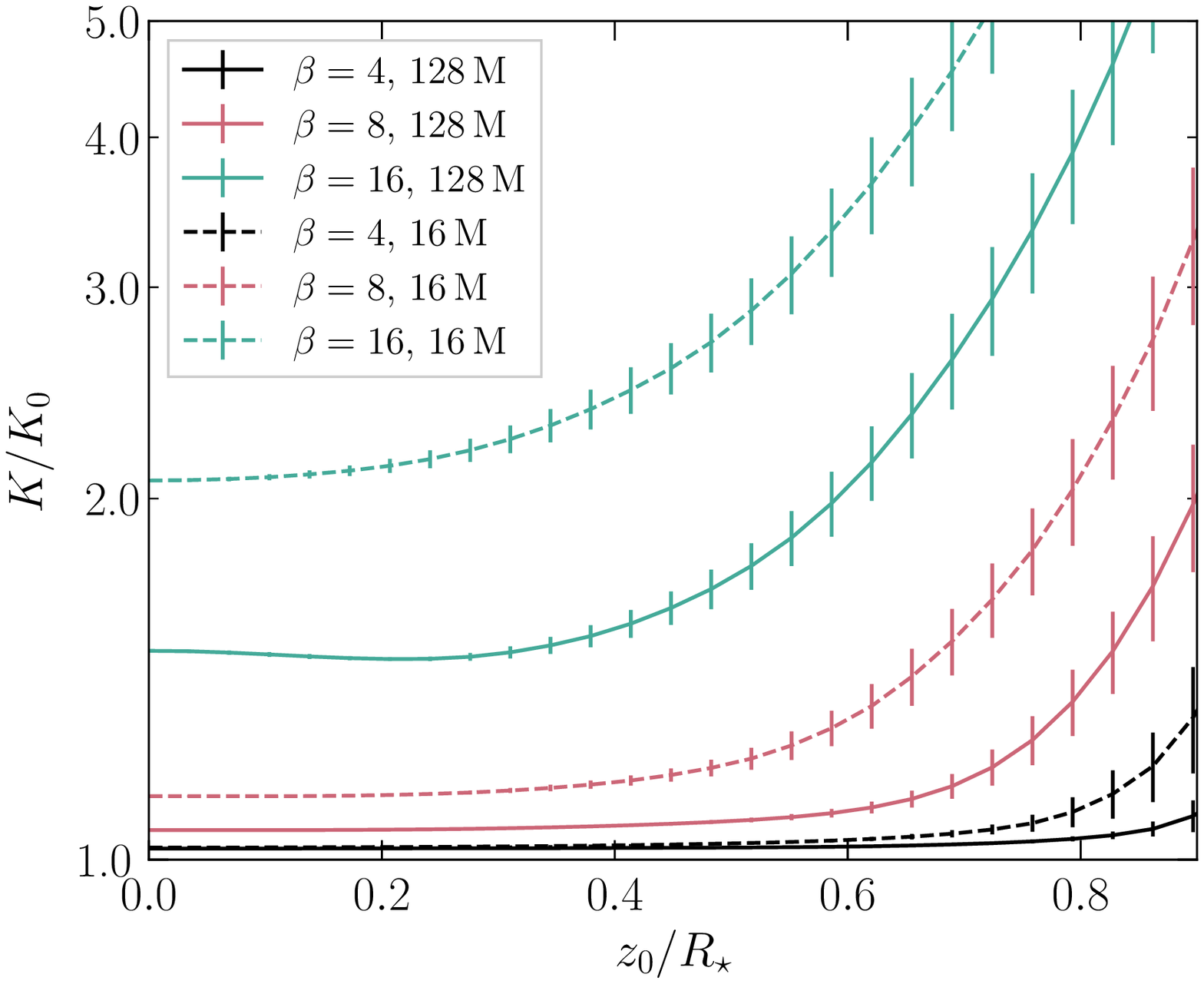} 
   \includegraphics[width=0.475\textwidth]{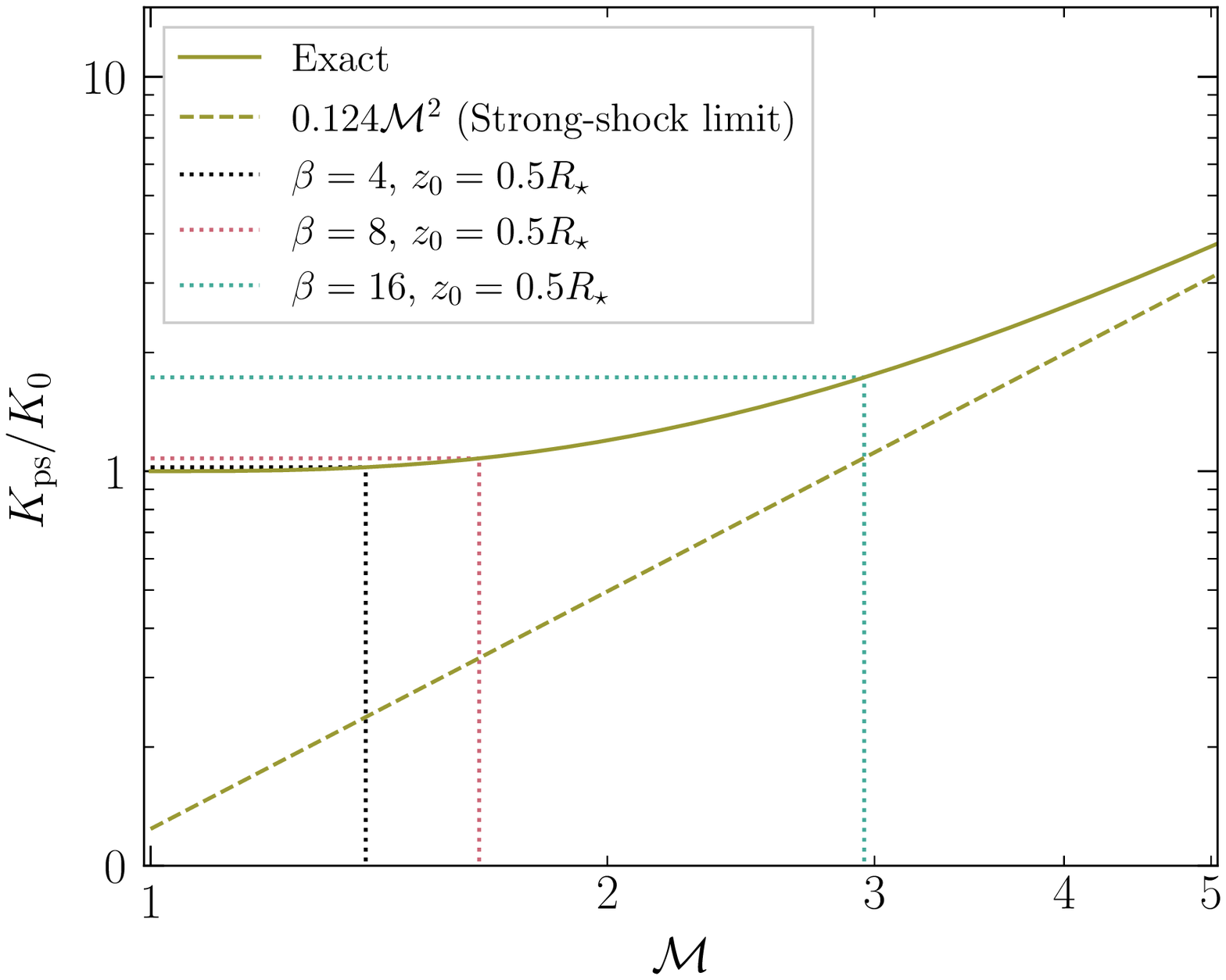} 
   \caption{Top: The entropy generated as a function of the initial height above the plane for the $\beta$ values listed in the legend and at 128M particles (solid) and 16M particles (dashed). Here $K = P/\rho^\gamma$ is the entropy function evaluated when the centre of mass has reached $5 r_{\rm t}$ from the black hole, and $K_0$ is the initial entropy function. Bottom: The relative jump in entropy function across a shock with arbitrary Mach number (i.e., including the limit when $\mathcal{M}$ is on the order of unity). The horizontal lines show the magnitude of the increase in the entropy function from the simulation measured at $z_0 = 0.5$, and the vertical lines give the corresponding Mach number of the shock that would yield that large of a change in entropy function. This figure demonstrates that any putative shocks formed during the compression of the gas must be weak.}
   \label{fig5}
\end{figure}

In the recent analysis of \citet{Coughlin:2021aa} we demonstrated that even in high-$\beta$ encounters (i.e., those with $\beta \gtrsim 10$) the shocks generated during the compression of the star were weak, with Mach numbers $\lesssim 1.5$ (see their Figure 16). We can quantitatively assess the degree to which shocks modify the thermodynamics of the compressing gas -- from the standpoint of the numerical simulations presented here -- by analyzing the entropy generated as a function of initial height above the plane. Almost all of the entropy generation occurs during the pericentre passage of the star. We therefore analyse the snapshot corresponding to when the star has receded to $5r_{\rm t}$ from the black hole, corresponding to a time of $\approx 2.4-2.9$\,hr. In terms of the cylindrical distance, $s_0$, and vertical distance measured out of the orbital plane, $z_0$, both measured from the centre of mass of the star at the start of the simulation, we restrict our analysis to particles with $s_0 < 0.03R_\star$. We then bin the particles with height above the plane from $z_0 = 0$ to $z_0 = 0.9R_\star$ with a bin width of $\delta z_0 = 0.03R_\star$. We then take the average of the entropy function $K=P/\rho^\gamma$ within each bin, and the standard deviation of the values within each bin as a measure of the error associated with the binning procedure. The top panel of Fig.~\ref{fig5} shows the ratio of the entropy function of the gas post-pericentre (at a distance of $5r_{\rm t}$) to its initial value as a function of initial height above the plane, where the black curve is for $\beta = 4$, the red curve is for $\beta = 8$, and the blue curve is for $\beta = 16$, all at 128M particles. The dashed curves are for the same values of $\beta$ but with 16M particles, i.e., at reduced resolution compared to the solid curves. The overall result to be inferred from this figure is that the entropy generation is globally small for all values of $\beta$, and that there is a slight positive correlation between initial height above the plane and the amount of entropy generated for $z_0 \lesssim 0.6$ (except for $z_0 \lesssim 0.3$ for $\beta=16$, where the entropy generated decreases with increasing $z_0$). We caution, however, that there are noticeable differences between the results of the simulations as we go from 16M to 128M particles, and hence these trends may not be representative of the solution in the limit of infinite resolution. Instead, we infer that the relative entropy change at 128M particles is an upper limit (for a given value of $z_0$) on the amount of entropy generated. 

For a shock with Mach number $\mathcal{M}$, the post-shock density and pressure are given by
\begin{equation}
\rho_{\rm ps} = \frac{\gamma+1}{\gamma-1}\left(1 +\frac{2}{\gamma-1}\mathcal{M}^{-2}\right)^{-1}\rho\,,
\end{equation}
and 
\begin{equation}
P_{\rm ps} = \frac{2}{\gamma+1}\left\{1-\frac{\gamma-1}{2\gamma}\mathcal{M}^{-2}\right\}\rho\left(v_{\rm sh}-v\right)^2\,,
\end{equation}
respectively. Note that these equations are appropriate for arbitrary Mach numbers (i.e., they do not adopt the strong-shock limit in which the Mach number is $\gg 1$). From these equations we can calculate the ratio of the post- and pre-shock entropy function (recalling that $K_{\rm ps} = P_{\rm ps}/\rho_{\rm ps}^\gamma$, the sound speed is given by $c_{\rm s}^2 = \gamma P/\rho$, and the Mach number is $\mathcal{M} = \left(v_{\rm sh}-v\right)/c_{\rm s}$), as
\begin{multline}
\frac{K_{\rm ps}}{K_0} = \frac{2\gamma}{\gamma+1}\left(\frac{\gamma-1}{\gamma+1}\right)^{\gamma} \\ 
\times \left(1-\frac{\gamma-1}{2\gamma}\mathcal{M}^{-2}\right)\left(1+\frac{2}{\gamma-1}\mathcal{M}^{-2}\right)^{\gamma}\mathcal{M}^2. \label{Ksh}
\end{multline}
Here $K_{\rm ps}$ is the post-shock entropy function and $K_0$ is the pre-shock entropy function of the ambient gas\footnote{Recall that the entropy is related to the entropy function via $S \propto \ln K$.}. In the bottom panel of Figure \ref{fig5} we plot this ratio (equation~\ref{Ksh}) as a function of the Mach number. The horizontal/vertical lines in the bottom panel of Fig.~\ref{fig5} correspond to the values of the entropy function at an initial height of $z_0 = 0.5R_{\star}$, and the vertical lines delimit the Mach number that a putative shock\footnote{Note that we are not necessarily indicating that a shock has passed through the gas at this height, even though the model in \citet{Coughlin:2021aa} predicts that there is one for $\beta \gtrsim 3$; we are merely investigating the Mach number that a shock would need to have had in order to produce the amount of entropy generation seen in the simulations.} must have had in order to generate that amount of entropy. This figure, in conjunction with the results of the numerical simulations, therefore shows that the Mach number of any shock potentially generated during the compression of the star for $\beta = 4$, 8, and 16 must have had a Mach number no greater than 1.4, 1.6 and 2.9 respectively. Again, we argue that these are upper limits owing to the fact that the simulations have not yet converged in their measurement of this quantity, as the top panel of this figure demonstrates; artificial heating due to numerical dissipation (i.e., the finite numerical viscosity) would, and likely does, artificially inflate the amount of entropy generation that occurs during these deep-$\beta$ encounters. Nonetheless, these results demonstrate that any shocks produced during the compression of the star -- even in high-$\beta$ encounters -- must be weak. 

\subsubsection{Time-dependence of the energy distribution}
\label{sec:time}
For partial disruptions (not simulated here; see \citealt{Nixon:2021ab}) \cite{Coughlin:2019aa} point out that the time dependence of the gravitational potential from the surviving stellar core means that the orbital energy for each gas parcel in the stream is not a conserved quantity, and thus measuring the Keplerian orbital energy with respect to the black hole at any time is not a reliable indicator of the fallback rate. However, for full disruptions in which there is no surviving core it is generally accepted that the fallback rate may be predicted using the Keplerian orbital energies of each gas parcel in the debris stream provided that the energies are measured at a sufficiently late time. This is the standard method for calculating the fallback rate from TDE simulations, and indeed is still used by some authors for the partial disruption case as well. 

\begin{figure*}
	\centering
	\includegraphics[width=0.33\textwidth]{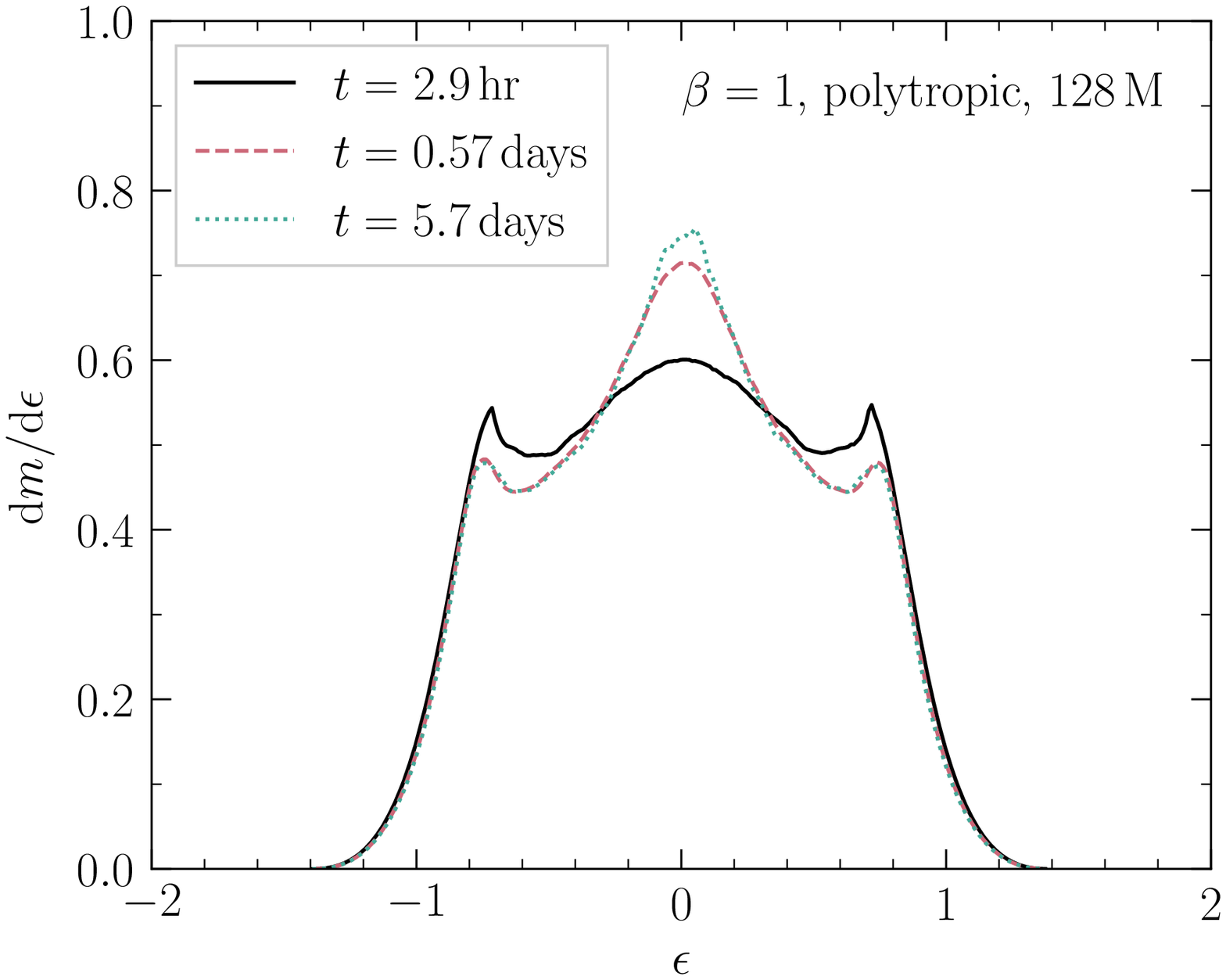}\hfill
	\includegraphics[width=0.33\textwidth]{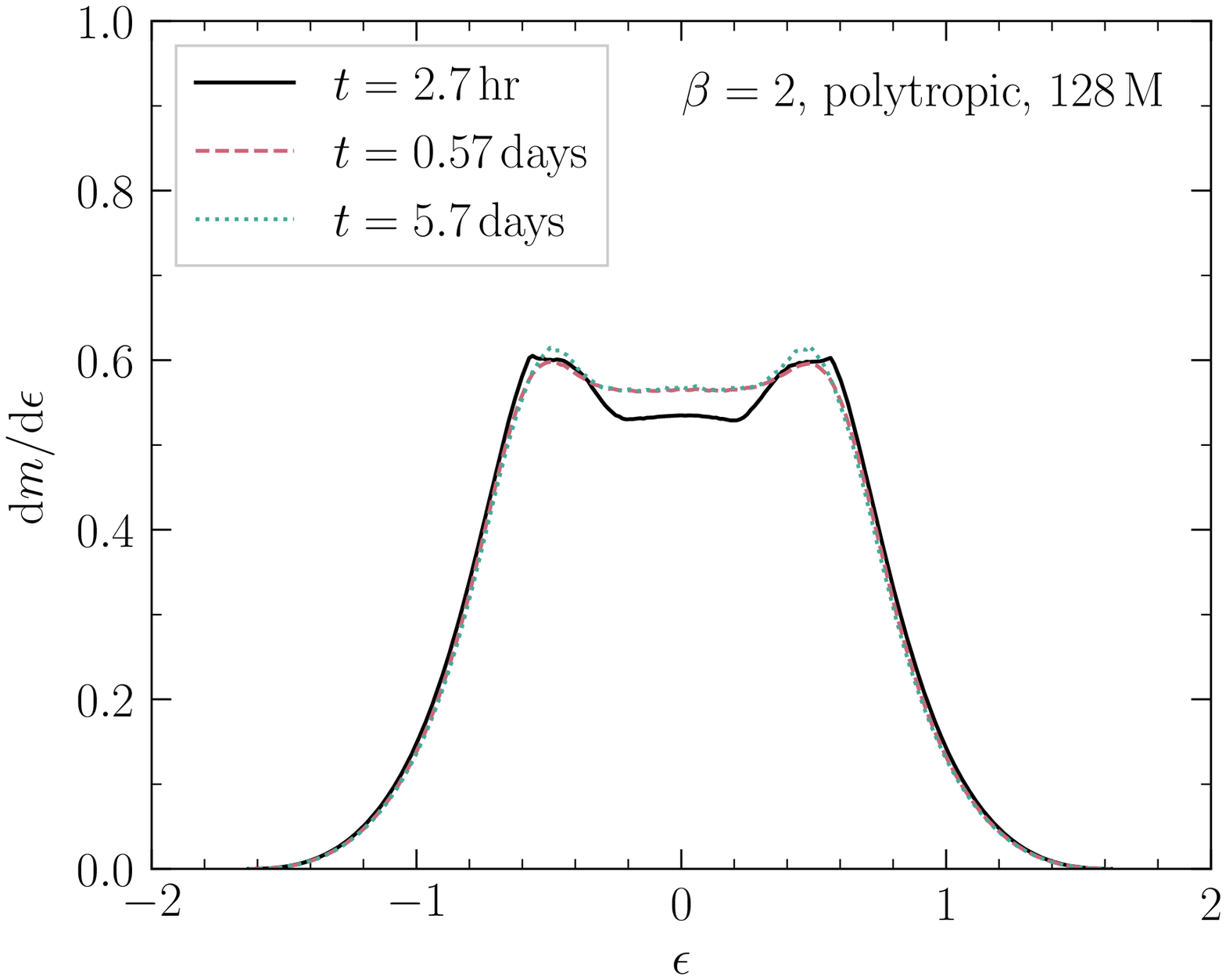}\hfill
	\includegraphics[width=0.33\textwidth]{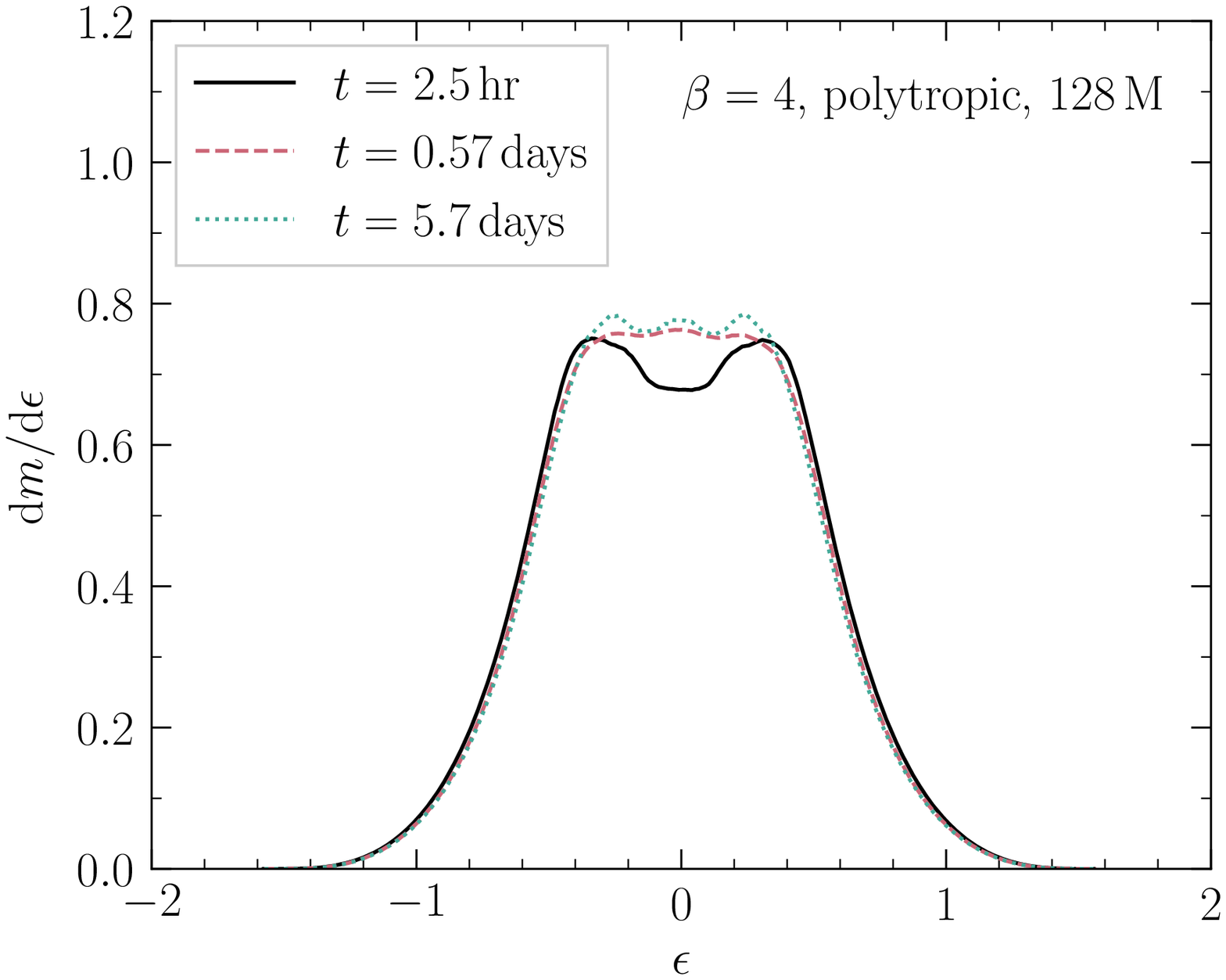}
	\includegraphics[width=0.33\textwidth]{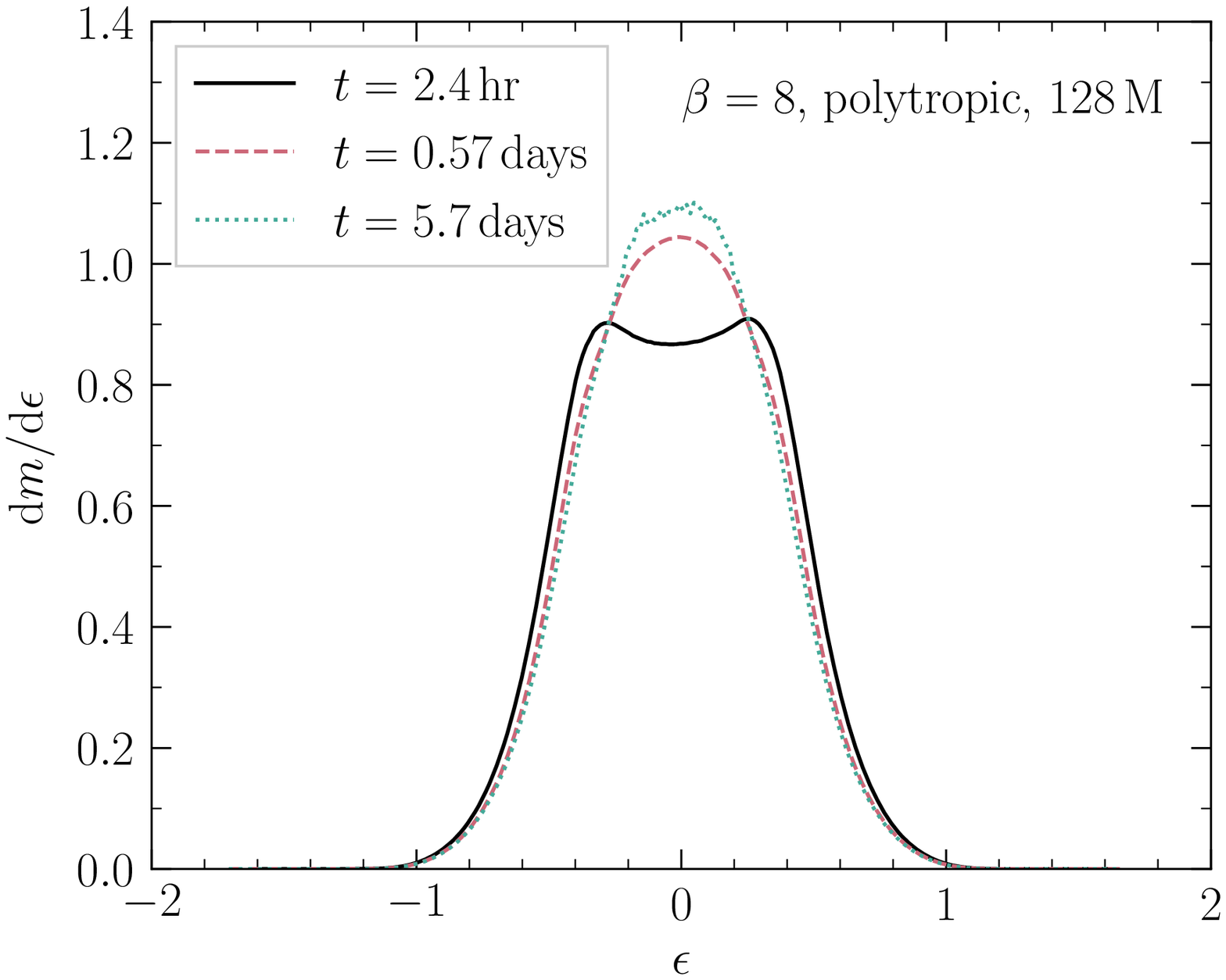}\hspace{0.35in}
	\includegraphics[width=0.33\textwidth]{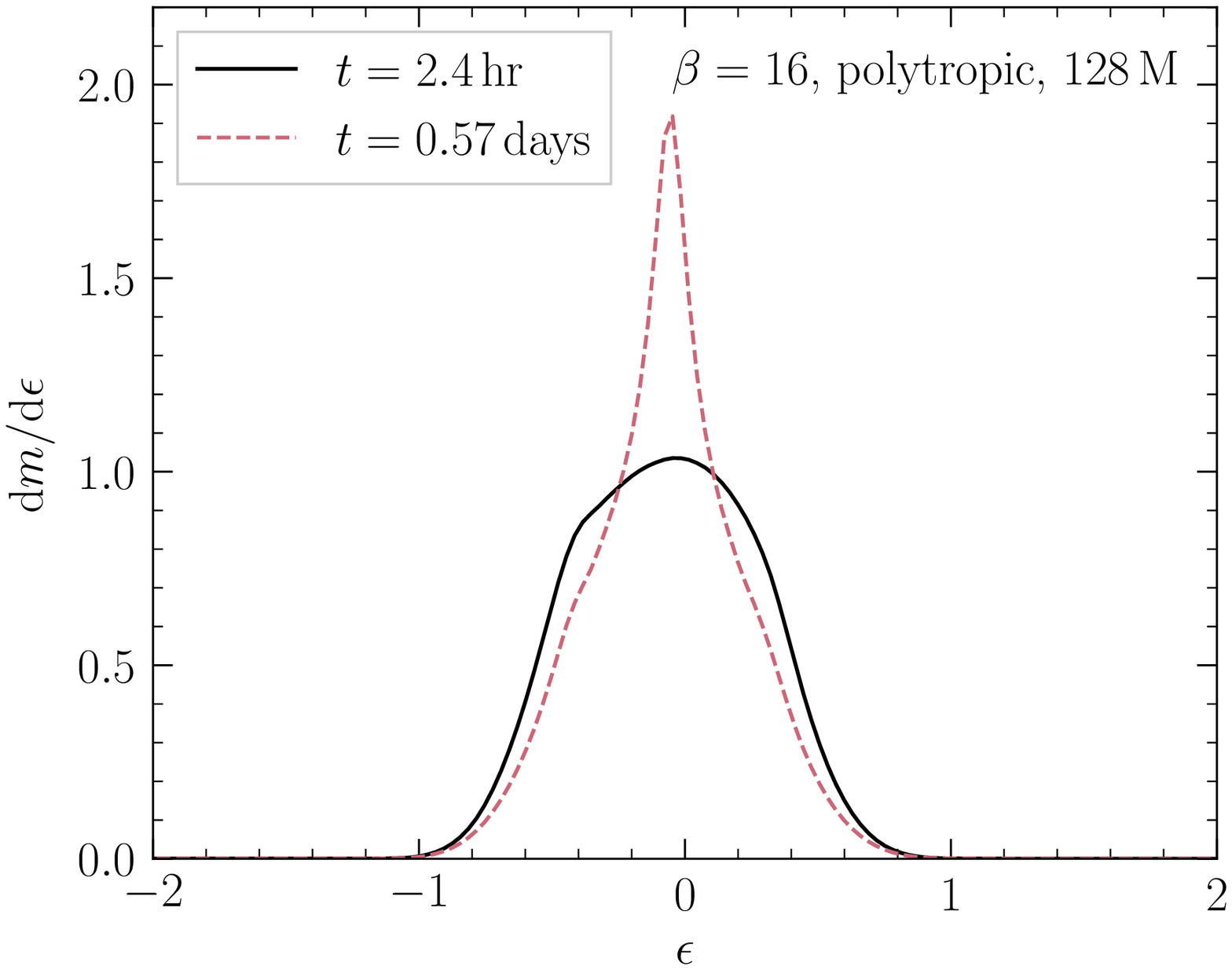}
	\caption{Energy distributions for the simulations with $N_{\rm p} = 128$M and a polytropic equation of state for different $\beta$ values. In each panel the value of $\beta$ is given at the top of the panel, and the different lines correspond to different times post pericentre passage. The solid line corresponds to the time when the zero-energy orbit has receded to a distance of $5r_{\rm t}$ from the black hole. The dashed line corresponds to a time 0.57\,d post-pericentre and the dotted line to a time of 5.7\,d post-pericentre. For $\beta=16$ we do not show the energy distribution at the latest time as matter has already started returning to the black hole in this case. In each case the energy distribution evolves significantly going from a few hours to approximately half a day. However, there is also continued evolution of the energy distribution over later times due to the influence of self-gravity which acts to re-distribute the debris along the debris stream.}
	\label{fig4}
\end{figure*}

In Fig.~\ref{fig4} we plot the Keplerian energy distributions for the stellar debris in our simulations at several times. Here we show only the highest resolution simulation (corresponding to $128$M particles) with a polytropic equation of state, for each $\beta$ value. The earliest time shown corresponds to when the zero-energy orbit reaches a radius of $5r_{\rm t}$ from the black hole (black line), and two later times are shown corresponding to 0.57\,days and 5.7\,days post pericentre. From these plots it is clear that the energy distribution is still evolving, particularly for the high-density gas where ${\rm d}m/{\rm d}\epsilon$ is large. This evolution is driven by self-gravity in the debris stream, which leads to a re-distribution of mass along the stream. In the $\beta=1$ and $\beta=8$ cases the data on the plot appear to show increased noise at later times; this is due to the onset of gravitational instability in the high-density regions of the stream, which is resolved in our numerical simulations but appears as noise in the more coarsely binned energy distributions. As the stream densities are typically highest around $\epsilon\approx 0$, there is a tendency for mass to accumulate there over time. This is particularly evident for $\beta=1$ and $\beta=8$, but can also be seen in the `shoulders' located away from $\epsilon=0$ for $\beta=2$ and $\beta=4$. The fact that the debris stream remains self-gravitating after the disruption has been shown by \cite{Coughlin:2015aa} and \cite{Coughlin:2016ab,Coughlin:2016aa} for $\beta=1$, and by \cite{Steinberg:2019aa} and \cite{Coughlin:2020ac} for $\beta \gg 1$. In some cases the differences in the energy distributions at late times may not strongly affect the forward predictions of the fallback rates; for example, in the $\beta=1$ case the differences are small for energies that are not close to $\epsilon=0$. However, it is also clear that in some cases these differences lead to inaccurate estimates of the fallback rate. Therefore we suggest that if accurate fallback rates are required, they should be obtained from numerical simulations that follow the debris orbits---including the gas hydrodynamics and self-gravity---until they return to the vicinity of the black hole \citep[see, e.g.,][]{Coughlin:2015aa}.

\section{Discussion}
\label{sec:discussion}
\subsection{Physical origin of the energy spread}
The frozen-in model posits that the energy spread imparted to the gas is established when the star crosses the tidal radius, and hence predicts that $\Delta E$ is independent of $\beta$. To our knowledge, this was first understood and demonstrated by \cite{Lacy:1982aa}, and this point has since been revisited by other authors \citep[e.g.,][]{Stone:2013aa}. Other authors have assumed that the conditions that regulate the evolution of the tidally disrupted debris are set at the pericenter distance of the centre of mass, in which case the energy spread would be proportional to $\beta^2$ \citep[e.g.][]{Evans:1989aa,Ulmer:1999aa,Lodato:2009aa,Strubbe:2009aa}. Our results demonstrate that, in the limit of large-$\beta$, the energy spread is nearly independent of $\beta$ (as seen qualitatively in Figure \ref{fig1}), but that there is a small, but noticeable, {\it inverted} dependence for $\beta \gtrsim 1$ -- the energy spread decreases as $\beta$ increases. This inverted dependence was also found by \citet{Steinberg:2019aa}, who attributed it to the influence of self-gravity just outside the tidal sphere of the black hole and the increased amount of time spent by the star near the tidal radius for low-$\beta$ encounters (compared to high-$\beta$ encounters). 

We note here two additional effects that further complicate the precise dependence of the energy distribution on $\beta$, the first of which is that the central density of a star is larger than its average density, and hence the effective tidal radius to which one would have to go if one equated the central density to the black hole density (i.e., $\rho_{\bullet} \sim M_{\bullet}/r^3$), and hence completely destroy the star\footnote{Meaning that the tidal density of the black hole is larger than the largest stellar density.}, is a factor of $(\rho_{\rm c}/\rho_{\star})^{1/3}$ smaller than the fiducial tidal radius. For a $\gamma = 5/3$ polytrope, $\rho_{\rm c}/\rho_{\star} \simeq 5.99$, which leads to a factor of $\simeq 1.82$ reduction in the distance of the center of mass, or $\beta \simeq 1.82$. For this value of $\beta$, one would expect the usual frozen-in argument to apply: the tidal force of the black hole has done enough work to completely unbind the star, and the energies of the fluid elements, and the corresponding spread in the energy, should be frozen-in and should no longer evolve with time. Investigating Figure 2 of \citet{Steinberg:2019aa}, this distance ($\simeq 55$ in the units on their horizontal axis) roughly corresponds to the time at which their $\Delta E(90\%)$ asymptotes to a constant during the stellar ingress.

The second effect arises from the fact that the material is strongly vertically compressed by the tidal force as it nears the pericenter of its orbit. \citet{Carter:1983aa}, as well as \cite{Stone:2013aa}, argued that the equivalence of the gas pressure and the ram pressure leads to an increase in the central density by a factor of $\propto \beta^3$, though the model of \citet{Coughlin:2021aa} (alongside the evidence of the simulations described in detail here) demonstrated that the increase in the density is not this strong (see their Figure 17). Nonetheless, by $\beta \simeq 8$, the central density at the point of peak compression is a factor of $\sim 30$ increased over the original central density of the star. Arguing analogously to the previous paragraph, therefore, the density near the center of the star is well-above the self-gravitating limit at the point that maximum compression is reached (the criterion for this being that $\rho / \rho_{\bullet} \gtrsim 1$; \citealt{Coughlin:2016ab}), and the sudden increase in the importance of self-gravity at this point results in a corresponding {drop} in the energy spread; this drop is apparent near the pericenter distance of the star in Figure 2 of \citet{Steinberg:2019aa}. The black hole can then continue to exert a tidal potential on the material -- the self-gravity of which has now been refreshed -- as it recedes from pericenter; this is also apparent from the increase in $\Delta E$ as seen in Figure 2 of \citet{Steinberg:2019aa}. That $\Delta E$ continues to rise above the value at which it plateaus on the stellar ingress is because the material is now even more self-gravitating than it was initially (i.e., the peak density achieved shortly after pericenter is larger than the original central stellar density by a factor of $\sim 25$ for $\beta = 7$; see Figure 17 of \citealt{Coughlin:2021aa}). 

Given the complexity of these effects, and additional effects such as the spin of the star prior to the encounter with the black hole \citep{Golightly:2019aa} and the details of the stellar structure \citep{Golightly:2019ab}, it is currently necessary to use numerical simulations (rather than a comparatively simple analytical model) to determine the energy distribution accurately. An analytical framework, that is more physically accurate than the frozen-in model, with which to test the numerical simulations would be highly desirable. Furthermore, we note that while the energy spread is an important characteristic of the distribution of the debris and that this spread appears relatively insensitive to $\beta$, different values of $\beta$ can lead to different functional forms for the energy distribution (see Fig.~\ref{fig1}). It is this functional form that is important in determining the shape of the fallback curve. However, as we have seen the energy distribution can evolve significantly over time (see Fig.~\ref{fig4}), and thus the most accurate approach to determining the fallback rate is to directly measure the return of debris to the vicinity of the black hole \citep{Coughlin:2015aa}.

\subsection{Validity and importance of assumptions}
\label{sec:assumptions}
The set of numerical simulations that we have carried out here make a number of simplifying and/or model specific assumptions that warrant further, but limited, discussion that we provide here. One such assumption is that the star being disrupted is a $\gamma = 5/3$ polytrope modeled to match the Sun in its bulk properties, i.e., its mass and radius. There are multiple motivating factors for this assumption, including that this type of star has been simulated by a large number of authors (and, indeed, appears to be the canonical stellar-type employed in TDE simulations; e.g., \citealt{Bicknell:1983aa, Evans:1989aa, Lodato:2009aa, Guillochon:2013aa, Coughlin:2015aa}) and therefore provides us with the ability to compare our results to others in the field. Another reason for this choice is that the majority of stars are low in mass, and hence a $\gamma = 5/3$ polytrope -- accurately descriptive of low-mass stars owing to their fully convective nature -- is representative of the typical encounter to be actualized in nature, though the Sun (or at least the radiative interior) is better-modeled by a $\Gamma = 4/3$ polytrope with a $\gamma = 5/3$ equation of state (where $p \propto \rho^{\Gamma}$ as concerns the density and pressure profiles of the star; e.g., \citealt{Hansen:2004aa}). Simulating the deep disruption of a $\Gamma = 4/3$ polytrope (with $\gamma = 5/3$) would likely result in our better resolving the compression of the star, as the higher central density naturally enforces a smaller smoothing length, though the results in \citet{Coughlin:2021aa} suggest that one must go to much larger values of $\beta$ to achieve the same level of relative compression as compared to a $\gamma = 5/3$ polytrope. We leave a detailed study of high-$\beta$ encounters of $\Gamma = 4/3$ polytropes with $\gamma = 5/3$, or more general ``real stars,'' \citep[cf.][]{Golightly:2019ab} to a future investigation. 

The second assumption we made was to model the gravitational field of the supermassive black hole as that of a Newtonian point mass. The inclusion of general relativistic effects, and a more accurate description of the gravitational field in the immediate vicinity of the black hole compared to the purely Newtonian field employed here, becomes necessary for modeling the dynamics once the $\beta$ of the encounter becomes modestly large (we would argue $\gtrsim 10$) for a $10^6M_{\odot}$ supermassive black hole, because the ratio $r_{\rm t}/R_{\rm G} \simeq 47$ in this case (where $R_{\rm G} = GM_{\bullet}/c^2$). Therefore, once $\beta \gtrsim 10$, the direct capture of a nontrivial fraction of the stellar debris becomes possible, and the large periapsis advance angle means that direct collisions may occur almost immediately after pericentre is reached \citep{Darbha:2019aa}. The tidal effects of the supermassive black hole are also enhanced when general relativistic effects are taken into account (e.g., \citealt{Stone:2019aa, Kesden:2012aa}), and hence the tidal effects of what might be considered a relatively small-$\beta$ encounter in Newtonian gravity may behave effectively as a larger-$\beta$ encounter (again, in Newtonian gravity) when general relativistic effects are included. These considerations become increasingly important as the black hole mass increases for smaller $\beta$, but the neglect of relativistic gravity -- even in very high-$\beta$ encounters -- is even more justified as the black hole mass is reduced. We leave an analysis of the effects of relativity on the behaviour of (e.g.) the maximum-achievable density within the compressing star to future work (see also \citealt{Gafton:2019aa} for an investigation of highly relativistic TDEs).

\subsection{The maximum density and temperature}
\label{sec:max}
The density and temperature of the star, and in particular their maximum values achieved as the star nears and recedes from its pericenter, have received a great deal of attention since the suggestion of \citet[][rebuked by \citealt{Bicknell:1983aa}]{Carter:1983aa} that the compression induced by the tidal field may lead to induced nuclear detonation in the core of the star. Various authors have found agreement or disagreement with the predictions that result from the assumption of adiabatic compression and the $\sim$ equivalence of the gas pressure and the infalling ram pressure (assuming freefall collapse for the latter) at the point of maximum compression; these predictions are that $\rho_{\rm max}/\rho_{\rm c} = 0.22\beta^3$ and $T_{\rm max}/T_{\rm c} = 0.37\beta^2$ (assuming an adiabatic index of $\gamma = 5/3$; \citealt{Luminet:1986aa}), where $\rho_{\rm max}$ and $T_{\rm max}$ are the maximum density and temperature and $\rho_{\rm c}$ and $T_{\rm c}$ are their values at the center of the original star, respectively.

In \citet{Coughlin:2021aa} we developed a model that permitted a deeper understanding of the evolution of these fluid quantities, and in that same paper we compared our predictions to and found excellent agreement with the simulations described here (see Figure 17 of that paper in particular). Rather than reiterate the findings in full, we summarize them here by noting that the central density and temperature of the star effectively never conform to their $\propto \beta^3$ and $\propto \beta^2$ scalings, because at low $\beta$ ($\beta \lesssim 10$) the pressure gradient within the star becomes dynamically significant well before the gas pressure equals the ram pressure of the fluid. The importance of the pressure gradient at these low values of $\beta$ results in a much more gradual increase in $\rho_{\rm max}$ and $T_{\rm max}$ as $\beta$ increases, with their functional forms not being particularly well-matched by power-laws (again, see Figure 17 of \citealt{Coughlin:2021aa}). At large $\beta$ ($\beta \gtrsim 10$), the formation of a weak shock (with Mach number $\lesssim 1.5$; see the discussion in Section \ref{sec:shocks} above and \citealt{Coughlin:2021aa}) prematurely halts the adiabatic increase in the density when the shock reaches the midplane, in agreement with the arguments of \citet{Bicknell:1983aa}, and results in the much shallower scalings $\rho_{\rm max}/\rho_{\rm c} \propto \beta^{1.62}$ and $T_{\rm max}/T_{\rm c} \propto \beta^{1.12}$. We refer the interested reader to \citet{Coughlin:2021aa} for more details.

\subsection{The importance of shock heating}
\label{sec:shocks2}
We adopted two thermodynamic prescriptions for the evolution of the gas in our simulations, one with shock heating included in the gas-energy equation (the ``w/ heating'' set), and the other in which the entropy is forced to be a global constant (termed ``polytropic''). The latter prescription implies that the shock is, in a way, radiative, and that energy is not conserved as we move across the shock and the dissipated kinetic energy is instead lost from the system. In this scenario, the post-shock pressure is reduced compared to its value in the case where shock heating is included, and the increase in the density is correspondingly larger.

In reality, since the shock that we are analyzing here forms deep within the interior of the star, any additional heat generated must diffuse outward on a timescale that is comparable to the Kelvin-Helmholtz timescale of the original star. For typical stellar progenitors, this timescale is millions of years, and hence the energy lost by the star (via any process) is extremely tiny over the duration of the tidal encounter (which is roughly one dynamical time of the star by construction). The most realistic scenario is therefore likely given by the one in which any heat generated by the shock is retained by the system.

We emphasize, however, that the simulations that maintain the effects of shock heating on the gas may actually be less representative of the true solution -- the one achieved by going to infinite resolution -- as the amount of entropy generation may be overestimated due to numerical effects, which is apparent as the resolution increases in Figure \ref{fig5}. In fact, if the model in \citet{Coughlin:2021aa} is representative of the true level of entropy generation at at least the order of magnitude level, then -- perhaps contrary to expectations given that the star is highly compressed and the conditions are rather extreme for high $\beta$ (e.g., the increase in the central density is by a factor of $\sim 110$ for $\beta = 16$) -- the numerical simulations \emph{without shock heating} are likely more representative of reality than those that do. In other words, enforcing that the entropy be unchanged during the tidal compression is more accurate given the overproduction of the entropy from spurious numerical heating with shock-heating enabled (which we infer given the comparison between 16M and 128M particles in Figure \ref{fig5}). 

\begin{figure*}
	\centering
	\includegraphics[width=0.33\textwidth]{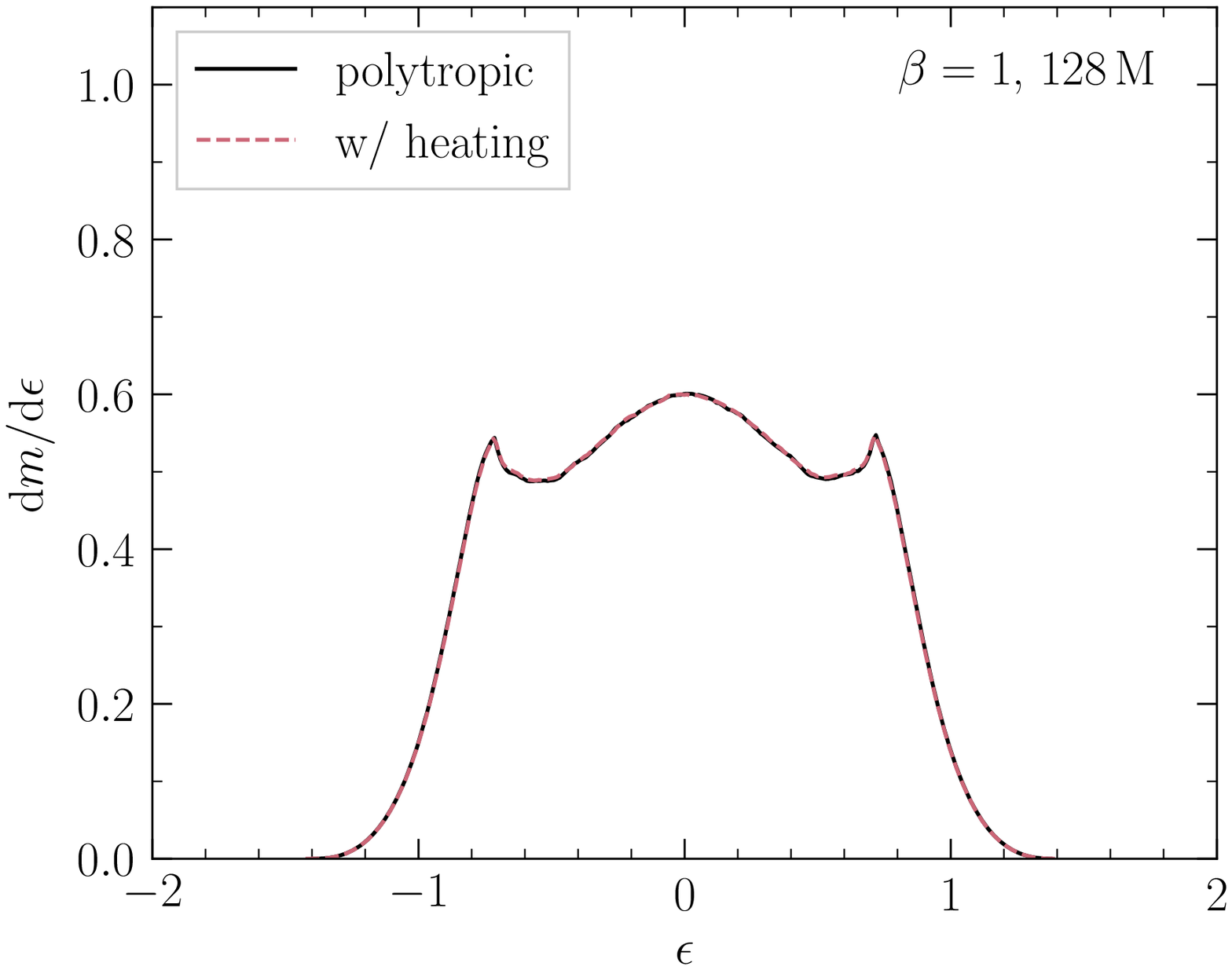}\hfill
	\includegraphics[width=0.33\textwidth]{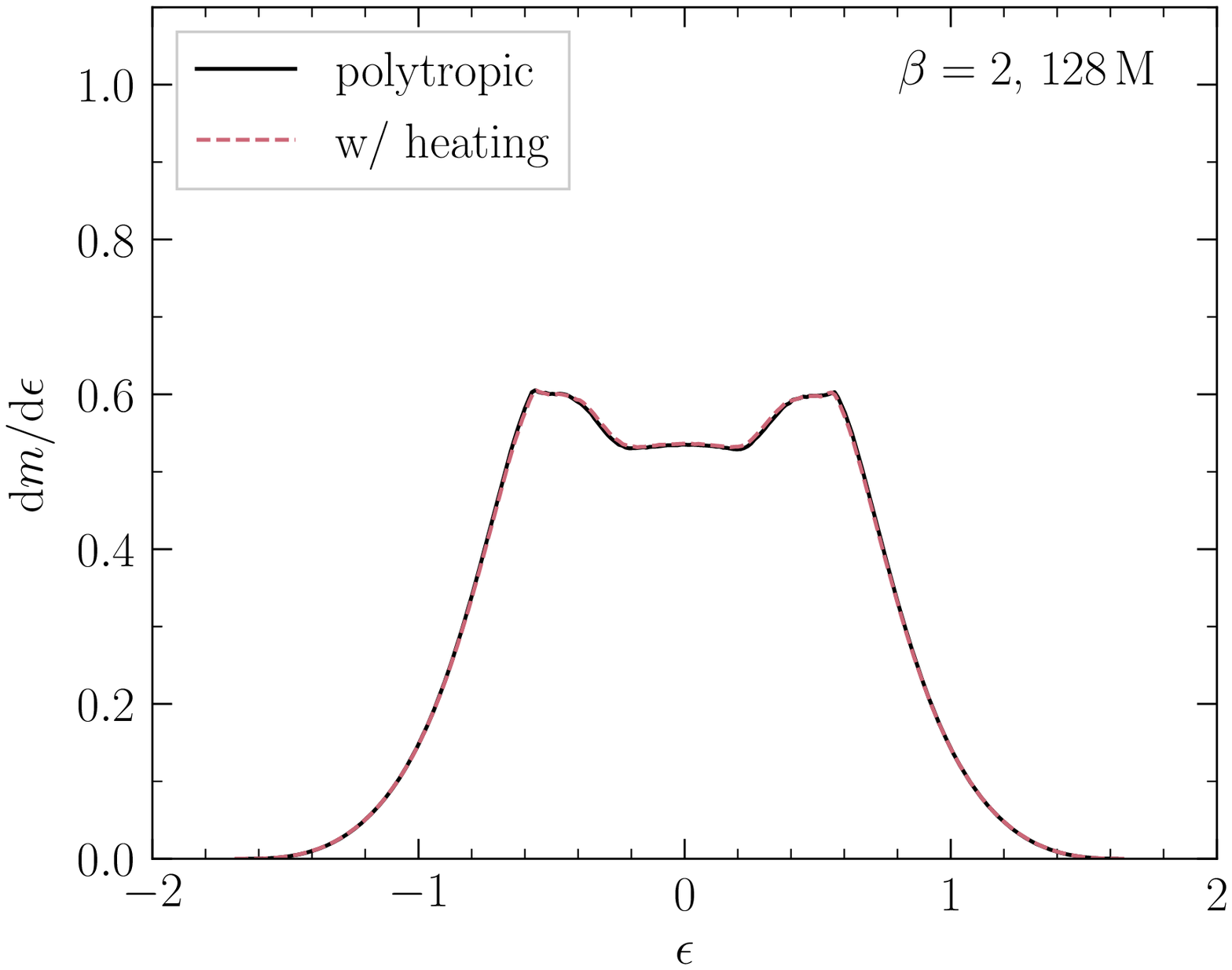}\hfill
	\includegraphics[width=0.33\textwidth]{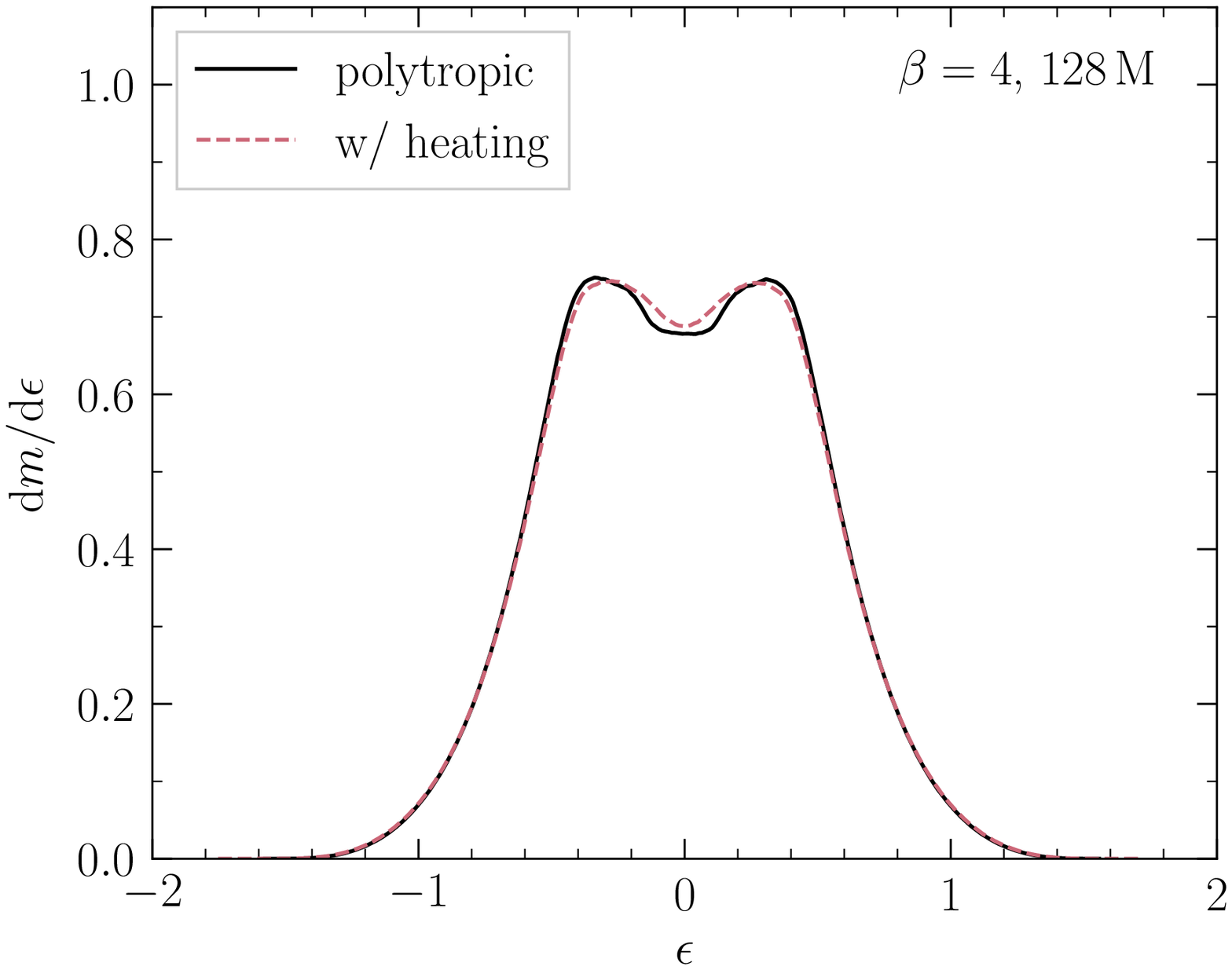}
	\includegraphics[width=0.33\textwidth]{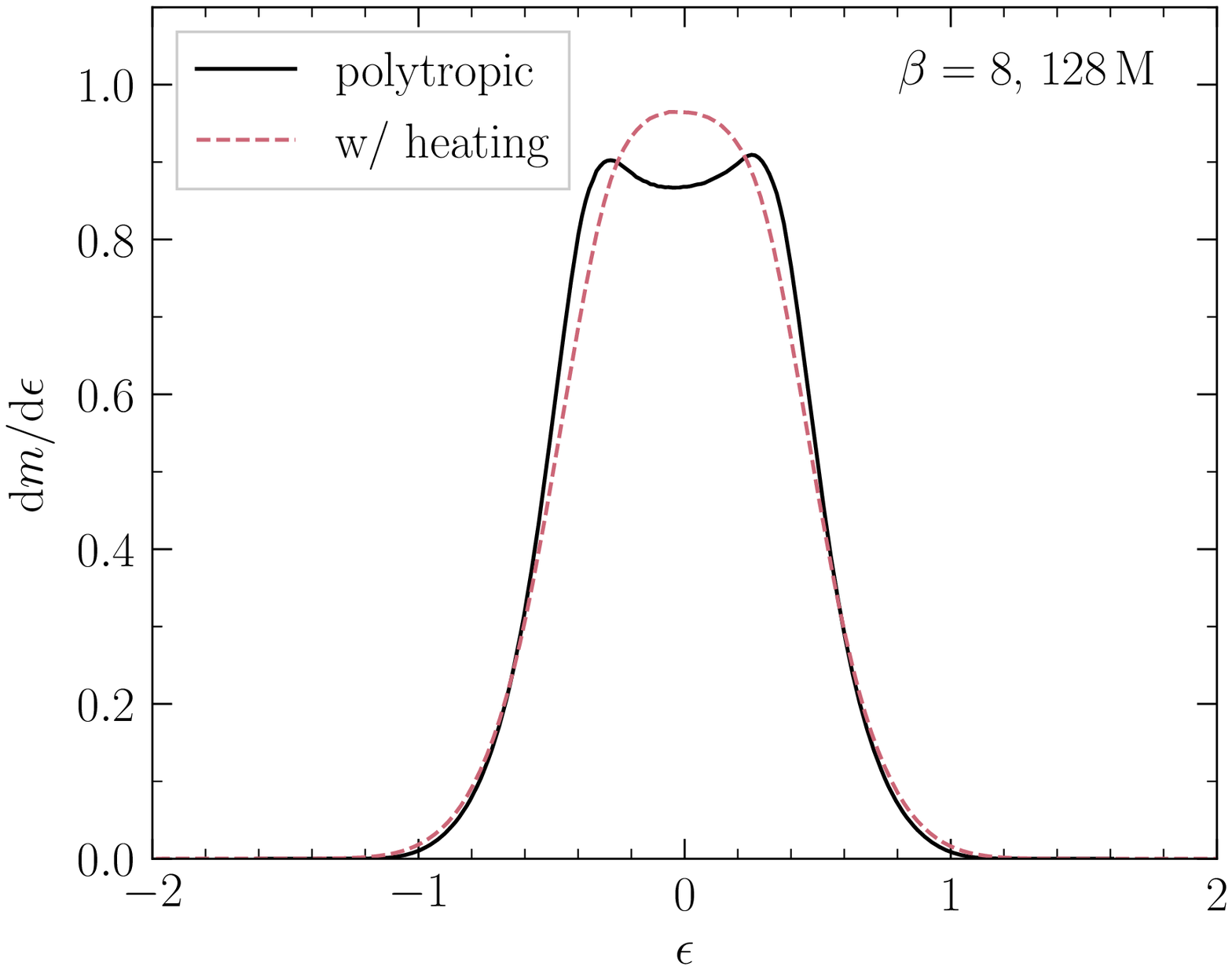}\hspace{0.35in}
	\includegraphics[width=0.33\textwidth]{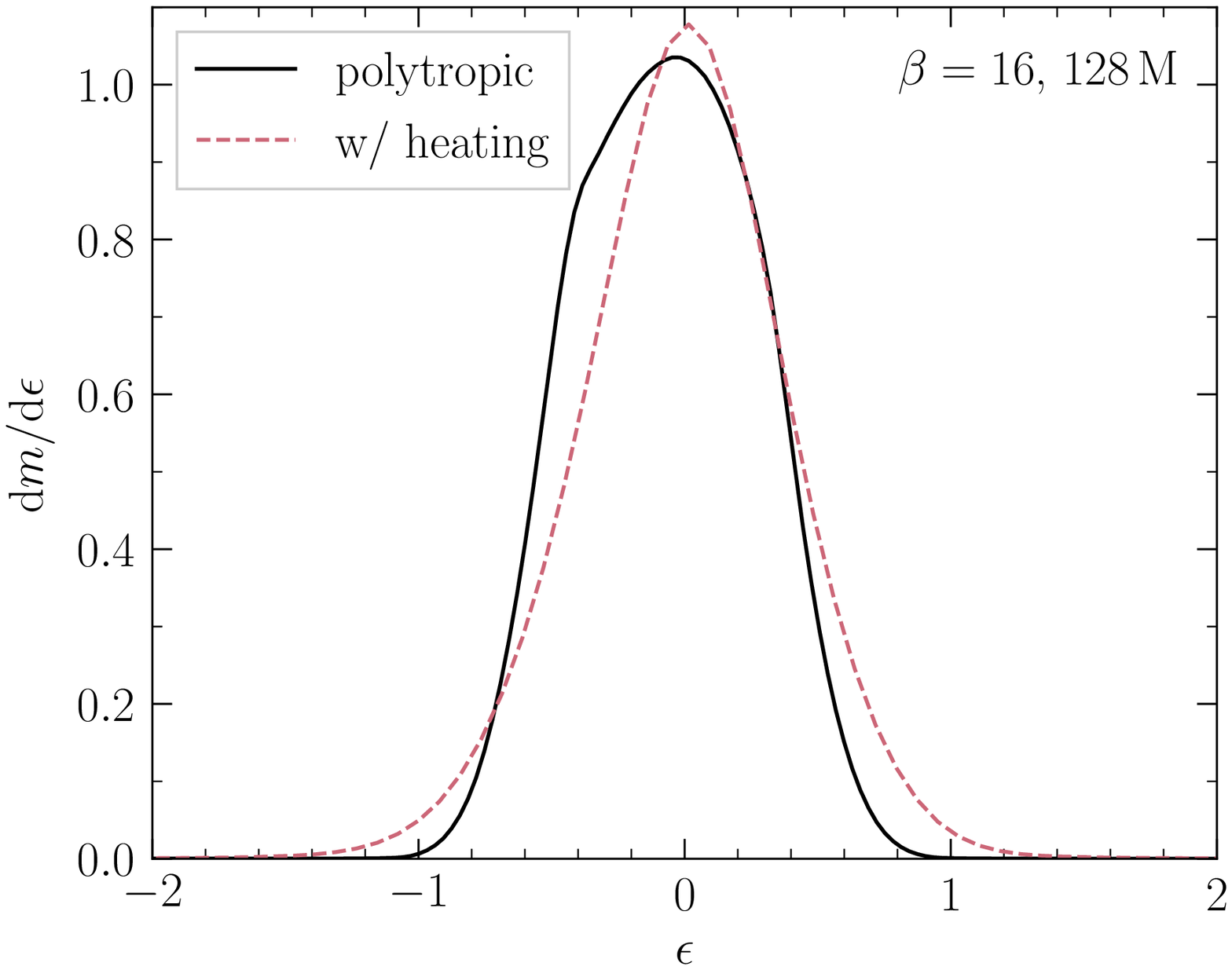}
	\caption{Energy distributions for the simulations with $N_{\rm p} = 128$M, comparing the effect of excluding (polytropic) and including heating of the debris (w/ heating). In each panel the value of $\beta$ is given at the top right of the panel. These energy distributions are calculated when the centre-of-mass orbit has receded to a distance of $5r_{\rm t}$ (e.g. Figs~\ref{fig2} \& \ref{fig3}). The heating of the gas is mediated in the simulations by the numerical viscosity, which captures the heating due to shocks. \cite{Coughlin:2021aa} show that shocks should occur for $\beta\gtrsim 3$, and that the shocks are typically weak, suggesting that their effect on the entropy of the gas is mild. This figure shows that the effect of shock heating on the energy distribution is typically small.}
	\label{fig6}
\end{figure*}

In addition, assuming that the model of \citet{Coughlin:2021aa} predicts at least roughly the right amount of entropy generated through the formation of shocks, our two prescriptions for the effects of shock heating should effectively converge to the same one in the limit of infinite resolution. Figure \ref{fig6} gives a comparison of the energy spread measured when the centre-of-mass orbit has receded to a distance of $5r_{\rm t}$ for $\beta = 1$, 2, 4, 8, and 16, all at 128M particles and including and ignoring the effects of shock heating. We see that the two curves are nearly indistinguishable for $\beta = 1$, 2, and 4, there are small differences for $\beta = 8$ (including the structure around the peak, and the high-energy wings present in the simulation that includes shock heating), and for $\beta = 16$ the differences between the simulations are more apparent. In particular, it is worth noting that the significant asymmetry in the energy distribution for the $\beta=16$ polytropic case is largely removed when shock heating is included. This occurs as the additional heating provides pressure support against the self-gravity that is responsible for generating the asymmetry as the debris moves through the point of maximal compression (as discussed in Section~\ref{sec:energy}). However, we reiterate that the heating in this case is significantly stronger than predicted by the analytical model of \citet{Coughlin:2021aa} and is not yet converged in the simulations (see Fig.~\ref{fig5}), and thus we expect that simulations performed at even higher resolution with shock heating included would show less entropy generation and would maintain the asymmetry.

\subsection{Nuclear energy generation}
\label{sec:nuclear}
We conclude this section by saying that we have ignored any energy liberated by nuclear fusion as the gas is intensely heated and compressed during the tidal encounter. This possibility was, as we noted above, the original motivation of \citet{Carter:1983aa} when investigating the nature of deeply plunging tidal disruption events. 

We already noted in \citet{Coughlin:2021aa} that the likelihood of producing substantial energy through the triple-$\alpha$ process is small until $\beta \simeq 10$, as one needs a central temperature of $\gtrsim 10^8$ K in order to generate substantial rates of energy production through this process. Even then, the amount of time spent at these temperatures is extremely small, which \citet{Bicknell:1983aa} argued would further reduce the possibility of significant energy production through Helium fusion. However, it seems possible that by pushing the temperature of the star to $\sim 10^{8}$ K and igniting the process in earnest, one could initiate a runaway process, but the situation is very dissimilar from the one presumably encountered during the standard picture of runaway Carbon burning that occurs in type Ia supernovae. In particular, the gas is highly dynamic and out of dynamical equilibrium, and is rapidly decompressed shortly after the burning starts. It is therefore unclear whether or not the initial burst of energy generation could maintain sufficiently high temperatures in the expanding debris to allow the process to continue.

Finally, there is also the possibility of igniting shell burning in the outer layers of the star by nuclear reactions that require less extreme temperatures and densities than the triple-$\alpha$ process. Whether or not nuclear burning occurs (and, if so, where in the star) requires further analysis, and determining the answer to this question is outside the scope of the present paper.

\section{Summary and Conclusions}
\label{sec:conclusions}
We have presented the results of numerical simulations of tidal disruption events in which we varied the pericentre of the stellar orbit, the spatial resolution, and performed the simulations with and without the effects of shock heating included in the gas dynamics. From these results we are able to draw the following conclusions:
\begin{enumerate}
\item The width of the energy distribution for the bulk of the stellar debris is essentially independent of the pericentre distance for tidal disruption events where the star passes close to or within the tidal sphere, and is close to the canonical spread of $\Delta E = GM_\bullet R_\star/r_{\rm t}^2$ predicted by \cite{Lacy:1982aa}. Close inspection of the energy distributions shows a weak inverse dependence of the width on $\beta$ for modest $\beta$ values which we attribute to a combination of the increased time spent near the tidal radius for these events compared to their higher $\beta$ counterparts and the effects of self-gravity. However, for the $\beta$ values we have simulated the overall shape of the energy distribution changes significantly for different $\beta$ values.
\item We have demonstrated that for $\beta \lesssim 2$ the energy distributions are accurate when the event is modelled with modest numbers of particles. For $\beta \lesssim 8$, the simulations show a good level of convergence between 16\,M and 128\,M particles, indicating that $\sim 10$\,M particles is sufficient in this case. For $\beta = 16$ there remain differences in the energy distributions between 16\,M and 128\,M, although in this case the width of the distribution for the bulk of the debris agrees closely with the expected value, the maximum density and temperature achieved agree closely with the predictions of \citet[][see their Fig.~17]{Coughlin:2021aa}, and our analysis of the errors at different resolutions suggest that all of the simulations are converging appropriately. We therefore conclude that the results of the $\beta=16$ simulation performed at $128$\,M particles are acceptable. 
\item We have demonstrated that any shocks occurring as the star passes through the tidal sphere are of low-Mach number by analysing the entropy generation in the simulations that include heating of the gas. Only the simulations with $\beta \gtrsim 4$ show any indication that shocks are present. In contrast, the simulations with $\beta= 1$ and $\beta=2$ show the same energy distribution whether heating is included or not (Fig.~\ref{fig6}), suggesting that if any shocks are present they are sufficiently weak as to play essentially no role in the gas thermodynamics. For larger $\beta$ we find that the effects of shock heating can be pronounced only when the simulation resolution is inadequate. Once sufficiently high resolution is reached the energy distributions are similar whether shock heating is included in the gas dynamics or not. Analysis of the entropy generation as a function of spatial resolution suggests that the Mach numbers of any shocks present are limited to $\mathcal{M} < 1.4$ for $\beta=4$, $\mathcal{M} < 1.6$ for $\beta=8$ and $\mathcal{M} < 2.9$ for $\beta=16$. These findings are in agreement with the predictions of shock properties in TDEs made by \cite{Coughlin:2021aa}.
\item We have shown that the energy distribution continues to evolve with time as the debris recedes to distances much greater than the tidal radius (see Fig.~\ref{fig5}). This evolution is enacted by the gas self-gravity, which can redistribute mass along the debris stream; in some cases this leads to widespread fragmentation of the debris stream \citep{Coughlin:2015aa}, and in other cases can result in the formation of a single, dominant `zombie' core within the stream \citep{Nixon:2021ab}. This result suggests that care should be taken when predicting the future fallback rate of stellar debris from the energy distribution at an early time. We advocate that the safest way to measure the fallback rate accurately from a numerical simulation is to follow the return of the debris to pericentre and measure the fallback rate directly \citep{Coughlin:2015aa}.
\end{enumerate}

\begin{acknowledgements}
We are grateful to the referee for a helpful report. CJN acknowledges funding from the European Union’s Horizon 2020 research and innovation program under the Marie Sk\l{}odowska-Curie grant agreement No 823823 (Dustbusters RISE project). E.R.C. acknowledges support from the National Science Foundation through grant AST-2006684. This research used the ALICE High Performance Computing Facility at the University of Leicester. This work was performed using the DiRAC Data Intensive service at Leicester, operated by the University of Leicester IT Services, which forms part of the STFC DiRAC HPC Facility (\url{www.dirac.ac.uk}). The equipment was funded by BEIS capital funding via STFC capital grants ST/K000373/1 and ST/R002363/1 and STFC DiRAC Operations grant ST/R001014/1. DiRAC is part of the National e-Infrastructure.
\end{acknowledgements}

\bibliographystyle{aasjournal}
\bibliography{nixon}


\end{document}